 \documentclass[final,5p,times]{elsarticle}
\usepackage{graphicx}
\usepackage{amssymb}
\usepackage{amsmath}
\usepackage{breqn}
\usepackage{longtable}
\usepackage{supertabular}

\usepackage[colorlinks=true,linkcolor=black, citecolor=blue, urlcolor=blue]{hyperref}

\journal{Astronomy and Computing}

\begin{document}
\newcommand{\torus}{\textsc{torus} }
\newcommand\mnras{MNRAS}  
\newcommand\aap{A\&A}                % Astronomy and Astrophysics
\let\astap=\aap                          % alternative shortcut
\newcommand\aapr{A\&ARv}             % Astronomy and Astrophysics Review (the)
\newcommand\aaps{A\&AS}              % Astronomy and Astrophysics 
\newcommand\apj{ApJ}                 % Astrophysical Journal
\newcommand\apjl{ApJ}                % Astrophysical Journal, Letters
\newcommand\aj{AJ}                   % Astronomical Journal (the)
\let\apjlett=\apjl   
\newcommand\pasa{Publ. Astron. Soc. Australia}  % Publications of the Astronomical Society of Australia
\newcommand\pasp{PASP}               % Publications of the Astronomical Society of the Pacific
\newcommand\pasj{PASJ}               % Publications of the Astronomical Society of Japan
\newcommand\araa{ARA\&A}             % Annual Review of Astronomy and Astrophysics
\newcommand{\angstrom}{\text{\normalfont\AA}}
\newcommand\apjs{ApJS}               % Astrophysical Journal, Supplement
\let\apjsupp=\apjs                       % alternative shortcut
\newcommand\planss{Planet. Space~Sci.} % Planetary Space Science
\newcommand\nar{New A Rev.}
\begin{frontmatter}

\title{The TORUS radiation transfer code}

\author[label1]{Tim J. Harries\corref{cor1}\fnref{label3}}
\address[label1]{Department of Physics and Astronomy, University of Exeter, Stocker Road, Exeter EX4 4QL.}

%\cortext[cor1]{I am corresponding author}
%\fntext[label3]{I also want to inform about\ldots}
%\fntext[label4]{Small city}

\ead{T.J.Harries@exeter.ac.uk}
\ead[url]{http://emps.exeter.ac.uk/physics-astronomy/staff/tjharrie}

\author[label5]{Thomas J. Haworth}
\address[label5]{ Astrophysics Group, Imperial College London, Blackett Laboratory, Prince Consort Road, London SW7 2AZ, UK}

\author[label1]{David Acreman}
\author[label1]{Ahmad Ali}
\author[label1]{Tom Douglas}
%\ead{author.three@mail.com}

\begin{abstract}

We present a review of the {\sc torus} radiation transfer and hydrodynamics code. {\sc torus} uses a 1-D, 2-D or 3-D adaptive mesh refinement scheme to store and manipulate the state variables, and solves the equation of radiative transfer using Monte Carlo techniques. A  framework of microphysics modules is described, including atomic and molecular line transport in moving media, dust radiative equilibrium, photoionisation equilibrium, and time-dependent radiative transfer. These modules provide a flexible scheme for producing synthetic observations, either from analytical models or as post-processing of hydrodynamical simulations (both grid-based and Lagrangian). A hydrodynamics module is also presented, which may be used in combination with the radiation-transport modules to perform radiation-hydrodynamics simulations.  Benchmarking and validation tests of each major mode of operation are detailed, along with descriptions and performance/scaling tests of the various parallelisation schemes.  We give examples the uses of the code in the  literature, including applications to low- and high-mass star formation, cluster feedback, and stellar winds, along with an Appendix listing the refereed papers that have used {\sc torus}. 

\end{abstract}

\begin{keyword}
radiative transfer \sep hydrodynamics  \sep methods: numerical
%example \sep \LaTeX \sep template
\end{keyword}

\end{frontmatter}

% \linenumbers

\tableofcontents

\newpage

\section{Introduction}

Radiation transfer (RT) is the principal mode of energy transport in the Universe. The microphysical interactions of light with matter control the formation and evolution of planets, stars, and galaxies, and also provide the mechanisms with which we observe them. Modern RT codes can capture the complex interplay between light and matter, and provide new insights into astrophysical phenomena, from modelling the atmospheres of exoplanets to emission lines from core collapse supernovae. 

All RT codes are fundamentally attempting to solve the RT equation:
\begin{equation}
    \frac{dI_{\nu}}{d\tau_{\nu}} = S_{\nu} - I_{\nu}
    \label{eq:radeq}
\end{equation}
where $I_{\nu}$ is the specific intensity at frequency $\nu$, $\tau_{\nu}$ is the optical depth, and $S_{\nu}$ is the source function, which is the ratio of local emission and absorption  coefficients. This equation describes the attenuation or amplification of a pencil beam of light through a medium. Its complexity derives from the fact that the source function is usually a function of the radiation field, for example via the population densities of different quantum mechanical levels in an atom or molecule, or the temperature of dust grains. The calculation of the radiation field is therefore just one step in an iterative cycle that couples the state of the radiation field ($I_\nu$) to the state of the gas ($S_\nu$). The coupling to the detailed microphysical state of the gas (e.g. the temperature, excitation state, ionisation state) can be very difficult to solve \citep[e.g.][]{2005pcim.book.....T, 2006agna.book.....O} particularly because non-local effects may be important (physically very distinct regions may communicate via radiation).

Equation~\ref{eq:radeq} is posed in terms of a single beam of radiation at a single frequency. Solving for the radiation field in the {\em general case} is a formidable numerical problem. One must find a solution to a set of non-linear differential equations that  is valid over three dimensions of space, as well as direction, frequency and (potentially) time. The medium may have velocity fields, bringing  discrete states of matter into resonance with different frequencies, and if one needs to follow the polarisation state of the radiation a further three intensities must be solved.

Fortunately in many situations the problem may be simplified, for example by assuming an equilibrium state, or reducing the spatial dimensionality of the problem by assuming symmetries.
This complexity may be further reduced if  one can assume that the material is in local thermodynamic equilibrium (LTE),  thereby fixing the quantum states of the material by its temperature and reducing the source function to the Planck function. However iteration may still be necessary if computing radiative equilibrium, as the temperature of the material is coupled to the radiation field. 

So how does one calculate the radiation field? Numerical schemes are broadly divided into those that employ ray-tracing techniques and those that rely on Monte Carlo (MC) methods. 
Ray-tracing methods involve calculating the specific intensity by integrating the radiative transfer equation along a direction. This integration may be performed in a unique direction from one side of the computational domain to the other, while interpolating spatially for the opacities and emissivities. This is known as the long characteristic method, and the cost of  calculating the specific intensity for a single direction in a grid of $N$ points scales as $O(N^4)$. The short characteristic method requires a piecewise integration of the radiative transfer equation across each individual cell, with an interpolation of intensities from previously computed cells acting as a boundary condition for subsequent cells. The interpolation of the specific intensity leads to some numerical diffusion, but the computational cost of this method scales as $O(N^3)$.

Over the last few decades the Monte Carlo methods have started to replace ray-tracing algorithms for some RT problems, particularly those that involve 2- or 3-dimensions, heterogeneity of material, or anisotropic scattering. These methods randomly sample probability distributions that, when sufficiently sampled, yield a reproducible converged result. We will discuss this in much more detail throughout this paper, but typically a photon source energy output is divided into a number of packets. The frequency, direction and, for example, the propagation distance before the packet is absorbed are all random samplings of physically motivated probability distributions. Computing the evolution of large numbers of these packets hence builds up an estimate of the radiation field properties. Monte Carlo radiative transfer (MCRT) has many strengths, including the fact that the evolution of each packet is an independent event, meaning that it can be efficiently parallelized. Furthermore, MCRT is in the first instance easy to implement and naturally accounts for physics (e.g. multiple anisotropic scattering) that are not so trivially accounted for by ray tracing schemes. 

MCRT has become increasingly popular in recent years, due in part to advances in computer speed, in particular for dust continuum transfer calculations. A plethora of codes to solve radiative equilibrium in three dimensions now exist \citep{2013ARA&A..51...63S}.

In this paper we present the capabilities of version 4.0 of the \textsc{torus} Monte Carlo radiation transport and hydrodynamics code, which has evolved over more than a decade to be capable of radiation hydrodynamic modelling with microphysics at the level of sophistication of dedicated radiative transfer, photoionisation and photodissociation region modelling codes. We summarise the methods and algorithms, including the optimisations and parallelisation techniques required to make such a comprehensive approach computationally feasible. We also discuss testing and astrophysical applications of the code. Our intention is to provide a definitive description of the latest version code, for the combined benefit of users, or potential users, of the code, who may have a specific application in mind, to developers who may wish to know more about the code's operational framework and its scaling.

\section{Code summary}

 \torus\ is an acronym of  ``Transport Of Radiation Using Stokes (Intensities)''. The code itself is written in Fortran (using the 2003 standard), and version~4 of {\sc torus} comprises over 215,000 lines of code.
 
 {\sc torus} was original developed to model polarized line transfer in stellar winds, and the first use of the code was to model the structure winds of O-supergiant stars \citep{2000MNRAS.315..722H}. Dust radiative equilibrium was added in 2004 in order to model the dust-producing Wolf-Rayet binary WR~104 \citep{2004MNRAS.350..565H}. Further major developments include the implementation of molecular transport \citep{2010MNRAS.407..986R} and hydrodynamics \citep{2010MNRAS.403.1143A,2012MNRAS.420..562H}, radiation pressure and sink particles \citep{2015MNRAS.448.3156H, 2017MNRAS.471.4111H}.

\begin{figure*}
    \begin{center}
    \includegraphics[width=13cm]{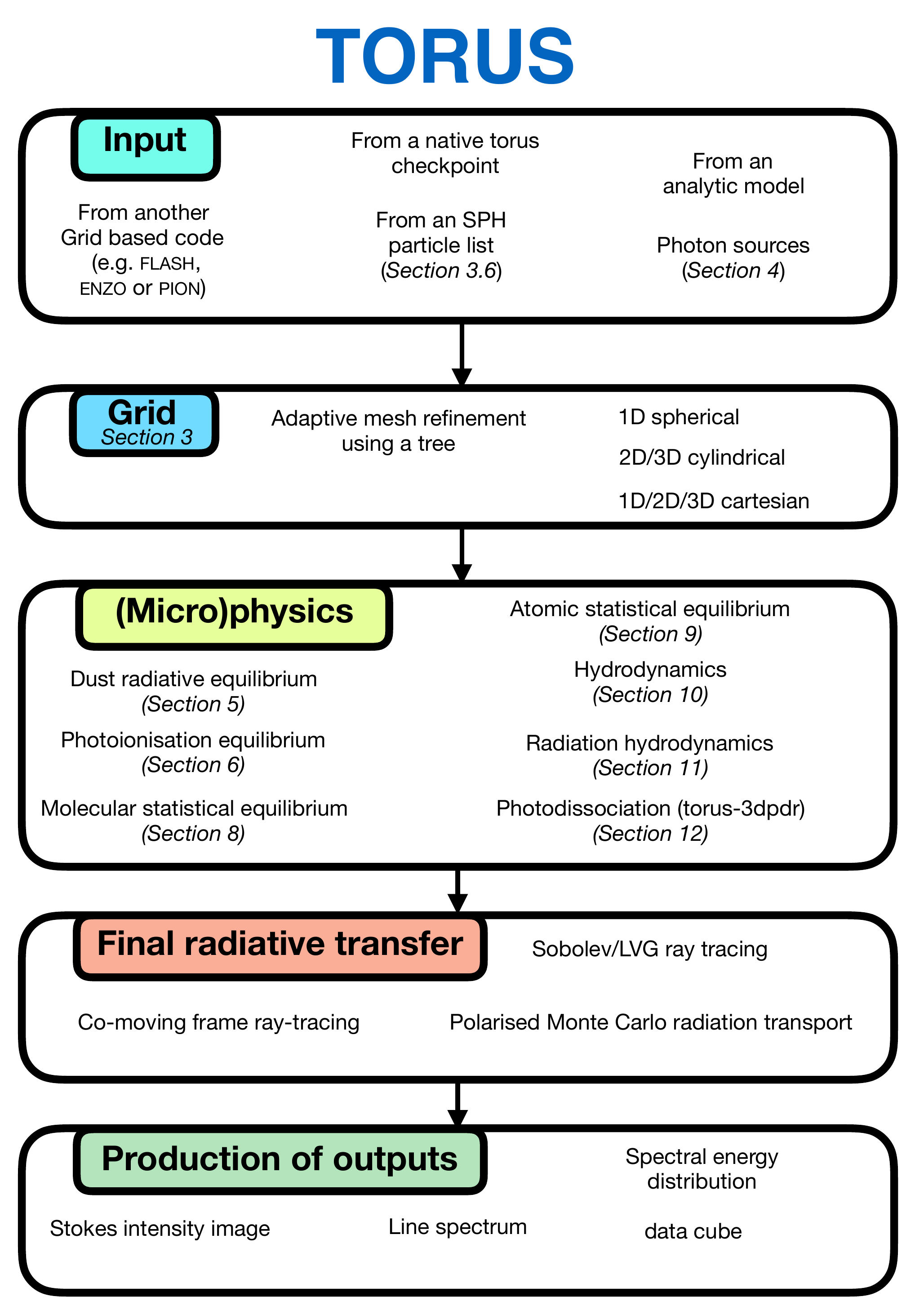}
    \caption{An overview of the capabilities and workflow of \textsc{torus}. Included are references to the components of this paper in which different features are discussed. }
    \label{fig:torus}
    \end{center}
\end{figure*}

We begin by providing a brief overview of the features of \textsc{torus}, which will be explored in more detail in subsequent sections. A schematic summary of the code's main features is given in Figure \ref{fig:torus}. \textsc{torus} is first and foremost a Monte Carlo radiative transfer code - propagating packets of photons over a computational domain to estimate the radiation energy density and hence radiatively determined gas and dust properties, as well as synthetic observables. Using a different approach, \textsc{torus} can also compute molecular level populations and, by coupling with the \textsc{3d-pdr} code can also compute the chemical and thermal properties of photodissociation regions.  In recent years it has also been coupled with hydrodynamics, uniquely offering the microphysics of a dedicated radiation transport code in dynamical applications.

\section{Grid architecture}

Physical variables in \textsc{torus} are represented by a grid of numerical values and the  accuracy of the solution generated depends on the size of the grid cells (the spatial resolution). For many calculations the required spatial resolution varies in different regions of the computational domain (e.g. in a circumstellar disc model the inner edge of the disc generally needs to be better resolved than regions in the optically thick inner disc). It can be inefficient or infeasible to use the highest required resolution for the whole grid so \textsc{torus} uses an adaptive grid which enables different regions of the computational domain to have different resolutions \citep{2005MNRAS.356.1489S}. The numerical grid is stored as a tree structure which provides excellent flexibility by allowing individual grid cells to be refined where required. The implementation of the tree structure for storing the grid is described in section~\ref{sec:octreegrid}.
\textsc{torus} can represent a variety of 1D, 2D and 3D geometries which are described in section~\ref{sec:gridgeometries}.
Methods for refining the grid are described in section~\ref{sec:adaptivemeshrefinement}.  In section~\ref{sec:nativegridsetup} we describe the construction of grids natively in \textsc{torus} and in section~\ref{sec:postprocessing} we describe the extensive capabilities of \textsc{torus} to generate grids from other numerical models for post processing purposes. 

\subsection{Tree structure}
\label{sec:octreegrid}

The numerical grid in \textsc{torus} is stored as a tree structure,  consisting of parent nodes connected to child nodes \citep[see][for an earlier application to astrophysical radiative transfer]{2001A&A...379..336K}. In a 1D geometry the tree is a full binary tree in which each node is connected to zero or two children. Correspondingly in 2D/3D each node is connected to zero or four/eight children  i.e. in 3D the tree is an octree \citep{meagher1982geometric}. 
\begin{figure*}
    \centering
    \includegraphics[width=14cm]{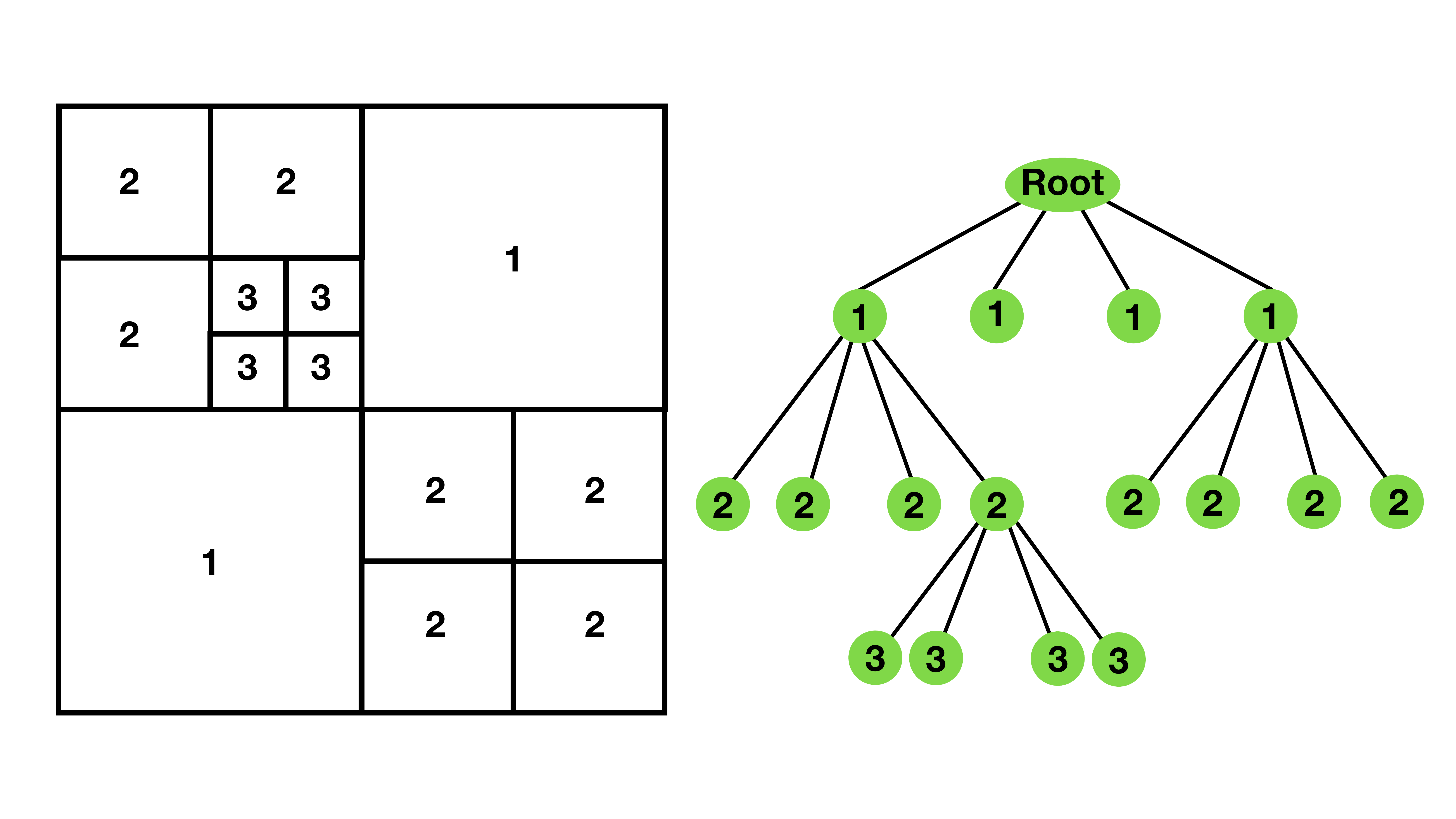}
    \caption{A schematic of a non-uniform grid and the corresponding tree structure. Nodes on the tree are represented by the green circles, which are connected by branches to other (child) nodes. Deeper levels of the tree correspond to higher resolution parts of the grid. The text labels illustrate the correspondence between leaf nodes on the tree and grid cells.}
    \label{fig:octreeSchematic}
\end{figure*}
A schematic of a 2D tree, with the corresponding grid, is shown in Figure \ref{fig:octreeSchematic}.
The coarsest level of refinement (designated depth 1) is a level comprising $2^N$ cells (where $N$ is the number of spatial dimensions). Each cell below this level can split into further cells (determined by refinement conditions) until a level in the tree is reached at which there are no more children (this occurs when none of the refinement criteria are locally satisfied). Leaf nodes (nodes without children) hold information at the main grid points over which a computation takes place and leaf nodes at deeper levels in the tree correspond to higher resolution portions of the grid. 

The tree structure which represents the grid is constructed from a Fortran derived type referred to as an ``octal", although in practice an octal in \textsc{torus} can represent two, four or eight grid cells depending on the number of spatial dimensions. To represent the connectivity of the tree each octal contains an array of pointers in which each element either points to a child octal or is null. Physical variables to be stored on the grid are held as pointer arrays\footnote{Fortran standards prior to Fortran 2003 do not allow allocatable components of derived types and pointer arrays are retained for compatibility with older compilers} which are components of the octal type. In order to minimize the memory footprint these arrays are allocated dynamically at run time so that only the required variables are allocated based on the physics included in the calculation (e.g. hydrodynamic attributes such as the pressure need not be allocated for a pure radiative transfer calculation). 

Operations on the grid typically involve following pointers up and down the tree structure to identify and operate on leaf nodes. These operations are implemented as recursive subroutines which loop over an array of pointers to child octals and pass a pointer to each child octal as an argument to a recursive subroutine call. 

\subsection{Grid geometries}
\label{sec:gridgeometries}

\begin{figure}
 %   \centering
%    \hspace{-0.6cm}
    \includegraphics[width=7cm,angle=-90]{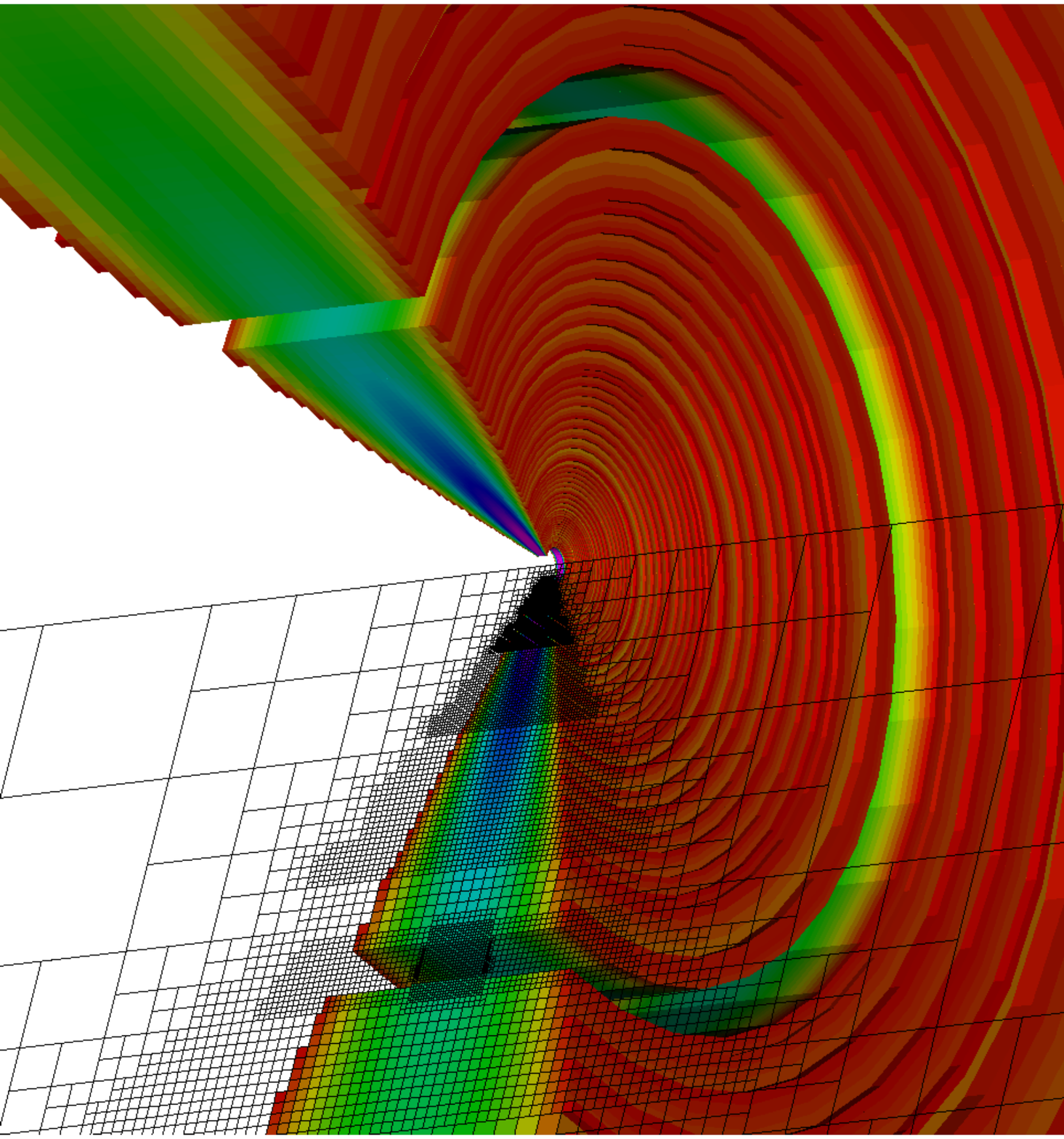}
    \caption{An example of the 3D cylindrical adaptive mesh refinement (AMR) coordinate system. The colour scale represents mass density in a flared, protoplanetary disc into which a planet has carved a gap. A segment of the disc has been removed to reveal its internal structure, and a slice through the AMR mesh is also shown (black lines) in order to illustrate the enhancement of the AMR resolution in the gap around the planet (not shown).}
    \label{fig:planet-disc}
\end{figure}

The tree structure describes the connectivity between grid cells but additional information is required to specify how the tree structure maps to a physical geometry. \torus can represent a number of physical geometries (see Table~\ref{tab:gridgeometries}) which enables a grid to be chosen which matches the geometry of the system being studied.
\begin{table}
    \centering
    \begin{tabular}{|c|c|c|}
    \hline
      No. of    &  Geometry & Co-ordinate \\
      dimensions&   type    & variables   \\ 
      \hline
    1           &  Cartesian   & $x$   \\
    1           &  Spherical   & $r$    \\
    2           &  Cartesian   & $x$,$y$ \\ 
    2           &  Cylindrical & $r$,$z$ \\
    3           &  Cartesian   & $x$,$y$,$z$\\
    3           &  Cylindrical & $r$,$\phi$,$z$ \\
    \hline
    \end{tabular}
    \caption{Grid geometries supported by \textsc{torus}. }
    \label{tab:gridgeometries}
\end{table}
 For example circumstellar discs are well represented in a cylindrical polar co-ordinate system. \torus has been extensively used for circumstellar disc models and consequently the cylindrical polar co-ordinate capabilities are provided, allowing refinement of the grid in the azimuthal co-ordinate as well as $r$ and $z$. Refinement in azimuth is important for applications such as planet in disc models which must not only resolve the inner edge of the disc but also resolve the region around the planet (see Figure~\ref{fig:planet-disc}). 

The octal type contains logical flags which indicate the number of spatial dimensions and the geometry. By including geometry information in the octal data structure it is possible to determine the correct geometry terms without access to any other data items. The octal type also includes the octal centre and size which makes it easy to determine whether a given point lies within a given octal.

\subsection{Adaptive mesh refinement}
\label{sec:adaptivemeshrefinement}

The \textsc{torus} grid can be refined at the level of individual cells to provide a very flexible numerical representation of the system being studied. The grid is initialised using information available when the calculation starts but can be dynamically refined and coarsened as the calculation progresses. For example a hydrodynamics calculation could produce a shock front, requiring higher resolution to follow the shock over the grid as it advects, and ideally leaving lower resolution up/down stream of the shock. Similarly, for a pure radiative transfer calculations the grid might require higher refinement to capture an ionisation front or opacity gradient.

To determine whether an octal should be refined a pointer to the octal is passed to a function which returns true if the octal is to be refined. This means that it is relatively straightforward to decide whether to split an octal based on any of the data (including physical variables) stored within the octal type. Grid refinement conditions are very flexible and vary according to the configuration in use. A maximum and minimum depth of the tree (i.e. resolution) can be specified and a simple refinement scheme is to set the maximum and minimum refinement depth equal which produces a uniform grid. 

More sophisticated refinement schemes refine based on values of physical variables or gradients in physical variables. A frequently employed refinement condition is to split a cell if the fractional difference in quantity $q$ between cells $i$ and $i-1$ satisfies
\begin{equation}
    \left| \frac{q_i - q_{i-1}}{q_i}\right| - \delta_{lim} > 0
\end{equation}
where $\delta_{lim}$ is the critical fractional difference for refinement.
The act of refinement actually entails locally increasing the depth of the tree. For example, a leaf of a 2D grid gains 4 children. The new cells inherit the parent cell’s values, which are distributed among them using an interpolation based on the values in the cells  surrounding the corners of the parent.

In addition to adaptive mesh refinement (AMR), \textsc{torus} also automatically identifies cells that can be coarsened — reducing memory usage and computational expense of the calculation. A coarsening involves locally decreasing the depth of the tree. For example in a 2D geometry four cells would have their pointers nullified and their parent cell would become the leaf node at the corresponding points in space. A coarsening takes place if, (n the 2D example) of the four cells making up an octal and storing quantity $q$, the quantity ($[q_{max}-q_{min}]/q_{mean}$) is less than some user specified value. In dynamical applications we force the condition that no newly-refined set of children can be re-coarsened until the next hydrodynamic step is completed. This is to ensure that a refined cell is not immediately re-coarsened should the refinement/coarsening criteria overlap.

%\subsection{Adaptive mesh refinement}
%The non-uniform grid modifies its resolution adaptively to automatically capture regions of the grid where higher resolution is required. Precisely what such a region is defined as is expressed in terms of gradients of arbitrary quantities on the grid, such as density, velocity, pressure or ionisation fraction. \textsc{torus} imposes the condition that no two neighbouring cells can differ by more than one level of refinement. See section \ref{sec:adaptivemeshrefinement} for more information. 

\subsection{Native grid setup}
\label{sec:nativegridsetup}
%\textbf{In this section we will describe the process of setting up a grid without reading in fields from an external model.}

\textsc{torus} can run calculations from scratch  without reading in fields from an external model. In order to do so a description of the model, both in terms of grid refinement and the initial physical conditions, is required on a cell-by-cell basis. For example, if constructing a 2D cylindrical disc model the density might be described by $\rho(r,z)=\rho(r)\exp(-z^2/2H(r)^2)$, where $H(R)$ is the scale height. As \textsc{torus} initialises it traverses the tree and for a cell at position $(x, z)$ assigns a density according to the above function. A similar prescription is also required for other initial conditions. 

The initial grid can be refined according to a predetermined prescription, for example forcing the cells to vary in refinement radially, or be refined to the maximum depth in some component of the grid. Alternatively \textsc{torus} has an iterative grid refinement capability where it sets up the grid according to the analytic prescription and then automatically refines in the way it would for the adaptive mesh. It then iterates over this process, populating the refined grid and then re-applying the refinement criteria. In this way \textsc{torus} sets up an optimally refined grid for the initial conditions.

\subsection{Post processing}
\label{sec:postprocessing}

Outputs from a number of other numerical models can be read by \torus for post processing purposes e.g. generating synthetic observations. Table~\ref{tab:compatiblecodes} lists codes which are compatible with \torus and the corresponding file format which \torus is capable of reading. \torus has extensively been used for post-processing results from SPH (smoothed particle hydrodynamics) calculations and this functionality is described in detail in section~\ref{sec:gridfromsph}.

\begin{table*}
    \hspace{-14pt}
    \begin{tabular}{|c|c|c|c|c|}
    \hline
    Code                 &  Type & File format & Code reference & Examples of application with \textsc{torus}  \\
    \hline
    \textsc{phantom} &  SPH       & ASCII  & \cite{2017arXiv170203930P} & \cite{2014MNRAS.444..919P},  \cite{2015MNRAS.449.1996D}          \\
    \textsc{sph-ng} &  SPH       & Binary, ASCII & \cite{1990ApJ...348..647B}, \cite{1990nmns.work..269B} & \cite{2004MNRAS.351.1134K}, \cite{2010MNRAS.407..986R}    \\
     & & & \cite{1995MNRAS.277..362B} & \cite{2014MNRAS.444..919P}, \cite{2015MNRAS.447.2144D}  \\
    \textsc{gadget-2}    &  SPH       & Binary, ASCII  & \cite{2005MNRAS.364.1105S}  &  \\
    \textsc{dragon} & SPH & ASCII & {\cite{1995MNRAS.277..705T},} & \\
    & & & {\cite{2004A&A...414..633G}} \\
    \textsc{enzo}   &  Grid      & ASCII   & {\cite{2014ApJS..211...19B}}  &  \cite{2015MNRAS.450...10H}, \cite{2015MNRAS.454.1634H}       \\
    \textsc{flash}  &  Grid      & HDF5  & {\cite{2000ApJS..131..273F}} &            \\
    \textsc{mc-max}  &  Grid & ASCII  &  {\citet{2009A&A...497..155M}} & \citet{2016MNRAS.461..385B} \\ 
    \textsc{pion}    &  Grid      & FITS  & {\cite{2010MNRAS.403..714M}},   & {\cite{2016A&A...586A.114M}}, {\cite{2017MNRAS.466.1857G}} \\
    & & & {\cite{2012A&A...539A.147M}} & Green et al. (subm.) \\
   \textsc{vh-1}                 & Grid       & Binary & \cite{1984JCoPh..54..174C} &  \cite{2016MNRAS.456..136A}\\
    \hline
    \end{tabular}
    \caption{Codes which have output files that can be read by  \torus\ for post-processing.}
    \label{tab:compatiblecodes}
\end{table*}

\subsection{Generating a grid from SPH data}
\label{sec:gridfromsph}

The first use of \torus to post-process results from a SPH simulation was by \cite{2004MNRAS.351.1134K} with more recent applications by \citet{2010MNRAS.403.1143A}, \citet{2010MNRAS.406.1460A}, 
\citet{2010MNRAS.407..405D},
\citet{2010MNRAS.407..986R},
\citet{2015MNRAS.447.2144D}, and
\cite{2018MNRAS.474..800Y}. 
Significant development work has taken place since the initial application and this section describes the current capabilities of \torus to work with SPH data. Setting up a \torus grid from SPH particles comprises three steps:
\begin{enumerate}
    \item Read SPH data into \torus
    \item Refine the \torus grid based on the SPH particle data
    \item Initialise \torus grid by mapping physical variables from the SPH particles to the grid
\end{enumerate}

\torus can read binary dump files from the \textsc{sph-ng} and \textsc{gadget2} codes (see Table~\ref{tab:compatiblecodes} for references). Dump files from other SPH codes can be used by converting the binary file into ASCII format using \textsc{splash}  \citep{2007PASA...24..159P}. Although the results of converting an SPH dump to ASCII using \textsc{splash} are similar for different SPH codes there are some variations which need to be accounted for (e.g. differences in non-gas particle types such as sinks and dark matter particles). \torus includes switches to handle ASCII dump files from the \textsc{gadget2} and \textsc{dragon} codes.

The \torus grid is refined by considering the properties of the SPH particles which are located in a given grid cell. The primary method for refining the grid using SPH is to split a cell if the mass of gas particles in the cell exceeds a specified threshold. This produces a grid where resolution follows mass, which is similar to the way that the smoothing length (and hence effective resolution) varies in modern SPH formulations. The grid can also be split based on density differences (e.g. fractional difference between the most and least dense particles in the cell). Splitting based on density differences produces extra refinement in regions where the density gradient is high and can be used to capture edges and surfaces \citep{2010MNRAS.403.1143A}. Other refinement conditions have been used for specific applications, such as splitting to capture velocity gradients when calculating line emission \citep{2010MNRAS.407..986R}. Multiple refinement conditions can be combined and a cell will be split if any refinement condition is satisfied. 

In the SPH method physical quantities are represented by a sum over particles (a set of disordered points) with a smoothing kernel applied \cite[see][for a review]{1992ARA&A..30..543M}. The value of a function $A\left( \vec{r} \right)$ at position $\vec{r}$ is approximated by
\begin{equation}
    A\left( \vec{r} \right) \approx \sum_{i=1}^{N_\rm{sph}} A_i \frac{m_i}{\rho_i} W\left(\vec{r}-\vec{r}_i,h_i\right)
    \label{eqn:sph_summation}
\end{equation}
where the sum is over particle index $i$ and $N_\rm{sph}$ is the number of particles used to estimate $A\left( \vec{r} \right)$. $A_i$ is the value of function $A\left( \vec{r} \right)$ at the position ($\vec{r}_i$) of particle $i$, and $m_i$ and $\rho_i$ are the mass and density of particle $i$ respectively. $W\left(\vec{r}-\vec{r}_i,h_i\right)$ is a smoothing kernel characterised by a variable smoothing length $h_i$. The smoothing length is determined by 
\begin{equation}
h_i=\eta \left( \frac{m_i}{\rho_i} \right)^{1/\nu}
\label{eqn:sph_smoothing_length}
\end{equation}
where $\nu$ is the number of spatial dimensions in the SPH calculation \citep{2007MNRAS.374.1347P}. \torus assumes $\eta=1.2$ by default but can optionally calculate $\eta$ for each particle from the particle mass and density values. \torus performs calculations with a non-dimensional form of the smoothing kernel $\mathcal{W}_i\left(q_i\right)$ which is derived from  $W\left(q_i,h_i\right)$ and the smoothing length $h_i$ using 
\begin{equation}
\mathcal{W}_i\left(q_i\right) = W\left(\vec{r}-\vec{r}_i,h_i\right) h_i^{\nu}
\label{eqn:sph_nd_kernel}
\end{equation}
where
\begin{equation}
q_i=\frac{|\vec{r}-\vec{r}_i|}{h_i}
\end{equation}
When the smoothing kernel is expressed in the form of equation~\ref{eqn:sph_nd_kernel}, and the smoothing length is calculated according to equation~\ref{eqn:sph_smoothing_length}, the approximation to $A\left( \vec{r} \right)$ can be written as
\begin{equation}
A\left(\vec{r} \right) \approx \sum_{i=1}^{N_\rm{sph}} A_i \eta^{-3} \mathcal{W}_i \left(q\right)
\label{eqn:sph_summation_weighted_kernel}
\end{equation}
assuming three spatial dimensions. \torus calculates an array which holds values of $\eta^{-3}\mathcal{W}_i$ (referred to as the ``weights'') for all particles to be used in the kernel summation. 
% see sph_data_class::doWeights
The array of weights is then multiplied by different physical variables (e.g. density, velocity, abundances) and summed in order to calculate  equation~\ref{eqn:sph_summation_weighted_kernel}. 
% clusterParameter does the multiplication and summation

\torus can use either a Gaussian or spline kernel where the Gaussian kernel has the form
\begin{equation}
    \mathcal{W}\left(q\right) = \frac{1}{\pi^{3/2}} \exp{\left(-q\right)}
\end{equation}
and the spline kernel has the form
% The spline kernel is calculated in sph_data_class::SmoothingKernel3d 
\begin{equation}
    \mathcal{W}\left(q\right) = \frac{1}{\pi} \begin{cases} 
                                1-\frac{3}{2}q^2 + \frac{3}{4}q^3 & \text{if } 0 \leq q < 2\\
                                \frac{1}{4} \left( 2-q\right )^3 & \text{if } 1 \leq q < 2 \\
                                0 & \text{otherwise}
                                  \end{cases}
\end{equation}
again assuming three spatial dimensions. 

The values calculated from equation~\ref{eqn:sph_summation_weighted_kernel} are normalised by dividing by the sum of the weights in order to reduce noise due to the particle distribution. However this normalisation can introduce artefacts at free surfaces \citep{2007PASA...24..159P} so normalisation is only applied when the sum of the weights exceeds 0.5 (except for velocities which are always normalised). The threshold for normalisation can be modified if required.

The value of $N_\rm{sph}$ in equation~\ref{eqn:sph_summation_weighted_kernel} depends on both the properties of the smoothing lengths used in the underlying SPH calculation and on the method used by \torus to select SPH particles. \torus\ has two options for determining the list of particles to use when calculating the summation in equation~\ref{eqn:sph_summation_weighted_kernel} but both methods typically select approximately 200 particles within three smoothing lengths. The first method of selecting particles is to include all particles for which $q<3$ for the Gaussian kernel or $q<2$ for the spline kernel. This is a straightforward way to select particles (referred to as the ``simple'' method hereafter) and is parallelised with OpenMP to speed up processing of large particle lists. However the time taken to initialise the grid cells can still be excessively long, so the default behaviour is to use a more sophisticated and faster method for selecting particles, as described by \citet{2010MNRAS.407..986R}. 

In the Rundle et al. method the particles are initially sorted according to their x-values. When an interpolated value is required the particle with an x-value closest to the required point is located. All particles within a specified physical distance (rather than non-dimensional q-value) along the x-axis are then identified; as the particle list is sorted by x-value this is simply a matter of determining lower and upper indices of the array of x-values. For this sub-set of particles the q-value is only calculated if the particle is also within the specified physical distance along the y and z axes. The initial selection of particles greatly reduces the number of q-values which need to be calculated. However it is not trivial to determine the appropriate physical distance to use when locating particles, as the contribution of a given particle depends on its smoothing length as well as the physical distance. In an SPH simulation there are typically a small number of low density particles with very large smoothing lengths and the majority of particles have much smaller smoothing lengths. Consequently using the largest smoothing length as the physical distance to search within is not an efficient strategy. To determine which particles to use \torus makes three attempts to select the sub-set of particles for which the sum of the weights is greater than $10^{-3}$ using increasingly large search distances. The first search is carried out over a distance $r_1$ where 
\begin{equation}
r_1 = \min\left(4d,2h_{\rm{crit}}\right)
\end{equation}
and $h_{\rm{crit}}$ is a ``critical'' smoothing length (set by default at the 80th percentile of the smoothing length distribution). The parameter $d$ is related to the grid cell size and is given by 
\begin{equation}
    d = \left( \frac{\rho_{max}}{\rho_{min}} \right)^{1/3} \Delta_{cell}
\end{equation}
where $\rho_{max}$ and $\rho_{min}$ are the maximum and minimum particle densities in the cell and $\Delta_{cell}$ is the size of the grid cell. When the grid is split according to a maximum mass per cell threshold the grid resolution is related to the smoothing length and $4d$ provides a good estimate of the search distance in most cases. The density ratio term weights the value of $d$ to account for the range of smoothing lengths within a cell (recalling that smoothing length scales as $\rho^{-1/3}$ in equation~\ref{eqn:sph_smoothing_length}). When there is a large density range the value of $d$ is increased so that more of the lower density particles are included. 
If the first search does not return a sum of weights greater than $10^{-3}$ then the search is repeated using a larger search distance $r_2$
where
\begin{equation}
    r_2 =  \min\left(\max\left(4r_1, 4 h_{crit}\right), 0.2 h_{max}\right)
\end{equation}
where $h_{max}$ is the smoothing length at the 99th percentile of the smoothing length distribution by default. 
If required a third and final search is made over a search distance $h_{\rm{max}}$. If the sum of the weights is still less than $10^{-3}$ after the third search then the cell is declared empty and is populated with floor data values.

\subsubsection{Validation}
\label{section:gridToSphValidation}

\begin{figure}
    \centering
    \includegraphics[width=8cm]{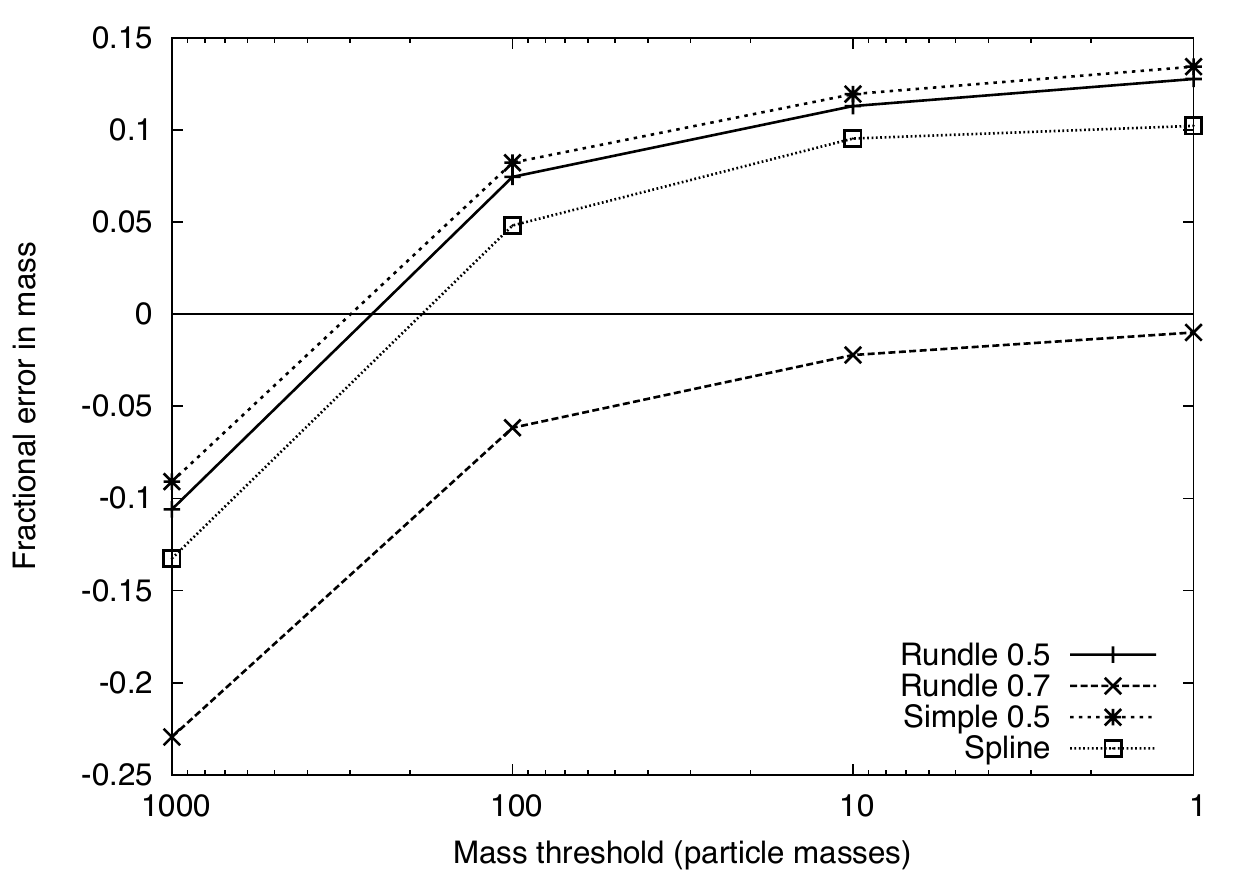}
    \caption{Fractional error in total mass resulting from the SPH to grid conversion. The conversion was performed for different mass per cell thresholds (x-axis) and different conversion parameters (lines).}
    \label{fig:sph_to_grid_frac_err}
\end{figure}

Tests of the mapping from SPH particles to the \torus grid were presented by \citet{2010MNRAS.403.1143A} (who studied the effects of the grid refinement criteria) and \citet{2010MNRAS.406.1460A} (who studied the effect of the normalisation threshold). In both cases the accuracy of the total mass on the \torus grid relative to the total mass of SPH particles was used as the figure of merit. Likewise in figure~\ref{fig:sph_to_grid_frac_err} we show the fractional error in mass resulting from the conversion from SPH particles to the \torus AMR grid for different mass per cell refinement thresholds (x-axis) and different conversion parameters (lines).
The SPH data set is a spiral galaxy simulation from \citet{2011MNRAS.417.1318D} comprising $10^6$ gas particles. The solid-line with pluses shows the fractional error using the default parameters (Rundle et al particle selection, 0.5 weighting threshold and a Gaussian kernel). The fractional mass error converges as the resolution of the grid increases but is systematically too large. The normalisation threshold can be modified to correct the total mass; the long dashed line with crosses shows the effect of changing the normalisation threshold to 0.7 which causes the fractional error to converge to a value nearer zero. The normalisation threshold has a larger impact on the total mass error than the particle selection method or the choice of kernel. The short dashed line with stars shows the simple particle selection method (with the default normalisation threshold of 0.5 and a Gaussian kernel) and the dotted line with squares shows the spline kernel (with the default normalisation threshold of 0.5 and Rundle et al particle selection). The particle selection method has little impact on the mass error, and although the choice of kernel has a larger impact it is not as significant as the weighting threshold.

\subsection{Checkpointing}

To enable a calculation to be restarted \torus can checkpoint by writing out a grid file, at an interval specified by the user, which includes all allocated grid variables. This allows a full grid to be read in from a file stored on disk and a calculation which has been interrupted to be resumed (see Section~\ref{sec:gridio} for details). Some physics modules (e.g. molecular physics and dust radiative equilibrium) write a small restart file which contains information about the last complete iteration to enable a calculation to be resumed at the appropriate point. The restart capability can also be used to generate multiple data products (e.g. images, data cubes, spectral energy distributions) from the same grid file without the need to repeat a radiative equilibrium calculation.
 
%%%%%%%%%%%%%%%%%%%%%%%%%%%%%%%%%%%%%%%%%%%%%%%%%%%%%%%%%%%%%%

\section{Photon sources}
\label{sec:sources}
 We are yet to discuss Monte Carlo radiative transfer in detail, but have noted that it proceeds by random sampling of physically motivated probability distributions. In particular the photon energy emitted by a star is distributed across $N$ energy packets representing collections of photons.

 Stellar photon sources in \textsc{torus} are described by a combination of two of the effective temperature, radius, or bolometric luminosity. A mass must also be specified, which is used for hydrodynamic calculations (where the sources are also sink particles) or models in which material is bound to the source (e.g. protostellar discs in Keplerian rotation). 

The spectral energy distribution emerging from the source can be defined by the user. The default is a blackbody, although it is also possible to use Kurucz LTE model atmospheres\footnote{\url{http://www.stsci.edu/hst/observatory/crds/k93models.html}} \citep{1993KurCD..13.....K} or the {\sc tlusty} grid of O star metal line blanketed non-LTE model atmospheres\footnote{\url{http://nova.astro.umd.edu/Tlusty2002/tlusty-frames-OS02.html}} 
\citep{2003ApJS..146..417L}. For the model atmospheres the surface gravity is calculated from the mass and radius, and bilinear interpolation is used in the appropriate model grid to find the correct SED. If an appropriate model cannot be found within the bounds of the grid of atmospheres a warning is given and a blackbody SED is used instead.

\torus\ supports surface temperature variations such as hot and cool spots. The source surface is divided up into a grid of surface elements, each of which may have their own SED. This has been used primarily for Classical T~Tauri star (CTTS) models, in which accretion produces hot spots on the protostellar surface \citep[e.g.][]{2009MNRAS.398..189H}. In the case of CTTS the mass flux immediately above each surface element is determined from the AMR mesh, and a fraction of this kinetic power is assigned as accretion luminosity to that element. The element area and luminosity are used to determine the accretion temperature and a blackbody SED is added to the photospheric SED for that element. Once the surface is set up the stellar luminosity is determined by integrating the SEDs over frequency and surface elements, and this is compared to the input source plus accretion luminosity as a sanity check.

For some models it may be necessary to evolve the star as the model proceeds. Specific cases where we have used this mode is for massive star formation simulations \citep{2017MNRAS.471.4111H} and for cluster gas dispersal models \citep{2018MNRAS.477.5422A}. Currently the evolutionary tracks by  \citet{1992A&AS...96..269S} are included, and the stellar zero age main sequence (ZAMS) mass plus the stellar age are used to interpolate in the model grid for the stellar radius and luminosity. A warning is given if the ZAMS mass and stellar age mean that the evolutionary state is beyond the end point of the tracks (for example if the star would have undergone core collapse). 

Alternatively the pre-main-sequence evolutionary models of \citet{2009ApJ...691..823H} may be used, although only the $\log(\dot{M}) = -3$ track has been included so far. The sink particle mass is used to interpolate in track, and the luminosity, radius and surface gravity are used to find the appropriate protostellar spectrum. For greater self-consistency the accretion luminosity, rather than being that of the model, is taken from the accretion rate onto the sink particle itself.

\begin{figure}
    \centering
    \includegraphics[width=8.8cm]{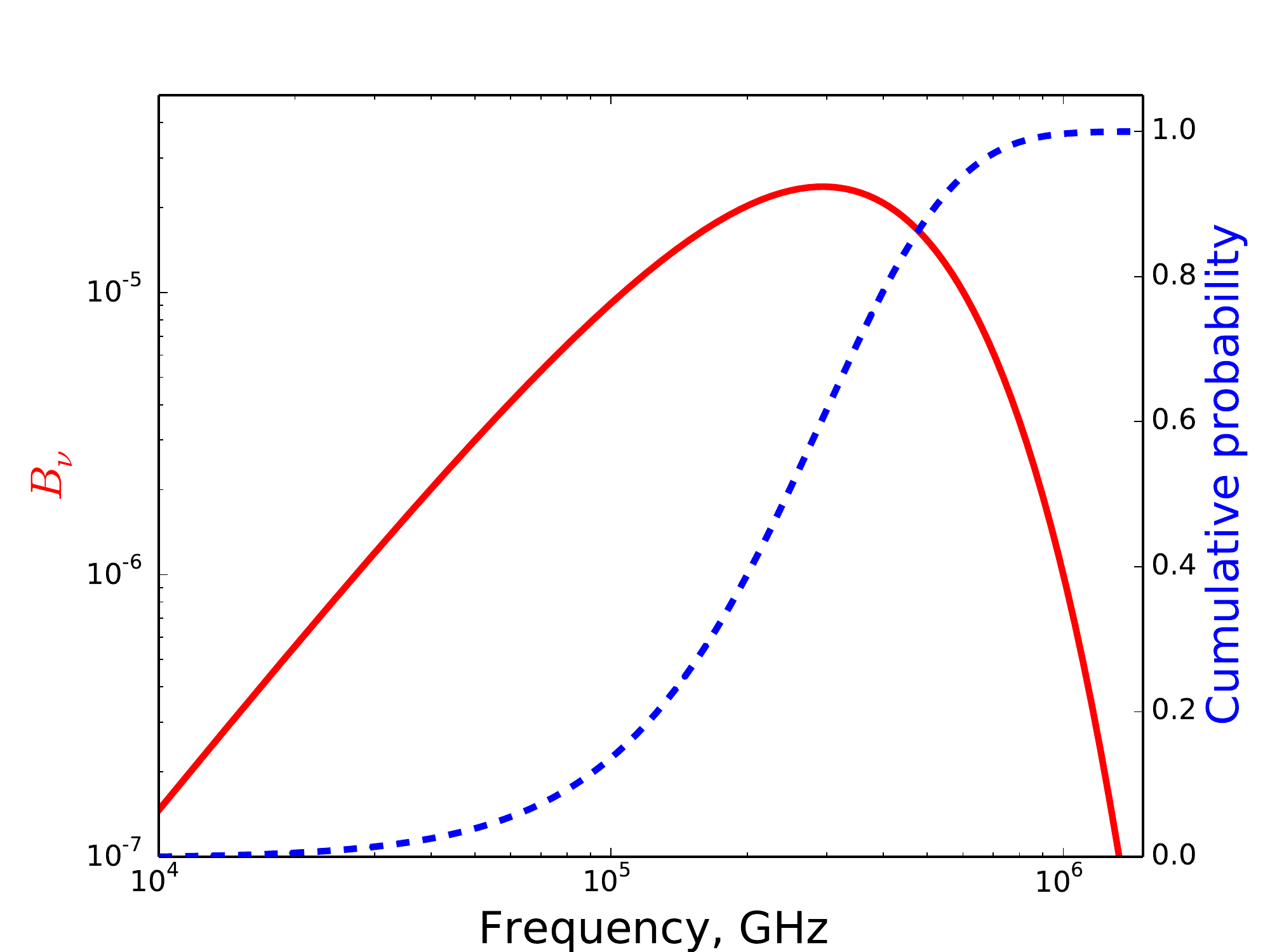}
    \caption{An example of a stellar (blackbody) spectrum (solid line, left axis) and the corresponding cumulative probability distribution (dashed line, right axis) from which numbers in the range [0:1] would randomly, but proportionally, sample the spectrum in frequency. The same approach holds for arbitrarily complex (singularly valued) spectral models. }
    \label{fig:spectrumSample}
\end{figure}

A photon packet is initiated at a random point on the surface of a photon source, with a frequency randomly sampled from the source spectrum. A cumulative probability distribution is created for the source spectrum, which allows the mapping of a random number in the range [0:1] onto a frequency, as illustrated in Figure \ref{fig:spectrumSample}. The spectrum is therefore sampled appropriately according to the relative intensity of its different frequencies. Photon packets are emitted with a random direction. If the star is a point source then this is usually with uniform probability of being emitted into any of 4$\pi$ steradians. If the photon source is not a point source then the packets are emitted from random points on the surface with trajectories uniformly sampled in $\cos \mu$ where $\mu$ is the angle between the trajectory and the local surface normal.

If a model has more than one source then it is possible to assign a probability that a photon packet is produced by a given source. When this probability is specified the photon packet energies from the sources are re-weighted appropriately. This may be important if the model contains sources that have a large range of luminosities since a constant photon packet energy would mean that the regions around low luminosity sources would be poorly sampled.

\section{Dust radiative equilibrium}
\label{sec:radeq}
One of the basic functionalities of \textsc{torus} is to calculate the temperature distribution of an arbitrary distribution of dust illuminated by photon sources (often stars, but also diffuse radiation such as the cosmic microwave background). 
This dust temperature distribution may then be used to calculate synthetic observables, including images and spectra.

At its core the radiative equilibrium routine is based on the path length algorithm described by \citet{1999A&A...344..282L}, and we refer the reader to this paper for a detailed description of the method. In summary, the total energy of the illuminating radiation from luminosity $L$ over the duration of the Monte Carlo simulation $\Delta t$ is divided into $N$ indivisible packets of energy $\epsilon$ defined by
\begin{equation}
\frac{\epsilon}{\Delta t} = \frac{L}{N}
\label{equn:ePacket}
\end{equation}
Since the energy of each packet is the same, packets of different frequency essentially just carry different \textit{numbers} of photons. 

\begin{figure}
    \hspace{-10pt}
    \includegraphics[width=10cm]{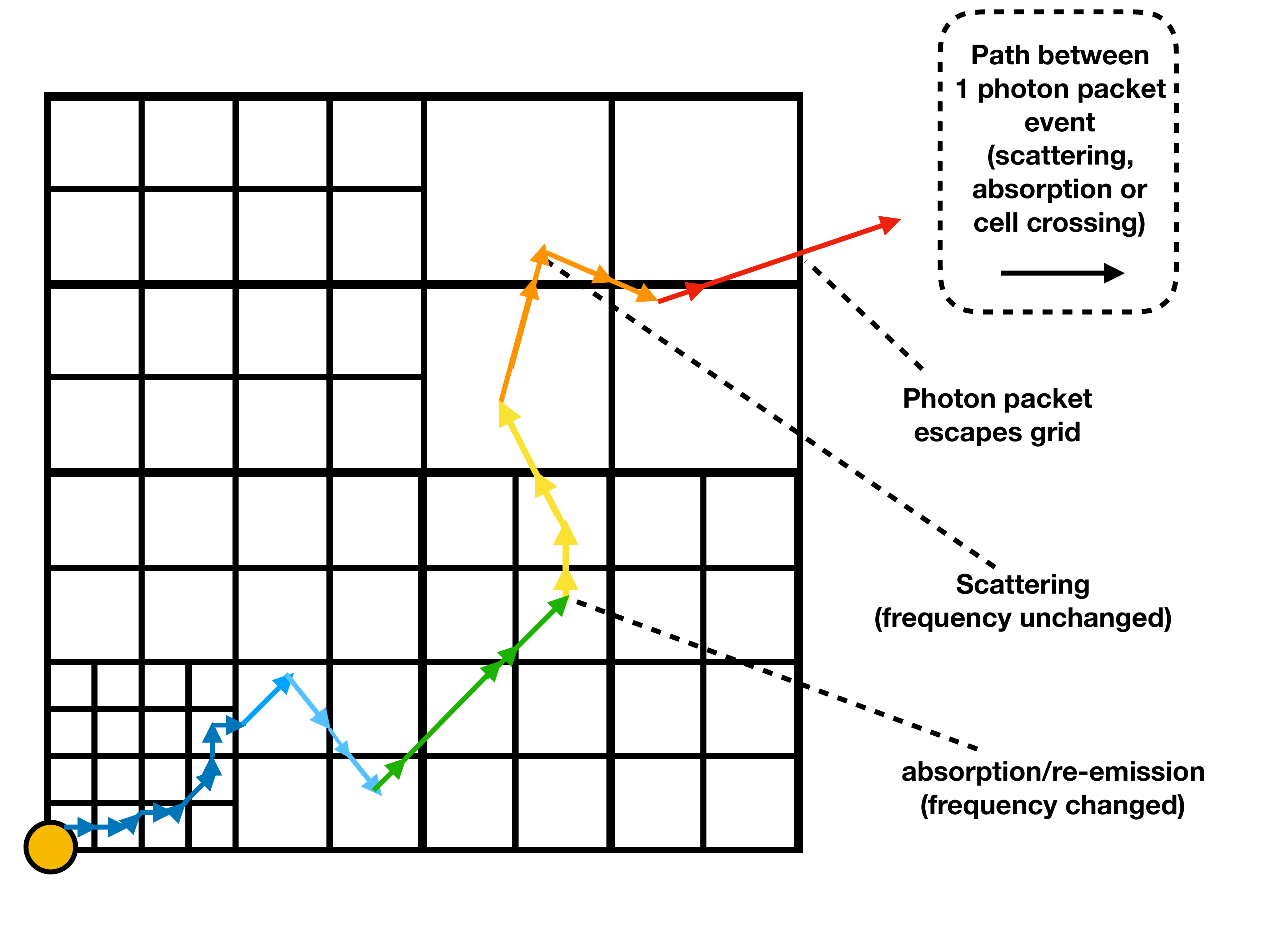}
    \caption{An illustration of the random walk of a photon packet through the computational grid. As the photon packet crosses each cell it traverses a path length, which contributes to the estimate of the energy density and hence mean intensity in that cell.  }
    \label{fig:randomWalk}
\end{figure}

Following emission, a photon packet is propagated for a random optical depth
\begin{equation}
    \tau = - \textrm{ln}(1-r)
    \label{eq:opticaldepth}
\end{equation}
where $r$ is a uniform random deviate. This translates into a distance dependent upon the opacity of the medium. After traversing this distance, the packet undergoes either an absorption or scattering event. To determine whether it is the former or the latter, a randomly generated number in the range [0:1] is compared with the albedo (the ratio of scattering to total, absorption plus scattering, opacities). In the event of absorption,  the photon packet is immediately emitted from the same location with a new random direction and frequency sampled from the local emissivity at the site of emission. In the event of scattering the photon frequency remains the same, but a new propagation trajectory is randomly computed using a Mie scattering phase matrix that is determined by the parameters of the scattering medium (i.e. the dust parameters). This process repeats, with the photon packet undergoing a random walk through the grid akin to the propagation of real photons through matter, until the packet escapes the grid. An illustration of this process for a single photon packet is given in Figure \ref{fig:randomWalk}.

As a photon packet propagates a distance $\ell$ through a cell, it contributes to the energy density $U$ in that cell by an amount $\epsilon \delta t / \Delta t$ where $\delta t = \ell/c$. By performing the above photon packet propagation procedure for a large number of photons this builds up a estimate of $U$ in each cell on the domain. That is, for a given cell of volume $V$ being traversed by photon packets in frequency range $\nu$ to $\nu+d\nu$ tracing path lengths $\ell$, the energy density is
\begin{equation}
    U_\nu {\rm d}\nu  = \frac{\epsilon}{c\Delta t}\frac{1}{V}\sum \ell. 
    \label{equn:MCgetJ}
\end{equation}
Furthermore since
\begin{equation}
U_\nu = 4 \pi \frac{J_\nu}{c}
\end{equation}
and the absorption rate is
\begin{equation}
    \dot{A} = 4 \pi \int_0^\infty J_\nu k_\nu  {\rm d}\nu
    \label{eq:absorption-rate}
\end{equation}
we see that
\begin{equation}
\dot{A} = \frac{\epsilon}{\Delta t} \frac{1}{V} \sum k_\nu \ell
\label{equn:adot}
\end{equation}
where $k_\nu$ is the absorption opacity per unit length. The emission rate of a cell is 
\begin{equation}
    \dot{E} = \int_0^\infty k_\nu B_\nu d\nu =4 \pi k_{\rm P} B
    \label{eq:dotE}
\end{equation}
where $k_{\rm P}$ is the Planck mean absorption coefficient and $B = (\sigma / \pi)T_d^4$.
Clearly at equilibrium we have $\dot{A}=\dot{E}$ and hence we can find an updated (dust) temperature of the cell from
\begin{equation}
T_d = \left( \frac{\dot{A}}{4 \sigma k_{\rm P}} \right)^{1/4}.
\end{equation}
This revised temperature is adopted for the next iteration, and gradually the dust temperatures will come into equilibrium with the radiation field. Note that this is effectively a $\Lambda$ iteration, but it has good convergence properties because the radiation field is divergence free (every energy packet that emerges from the photon source eventually makes it off the grid). Traditional $\Lambda$ iteration schemes do not have this energy conservation built in, and as such their convergence is notoriously poor.

The initial number of photon packets $N$ can be selected as a model parameter, but the default is 10 times the number of grid cells. The temperature of the entire grid is initially set to the floor temperature (3\,K), meaning that packets that are absorbed are remitted to very long wavelengths and immediately escape the grid. The new temperatures are then calculated and the iteration proceeds.

As the packets propagate through the grid the {\it number} of path lengths $\ell$ is stored for each cell. This is a measure of the quality of the MC estimator of the energy density for each cell. At the temperature correction stage if the number of path lengths is below some threshold value (the default is 200) then the cell is flagged as undersampled (in the sense that the absorption rate estimate has relatively low signal to noise). If the fraction of bad cells exceeds a set number (the default is zero), then the number of photon packets used in the next MC iteration is doubled. This adaptive modification of the photon packet number ensures that all cells in the grid have a good temperature estimate.

The iteration procedure stops when the total dust emissivity changes by less than some tolerance (the default is one per cent) between iterations. Typically this means that the mean temperature change within cells is on the order of $\pm 2$\,K. This is a relatively strict convergence criterion, and may not be reached if the photon packet number is too small (since the stochastic variations in dust emissivity alone may exceed the tolerance). A typical model will converge in around six or seven iterations, but this varies from model-to-model.

For some models there can be regions of the grid that are surrounded by very high optical depth (in particular the inner midplanes of protoplanetary discs). Photon packets are extremely unlikely to penetrate these deep regions, and we adopt a diffusion approximation in these volumes, using the temperatures of the surrounding cells as a boundary condition. The diffusion equation is solved using a Gauss-Seidel iteration.

If a photon penetrates into an extremely optically-thick environment it may get trapped, undergoing many tens of thousands of absorptions and re-emissions. We adopted a modified random walk (MRW) method to reduce the computational overhead and allow photon packets to escape more easily. The MRW algorithm is discussed in some detail in \cite{2009A&A...497..155M} and \cite{2010A&A...520A..70R}.

We also include an algorithm for dust sublimation. The temperature at which each dust species sublimates can be expressed as a power law function of density
\begin{equation}
    T_{\rm sub} = k_1 \rho ^{k_2}
\end{equation}
where $k_1$ and $k_2$ are constants the (defaults are 2000\,K and 0.0195 respectively), matching the expression in \cite{1994ApJ...421..615P}. In this mode the code first sets the dust fractions in each cell to a negligible value prior to computing radiative equilibrium. Once equilibrium is achieved the dust fraction in cells with temperatures below the local dust sublimation temperature is increased in each cell in order that the maximum optical depth across an individual cell does not exceed 0.01. Three radiative equilibrium iterations are then conducted and the dust fractions are then again increased in order that the optical depth across individual cells is 0.1 or less. Repeated radiative equilibrium iterations and dust fraction growth steps are made until a self-consistent dust distribution is found (see \citealt{2007ApJ...661..374T}).

Once the dust temperature distribution has been established it is possible to compute images and SEDs. Formally the exiting photon packets from the last radiative equilibrium step could be binned spatially and spectrally and used for this purpose, but this is not an optimal method. Instead we run a separate MC loop to calculate the observables (see section~\ref{sec:synthetic_obs}).

For disc models we may also solve for vertical hydrostatic equilibrium (HSE). Under the assumption that the disc mass is negligible compared to the central star the equation of HSE is
\begin{equation}
    \frac{dP}{dz} = - \rho g_z
\end{equation}
where $P$ is the pressure and $g_z$ is the local vertical component of the star's gravitational acceleration. Adopting an ideal gas equation of state ($P=\rho k T_g / \mu m_{\rm H}$) and assuming that the gas and dust are thermally coupled ($T_d=T_g=T$) we find
\begin{equation}
    \frac{d\rho}{dz}\frac{1}{\rho} = - \frac{1}{T} \left( \frac{\mu m_{\rm H}}{k} + \frac{dT}{dz} \right).
\end{equation}
The above equation is solved numerically for $\rho(z)$ on the AMR mesh since the temperature distribution is known from the radiative equilibrium calculation. The vertical density structure is then renormalised to conserve the radial surface density profile prescribed for the particular disc. Our implementation of vertical HSE was validated against an independently developed method by \cite{2007PhDT.......229W}.

\subsection{Grain description}
\label{sec:grainDescr}
The grain properties in a \textsc{torus} calculation are described in terms of their composition, minimum and maximum grain size ($a_{min}$, $a_{max}$), dust-to-gas mass ratio $\delta$ and power law of the distribution $q$
\begin{equation}
    \frac{dn(a)}{da}\propto a^{-q}
\end{equation}
\citep{1977ApJ...217..425M}. This collection of parameters defines a grain distribution (or dust type). Any number of grain distributions can be combined to have a spatially varying grain population. For example \cite {2018MNRAS.473..317H} use 10 grain distributions to construct a radially varying maximum grain size in models of the disc about the AGB star L$_2$ Pup. In addition to spatial variation in the dust properties, dust can be sublimated (i.e. reduced to a very low abundance) wherever the temperature rises above a user defined threshold. 

The file containing optical constants for the grain type is specified at run time, so a wide range of grain compositions are easily included. For example \textsc{torus} can read dust data from the Jena dust database\footnote{\url{http://www.astro.uni-jena.de/Laboratory/OCDB/index.html}}. 

\subsection{Validation}
\label{sec:radeq_validation}

We have benchmarked the dust radiative equilibrium module of \torus\ against several established codes in order to increase our confidence that the numerical methods have been correctly implemented. The first of these tests was against the one-dimensional {\sc dusty} code by \cite{dustycodepaper1997}, in which models of geometrically thick shells of varying optical depths were calculated. Excellent agreement was found for optical depths (at 5500\AA) of up to 100 (see \cite{2004MNRAS.350..565H} for details). We subsequently extended the testing to two dimensions by using the  disc benchmark described by \citet{2004A&A...417..793P}. Once again, excellent agreement between \torus\ and the benchmark temperature runs and SEDs was found (again see \cite{2004MNRAS.350..565H} for details).

A caveat of the \cite{2004A&A...417..793P} benchmark is that it does not really represent protostellar discs in terms of optical depth (the maximum midplane optical depth of the disc in the Pascucci benchmark is 100, whereas `real' discs have midplane depths approaching $10^6$). Also, due to limitations of some of the codes they benchmarked, their test problem was  restricted to isotropic scattering. We therefore developed a more challenging test based around an optically-thick flared disc containing anisotropically-scattering, micron-sized grains and compared temperature profiles, scattered light images and linear polarisation maps. This new benchmarked problem better represents real-world problems, in which it is very hard for radiation to penetrate the disc midplane, and where spatial resolution at the sharp disc inner-edge is important for correctly producing the SED. Seven codes were involved in the benchmarking, and good agreement was found. Details of this benchmark are presented in \citet{2009A&A...498..967P}, and the benchmark data is available online\footnote{{\url{ http://ipag.osug.fr/~pintec/benchmark/index.shtml}}}.

%\end{figure}

\section{Photoionisation}
\label{sec:photo}
In section \ref{sec:radeq} we discussed how \textsc{torus} uses Monte Carlo radiative transfer to solve for the radiative equilibrium temperature of dust grains. A very similar approach can be used to solve for the ionisation and thermal structure of gases that are exposed by a sufficiently high extreme ultraviolet (EUV) flux such that they are predominantly photoionised. 

The ionisation structure of such a gas is solved in equilibrium by considering the balance between the rates of photoionisation and recombination, i.e. 
\begin{equation}
	\frac{n(X^{i+1})}{n(X^i)} = \frac{1}{\alpha(X^i) n_e} \int_{\nu_1}^{\infty}\frac{4\pi J_{\nu} a_{\nu}(X^i) d\nu}{h\nu}
	\label{ionBalance}
\end{equation}
where $n(X^i)$, $\alpha(X^i)$, $a_{\nu}(X^i)$, $n_e$ and $\nu_1$ are the number density of the $i^{th}$ ionization state of species $X$, recombination coefficient, absorption cross section, electron number density and the threshold frequency for ionization of species $X^i$ respectively \citep{1989agna.book.....O}. So, just like for the radiative equilibrium calculation we require an estimate of the radiation energy density in each cell, which can similarly be made using a path length summation. 

The radiation energy from photon sources is hence split into  equal energy packets, probabilistically sampled based on the stellar properties (section \ref{sec:sources}) and propagated throughout the grid using a method akin to that described in section \ref{sec:radeq}. In terms of Monte Carlo estimators the ionisation balance then becomes
\begin{equation}
    \frac{n(X^{i+1})}{n(X^i)} = \frac{\epsilon}{\Delta t V \alpha(X^i)n_e}\sum\frac{\ell a_{\nu}(X^i)}{h\nu}.
\end{equation}
Note that solving the ionisation balance in this manner only works if one can solve for successive ionisation states. In higher energy photon (e.g. X--ray) regimes, a single photon can liberate more than a single electron (via so called ``inner shell ionisation'') which have to be probabilistically modelled using Auger yields \citep[e.g.][]{1998PASP..110..761F, 2008ApJS..175..534E}. This is not currently treated by \textsc{torus}, so the photoionisation components are not suited to X-ray dominated scenarios, such as the narrow line regions of active galactic nuclei. Note however that the parameterised X-ray heating scheme of \cite{2012MNRAS.422.1880O} is incorporated, but is only applicable for internal photoevaporation of protoplanetary discs around $\leq1$\,M$_\odot$ stars \citep{2016MNRAS.457.1905H}.  

The temperature is determined by balancing the heating and cooling rates. The heating contributors are from hydrogen and helium ionisation, as well as dust, while the cooling processes considered are those from free-free radiation,
hydrogen and helium recombination and collisional excitation of
hydrogen and metals. Additionally there is blackbody radiative cooling from dust. In locating the temperature, the heating and cooling rates are initially computed for the maximum and minimum allowed temperatures (10 and $3\times10^4$\,K by default) and then refined by bisection. Because the ionisation and thermal structure determines the opacity, the procedure discussed so far in this section has to be repeated iteratively until convergence, which we define as when the maximum fractional change in temperature on the grid is less than some threshold (typically 1\,per cent). Generally, \textsc{torus} photoionisation calculations start with a relatively modest number of photon packets that double with each iteration until convergence.

In 1D and 2D calculations, photon packets are still propagated in 3D space, utilising the symmetry of the problem to translate the packet trajectories back onto the grid. Almost exactly the same code is therefore used in each geometry.

The photoionisation implementation in \textsc{torus} was presented across \citet{2012MNRAS.420..562H}, \citet{2012MNRAS.426..203H} and \citet{2015MNRAS.453.2277H}.

\subsection{Photoionisation species and atomic data}
\label{sec:photoionization_species}
Presently \textsc{torus} includes the following states: H, He, C(I-IV), N(I-IV), O(I-III), Ne(II-III) and S(II-IV). The hydrogen, helium and C\,IV recombination rates used by \textsc{torus} are calculated based on \citet{1996ApJS..103..467V}. Other radiative recombination rates are calculated using fits to the results of \citet{1983A&A...126...75N}, \citet{1991A&A...251..680P} or \citet{1982ApJS...48...95S}. The photoionisation cross sections are calculated using the \textsc{phfit2} routine from \citet{1996ApJ...465..487V}. The abundances of each species (the sum of all ions of a given species) is assumed, rather than calculated explicitly (i.e. through reactions). By default the abundances are those used in the HII40 Lexington benchmark (see Table \ref{tab:photo_abundances}), however the abundances can be specified in the parameters file. 

\begin{table}
    \centering
    \begin{tabular}{|l|c|}
    \hline
    Species     & Default abundance  \\
    \hline
    log$_{10}$(He/H) & -1 \\
    log$_{10}$(C/H) & -3.66 \\
    log$_{10}$(N/H) & -4.40 \\
    log$_{10}$(O/H) & -3.48 \\
    log$_{10}$(Ne/H) & -4.30\\
    log$_{10}$(S/H) & -5.05 \\
    \hline
    \end{tabular}
    \caption{Default photoionisation species and abundances}
    \label{tab:photo_abundances}
\end{table}

\subsection{Dust in photoionisation models}
Dust is included in photoionisation calculations as an opacity source (see section \ref{sec:radeq}) and also contributes radiative heating (equation \ref{equn:adot})
and blackbody cooling (equation \ref{eq:dotE}). The gas and dust have the freedom to be thermally decoupled, with the dust temperature being the equilibrium value between the dust heating and cooling rates. An additional rate is added to account for collisional heat transfer between gas and dust, according to \cite{1979ApJS...41..555H}; this gas-dust transfer rate is
\begin{equation}
	\label{eq:gasgraincool}
	\Gamma_{\textrm{gas-dust}}= 2 f n_H n_d \sigma_d v_p  k_B (T_g - T_d)
\end{equation}
where $n_d$, $\sigma_d$, $T_d$ are the number density, cross-section and temperature of dust grains, $v_p$ is the thermal speed of protons at the gas temperature $T_g$, and $f$ depends on the ionization state and gas temperature. Depending on the sign of $(T_g - T_d)$, this can be a heating rate or a cooling rate for the gas (and vice versa for the dust). At present the dust and gas in \textsc{torus} are assumed to be dynamically well coupled (see section \ref{sec:hydro}). However, a range of grain types can be specified to give a spatially varying dust distribution, as discussed in section \ref{sec:grainDescr}.

%missing things like photoelectric heating and resonant line transfer (for now)

\subsection{Hybrid methods for optically thick regions}

In dust-only radiative equilibrium calculations, optically thick regions are treated with the diffusion approximation in place of Monte Carlo radiation transport. In photoionisation calculations a packet splitting approach is used \citep{2015MNRAS.448.3156H}. In optically thick regions small numbers of higher energy photon packets undergo the lengthy random walk. Once these packets reach an optically thin medium they are split into a large number of lower energy packets that (being in an optically thin region) will undergo few reprocessings but due to their larger numbers will still sample the optically thin region well.

\subsection{Validation}
\label{sec:photoionisation_validation}
In Figure 1 of \citet{2012MNRAS.420..562H} \textsc{torus} was shown to give agreement with other codes such as \textsc{cloudy} and \textsc{mocassin} in the Lexington HII40 benchmark of \citet{1995aelm.conf...83F}, in which the ionisation fraction and temperature profile about an O star is computed (neglecting dust). In Figure 2 of \cite{2015MNRAS.453.2277H} \textsc{torus} was also shown to give good agreement with \textsc{mocassin} for an otherwise identical benchmark, only with the inclusion of dust.

\section{Time dependent radiative transfer}
\label{sec:timedep}

\iffalse
For a gas at temperature $T$ the rate at which it
emits energy is given by
\begin{equation}
\dot{E} = 4 \pi \int_0^\infty k_\nu B_\nu \, d\nu,
\label{eq:emissivity}
\end{equation}
where $k_\nu$ is the absorption coefficient and $B_\nu$ is the Planck
function. The rate at which the same gas absorbs energy is given by
\begin{equation}
\dot{A} = 4 \pi \int_0^\infty k_\nu J_\nu \, d\nu,
\label{eq:absorption-rate}
\end{equation}
where $J_\nu$ is the mean intensity of the radiation field. Clearly if
the gas is in radiative equilibrium then $\dot{A}=\dot{E}$ and we find
\begin{equation}
T = \left( \frac{\dot{A}}{4 \sigma \kappa_P} \right)^{1/4}
\label{eq:rad-eq-time}
\end{equation}
where $\kappa_P$ is the Planck-mean absorption coefficient (these have repeated equations 7-9). 
\fi

In section \ref{sec:radeq} we discussed dust radiative equilibrium calculations. However, Monte Carlo radiative transfer can also be used in non-equilibrium scenarios, as introduced by \cite{2011MNRAS.416.1500H}. In the following the gas and dust are assumed to be thermally coupled ($T_g=T_d$=T). If
we consider gas that is not in radiative equilibrium then the net
change in energy density of the gas
\begin{equation}
\dot{u}_g = \dot{A} - \dot{E}
\label{eq:a-minus-e}
\end{equation}
and similarly the rate of change in energy density of
the radiation field is
\begin{equation}
\dot{u}_r = \dot{E} - \dot{A}.
\label{eq:e-minus-a}
\end{equation}
Now we consider a gas of volume $V$ at time $t$. Within this volume is
a star of luminosity $L_*$.  The luminosity of the gas is given by
\begin{equation}
L_g = \int_V \dot{E} \, dV.
\end{equation}
We assume that the temperature of the gas is constant over a single
timestep $\Delta t$. During this timestep we assume that the gas and
the star produce $N_g$ and $N_*$ new  photon packets
respectively. The gas and stellar individual photon packet energies are given by
\begin{equation}
\epsilon_g = \frac{L_g \Delta t}{N_g}, \quad  \epsilon_* =
\frac{L_* \Delta t}{N_*} \\
\end{equation}
respectively.
The energy density is related to the temperature by
\begin{equation}
u_g =  \frac{ R T \rho}{(\gamma-1) \mu}
\label{eq:gas-energy-density}
\end{equation}
where $R$ is the gas constant, $\rho$ is the mass density, $\gamma$ is
the ratio of specific heats and $\mu$ is the mean molecular weight. We
follow \cite{1999A&A...344..282L} and use the result that the energy density of
the radiation field (denoted by subscript $r$) in the interval $(\nu, \nu+d\nu)$ is given by
\begin{equation}
u_{r,\nu} = \frac{4 \pi J_\nu}{c} \textrm{d}\nu.
\label{eq:energy-density}
\end{equation}
A photon packet moving between events (scatterings, absorptions,
crossing grid-cell boundaries) contributes an energy $\epsilon_\nu$ for a
time $\ell / c$ (where $\ell$ is the distance between events) to the
local energy density. The photon energy density is therefore
\begin{equation}
u_{r,\nu} = \frac{1}{\Delta t} \frac{1}{V} \frac{1}{c} \sum \epsilon_{\nu} \ell.
\label{eq:energy-summation}
\end{equation}
Now combining equations \ref{eq:absorption-rate} and
\ref{eq:energy-density} with equation \ref{eq:energy-summation} we
obtain an expression for the energy absorption rate:
\begin{equation}
\dot{A} = \frac{1}{V} \frac{1}{\Delta t} \sum k_\nu  \epsilon \ell
\label{eq:new_absorption-rate}
\end{equation}
The new energy density of the gas may then be calculated
\begin{equation}
u^{n+1}_g = u^{n}_g + (\dot{A} - \dot{E}) \Delta t.
\label{eq:ug-update}
\end{equation}

The calculation proceeds by looping over photon packets, each of which has an individual energy $\epsilon_\nu$ and frequency $\nu$. The random walk of the each packet is followed until it is {\it (i)} absorbed, in which case it is removed from the calculation, {\it (ii)} passes across the edge of the grid, or {\it (iii)} has completed a time-of-flight of equivalent to $\Delta t$, in which case the packet is passed to a stack of packets to be processed along with the new photon packets generated at the start of the subsequent timestep. The number of packets on the stack is $N_s$. Clearly as the computation starts, $N_s$ grows, but eventually at equilibrium $N_s$ will reach an approximately constant value. If the absorption mean free path of photons is small (i.e. the material is optically-thick and highly absorbing) then $N_s$ will be small (most packets will be absorbed during the timestep). However, if the medium is highly scattering, or the computational domain is large compared to $c\Delta t$, then $N_s$ can be large.

The algorithm itself follows a sequence of steps:
\begin{enumerate}
    \item The energy density of the gas is used to compute the temperature distribution of the gas via equation~\ref{eq:gas-energy-density} and therefore $\dot{E}$ via equation~\ref{eq:dotE}. This can then be used to calculate the probability of a packet being emitted by gas in cell $i$ via
    \begin{equation}
        p_i = \frac{\sum_{j=1}^i \dot{E}_j V_j}{\sum_{j=1}^{N_{cells}} \dot{E_j}{V_j}}
    \end{equation}
    where $V_j$ is the volume of the $j$th cell.

\item The probability of a packet being produced by the gas is calculated by 
\begin{equation}
    p_g = \frac{L_g}{L_* + L_g}
\end{equation}
Taking $\eta$ to be a uniform random deviate, then if $\eta < p_g$ then the photon packet is emitted in the gas, otherwise it is produced by the star.

\item We now need to determine the frequency of the packet. For the gas this is found by
\begin{equation}
\eta = \int_0^\nu j_\nu d\nu \Bigg/ \int_0^\infty j_\nu d\nu
\end{equation}
where $j_\nu$ is the emissivity of the gas. Alternatively if the photon packet comes from the star, then its frequency can be determined from
\begin{equation}
    \eta = \int_0^\nu F_\nu d\nu \Bigg/ \int_0^\infty F_\nu d\nu
\end{equation}
where $F_\nu$ is the photospheric flux. The packet time $t_{p}$ is zeroed at this stage.

\item The photon packet is then propagated an optical depth given by equation~\ref{eq:opticaldepth}. The physical distance corresponding to this optical depth is then calculated: $\ell = \tau / (k_\nu + k_{{\rm sca}, \nu})$. We define the time to travel distance $\ell$ as $t_\ell = \ell/c$. If $t_\ell + t_p > \Delta t$ then the packet is propagated distance $\ell  (\Delta t - t_p)/t_\ell$ and the packet is added to the stack and $t_p$ is updated. If the distance to the cell boundary is less than $\ell$ then the packet is moved to  the cell boundary and $t_p$ is updated, a new random $\tau$ is then calculated. Alternatively the photon packet interacts with the gas. Defining the albedo as
\begin{equation}
    \alpha = \frac{k_{{\rm sca},\nu}}{k_\nu + k_{{\rm sca},\nu}}
\end{equation}
then if $\eta < \alpha$ the photon is scattered, and a new direction for the packet is chosen randomly from the appropriate phase function, and $t_p$ is updated. Otherwise the photon is absorbed and the next photon packet is selected.

\item Once all the packets have been processed (including those on the stack), the new absorption rates are computed and the energy density of the gas is updated. The calculation proceeds to the next time step.

\end{enumerate}

\subsection{Validation}
\cite{2011MNRAS.416.1500H} validated the time-dependent radiative transfer scheme across a series of tests. These included the transition to radiative equilibrium of a medium immersed in a higher radiation density bath, and a diffusion scenario, both of which have an analytic solution. \cite{2011MNRAS.416.1500H} also presented the time dependent evolution of the thermal properties of the \cite{2004A&A...417..793P} disc benchmark, through to the steady state of the benchmark solution.

\section{Molecular line transfer}
\label{sec:molecular}

Given a molecular abundance, \textsc{torus} can solve for the level populations of a molecule. The result of this can be used to produce synthetic position-position-velocity (PPV) molecular line data cubes, as we discuss in section \ref{sec:ppv}.

The abundance of the molecule is either assumed, derived from another code which \textsc{torus} is postprocessing, or can be computed explicitly using the \textsc{torus-3dpdr} tools that will be discussed more in section \ref{sec:pdr}. 

For the simplest case of an assumed abundance, processes such as freeze out and dissociation can be trivially approximated by specifying density/temperature regimes in which the molecule is depleted (i.e. the abundance set to a much lower value). For example for species like CO the molecule is expected to be depleted by freeze out onto grains for temperatures of around 20\,K and densities of $\sim10^4$\,cm$^{-3}$. It could also be dissociated above a threshold temperature. 

Some molecules such as CH$_3$CN have more complex dependencies that are sensitive to the coupled density-temperature (e.g. freeze out at different temperatures for different densities). More complex behaviour like this can still be accounted for in a straightforward manner. Generally, we construct a specific molecular depletion prescription for the specific molecule being used in a given application.

\subsection{Solving the level populations}
If the assumption of local thermodynamic equilibrium (LTE) is valid,  the fraction of a molecule in level $i$ with statistical weight $g_i$ in gas of temperature $T$ is trivially computed using a Boltzmann distribution, that is
\begin{equation}
    \frac{n_i}{\sum_i n_{i}} = \frac{g_i \exp\left(\frac{E_i}{k_BT}\right)}{z\left(T\right)}
\end{equation}
where $z$ is the partition function. LTE applies where local quantities (rather than gradients) dominate the level populations for example in dense, slowly varying environments. 

For many systems, the assumption of LTE is no longer valid. In this case we are required to solve the detailed balance equations
\begin{equation}
\begin{split}
    n_l\left[\sum_{l<u}A_{ul} + \sum_{l\neq u}\left(B_{ul}J_{\nu} + C_{ul}\right)\right] = \\
     \sum_{l>u}n_kA_{lu} + \sum_{l\neq l} n_l\left(B_{lu}J_{\nu} + C_{lu}\right)
\end{split}
\end{equation}
where $A_{ul}$, $B_{ul}$ and $C_{ul}$ are the Einstein A (for spontaneous emission), Einstein B (for stimulated absorption/emission) and collision rate coefficients for transitions from $u\rightarrow l$ respectively. $J_{\nu}$ is the mean intensity.  The coefficients are species specific constants and do not need to be computed at run time, owing to the efforts of laboratory and theoretical chemists/astrochemists. \textsc{torus} uses molecular data taken from the \textsc{lamda} database \citep{2005A&A...432..369S}.

Solving the detailed balance equation requires an estimate of the mean intensity. Unlike for most other modules, \textsc{torus} uses a cell-centric long characteristic ray tracing scheme (rather than propagation of photon packets) for molecular line transfer calculations. The methodology is based heavily on the accelerated Monte Carlo (AMC) scheme developed by \cite{2000A&A...362..697H} and was first presented in \textsc{torus} by \cite{2010MNRAS.407..986R}. Rays are propagated outwards from random positions in a given cell, with random frequency and direction. The radiative tranfser equation is solved along each ray with a boundary condition typically given by the cosmic microwave background (CMB, though other boundaries can be imposed, e.g. if there is a strong nearby photon source). 

The mean intensity is actually comprised of two components -- one from the ambient medium that doesn't change as the level population in the cell is being solved, and another from the cell itself which does change as the distribution of excited states alters. That is, over $i$ rays
\begin{equation}
    J_{\nu} = J^{ext}_{\nu} + J^{int}_{\nu} = \frac{\sum_i I_{\nu}^i\exp(-\tau_i\phi_{\nu})}{\sum_i\phi_{\nu}} + \frac{\sum_i S_{\nu}\left(1-\exp(-\tau_i)\phi_i\right)}{\sum_i \phi_{\nu}}
\end{equation}
where $S_{\nu}$ is the source function (ratio of emission and absorption coefficients)
\begin{equation}
    \begin{split}
        S_{\nu}=\frac{n_u A_{ul}}{n_lB_{lu} - n_uB_{ul}}
    \end{split}
\end{equation}
and $\tau_i$ the optical depth along ray $i$. $\phi_\nu$ is the line profile function, which is characterised by the microturbulent velocity broadening parameter $v_{turb}$
\begin{equation}
    \label{eqn:lineProfileFunction}
    \phi_{\nu} = \frac{c}{v_{turb}\nu_0\sqrt{\pi}}\exp\left(-\frac{\Delta v^2}{v_{\textrm{turb}}^2}\right).
\end{equation}
\textsc{torus} computes $J_{\nu}$ and hence the level populations in a two-stage process. In the first stage a set of random rays are generated (i.e. each ray has a point of origin in the cell, a frequency sampling the line profile function, and a direction). The radiation field/level populations are iteratively solved until convergence -- which we define as the point at which the root mean square fractional change between iterations in the level populations is less than 10$^{-2}$ for all relevant levels (the maximum relevant level can be user specified so that oscillations in energy levels much greater than the transition being observed do not affect convergence, defaults to all transitions being relevant). In this first phase the same set of initial random rays are used for all iterations. Convergence is therefore relatively fast.

In the second phase of the calculation every ray is random at each iteration. That is a whole new set of rays are chosen each time. This fills in the missing gaps in both frequency and spatial dimensions from the first phase. However, convergence towards a solution requires a large number of rays. \textsc{torus} doubles the number of rays after each iteration until convergence.

Calculations such as this which involve repeated function evaluation often benefit from vector sequence acceleration to reduce the computational time. To this end \textsc{torus} employs the \cite{NgAcceleration} scheme to extrapolate an updated set
of relative level populations from the previous four iterations
of $n_i$.

%LTE/NLTE level population solver given an abundance \citep{2010MNRAS.407..986R}. Drop model to account for freeze out. Abundances can be informed using the \textsc{torus-3dpdr} or \textsc{krome} tools. 
%Molecular line data from the LAMDA database \citet{2005A&A...432..369S}. 

%Pretty much the most popular collaborative part of the code in recent years. Has been used to support a number of ALMA proposals. 

%Can do Ro-vibrational lines now

\subsection{Validation}
In \cite{2010MNRAS.407..986R}, the molecular line transfer was benchmarked against a series of other codes in the \cite{2002A&A...395..373V} model of HCO$^+$ in a collapsing cloud. In this benchmark the radial profile of the level population is computed.

 \subsection{Example}
\begin{figure}
    \centering
    \includegraphics[width=8.5cm]{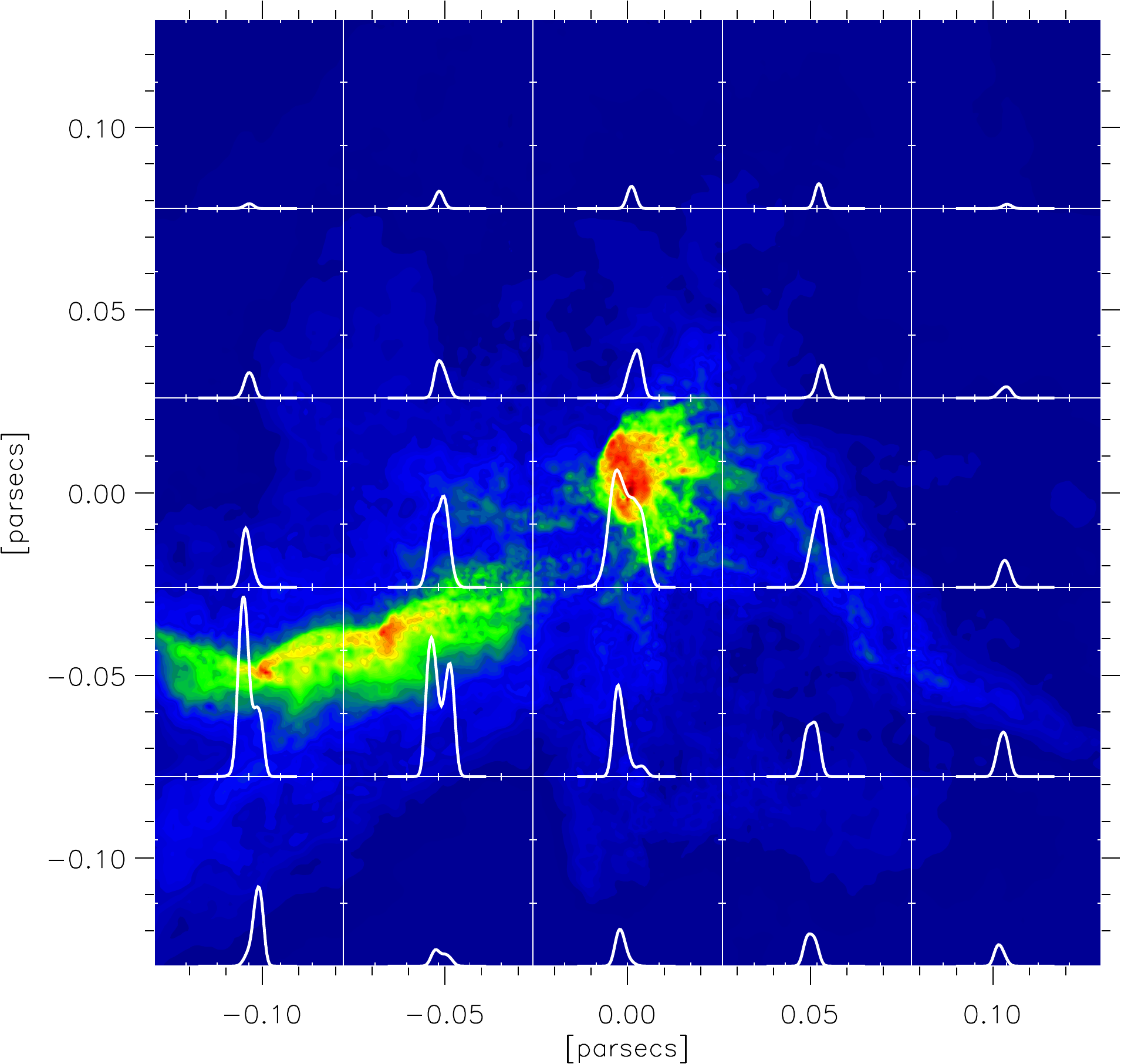}
    \caption{A synthetic N$_2$H$^+$ ($1\rightarrow0$) integrated intensity map with line profiles in each square region superimposed. This Figure originally appeared in Rundle et al. (2010)}
    \label{fig:RundleFigure}
\end{figure}

 An example application of the molecular line transfer from \cite{2010MNRAS.407..986R} is given in Figure \ref{fig:RundleFigure}. This is a NLTE N$_2$H$^+$ integrated intensity map, applied as a postprocessing of the \cite{2003MNRAS.339..577B} SPH calculations of giant molecular cloud collapse and star formation. Superimposed over the integrated intensity map are line profiles for each of the square zones, so this datacube contains both spatial and kinematic information. 
 
\section{Atomic line transfer}
The atomic line transfer method splits into two tasks. Firstly, determining the level populations by assuming statistical equilibrium, and secondly computing line profiles. Both these processes have much in common with the molecular line transport detailed in section~\ref{sec:molecular}. Currently the version of \torus\ described here is limited to pure hydrogen models. Although a helium atom has been implemented it is yet unvalidated. Note that the version of \torus\ used to compute helium emission line profiles by \cite{2012MNRAS.426.2901K} has diverged sufficiently from the `Exeter' version that they are essentially independent codes.

The initial method used for determining the level populations is a core-halo approximation based on the method described by \cite{1978ApJ...220..902K}, in which the radiative transfer is conducted under the Sobolev approximation \citep{1957SvA.....1..678S}. The equation for statistical equilibrium that is solved is the following, in which the populations are calculated up to a user-defined top level (the default is 15), with three levels above this held in LTE:

\begin{multline}
    \sum_{l<u} \left[ N_l \left(B_{lu}\mathcal{J}_{lu} + N_e C_{lu} \right) - N_u \left(  A_{ul} + B_{ul}\mathcal{J}_{lu} + N_e C_{ul} \right)\right] \\
    + \sum_{l>u}\left[N_l \left(A_{lu} + B_{lu}\mathcal{J}_{lu} + N_eC_{lu}\right) - N_u\left(B_{ul}\mathcal{J}_{lu}+N_eC_{ul}\right) \right] \\
    + N_u^*\left[\int_{\nu_u}^\infty\frac{4\pi}{h\nu}a_u\left(\nu\right) \left(\frac{2h\nu^3}{c^2} + J_{\nu}\right)\exp\left(-\frac{h\nu}{k_BT_g}\right)\textrm{d}\nu + N_eC_{uk}\right] \\
    - N_u\left(\int_{\nu_u}^\infty \frac{4\pi}{h\nu}a_u\left(\nu\right)J_\nu\textrm{d}\nu + N_eC_{uk}\right) = 0
\end{multline}
%\begin{multline}
 %   \sum_{m<n} \left[ N_m \left(B_{mn}\mathcal{J}_{mn} + N_e C_{mn} \right) - N_n \left(  A_{nm} + B_{nm}\mathcal{J}_{mn} + N_e C_{nm} \right)\right] \\
%    + \sum_{m>n}\left[N_m \left(A_{mn} + B_{mn}\mathcal{J}_{mn} + N_eC_{mn}\right) - N_n\left(B_{nm}\mathcal{J}_{mn}+N_eC_{nm}\right) \right] \\
%    + N_n^*\left[\int_{\nu_n}^\infty\frac{4\pi}{h\nu}a_n\left(\nu\right) \left(\frac{2h\nu^3}{c^2} + J_{\nu}\right)\exp\left(-\frac{h\nu}{k_BT}\right)\textrm{d}\nu + N_eC_{nk}\right] \\
%    - N_n\left(\int_{\nu_n}^\infty \frac{4\pi}{h\nu}a_n\left(\nu\right)J_\nu\textrm{d}\nu + N_eC_{nk}\right) = 0
%\end{multline}
Here $A_{lu}$ and $B_{lu}$ are the Einstein coefficients for quantum levels $l$ and $u$, and $C_{lu}$ is the collisional rate coefficient. The index $k$ refers to the continuum state. $N_u$ is the level population of level $u$, and $N_u^*$ is the level population for the $u^{th}$ level given by the Saha-Boltzmann equation in terms of the electron density $N_e$, gas temperature $T_g$, and $\alpha_u(\nu)$  is  the photoionisation cross section of level $u$. The angle-averaged, profile weighted intensity of the radiation field in the line transition between levels $m$ and $n$ is represented by $\mathcal{J}_{lu}$.

We can determine  $\mathcal{J}_{lu}$ using the Sobolev escape probability formalism:
\begin{equation}
    \beta_{lu} = \frac{1}{4\pi} \oint_{4\pi} e^{-\tau_{lu}({\bf n})}d\Omega
\end{equation}
where $\tau_{lu}({\bf n})$ is the Sobolev optical depth along unit vector ${\bf n}$ given by
\begin{equation}
    \tau_{lu}({\bf n}) = \frac{\pi e^2}{m_e c}\left(g_l f_{lu} \right) \frac{1}{\nu} \frac{1} {{\bf n}\cdot {\bf \nabla v}} \left(\frac{N_l}{g_l}-\frac{N_u}{g_u}\right)
\end{equation}
Here $g$ and $f$ are the statistical weights of the lower level and the oscillator strength of the line transition respectively. $e$ and $m_e$ are the electron charge and mass respectively. We also need
\begin{equation}
    \beta_{c,lu} = \frac{1}{\Omega_{disc}} \oint_{4\pi} \frac{1 - e^{-\tau_{lu}({\bf n})}}{\tau_{lu}(\bf n)} d\Omega
\end{equation}
where $\Omega_{disc}$ is the solid angle subtended by the stellar photosphere. This allows us to calculate
\begin{equation}
    \mathcal{J}_{lu} = \left(1-\beta_{lu}\right)\frac{2h\nu_{lu}^3}{c^2}\left[\frac{g_u}{g_l}\frac{N_l}{N_u}-1    \right]^{-1} + \beta_{c,lu}I_{c,lu} , \quad l<u
\end{equation}
where $g_l$ and $g_u$ are the statistical weights and $I_{c,lu}$ is the intensity of the continuum at the frequency of the transition. 

Since we assume pure hydrogen we have
\begin{equation}
N_{\rm e} = N({\rm H})^+
\end{equation}
and the conservation equation is 
\begin{equation}
\sum_{n=1}^{n_{\rm max}} N_n + \sum_{n=n_{\rm max}}^{n_{\rm max}+3} N_n^* + N({\rm H}^+) = \frac{\rho}{m_{\rm H}}
\end{equation}

The above system of equations is solved for each cell independently using a Newton-Raphson iterative scheme, yielding the line and continuum emissivity and opacity for each cell, which are subsequently used for line profile calculations. The line profiles may be computed using a co-moving frame formal solution to the RT equation, which allows for pressure  broadening. The method of calculating the level populations using the Sobolev approximation but the line profiles with co-moving frame is often called Sobolev with exact integration (SEI).The formal solution does not account for scattered radiation (either by dust or electrons) and only follows the total intensity $I$ rather than the Stokes intensities. If one is interested in the line polarization then it is possible to use the line emissivities and opacities to calculate the line profile using a MC loop in a very similar fashion to that used for computing the dust continuum. The Sobolev approximation is adopted for the line transfer in this case.

Alternatively one can use a co-moving transfer method to calculate the level populations. There is no restriction on the velocity fields in this mode, and the statistical equilibrium is performed using an accelerated Monte Carlo method which is the same as that used for the molecular transport in section~\ref{sec:molecular}, and is inevitably significantly more computationally expensive.

\subsection{Validation}

We have published several tests of the atomic line transport methods \citep{2000MNRAS.315..722H}. Specifically we have computed H$\alpha$ line profiles for a spherically symmetric stellar wind model using \torus\ and compared then with both a formal solution code {\sc linpro} \citep{1995PhDT.......168H} and the {\sc elec} Monte Carlo code \citep{1991A&A...247..455H}.

We have also tested the linear polarisation produced by the code using  a simple latitudinal density structure
\begin{equation}
    f(\mu)=k(1-x\mu^2)
\end{equation}
where $\mu$ is the cosine of the polar angle, $x$ is a density contrast factor, and $k$ is a normalization factor,
chosen so to conserve the mass-loss rate compared to the
undistorted model. The line emissivities and opacities (which result from recombination processes) were scaled by $f(\mu)$. We computed the continuum polarisation of the models as a function of $x$ and compared the results with the analytical formalism of \cite{1991ApJ...379..663F} and {\sc elec}. We found good agreement with the {\sc elec} results for all $x$, but that the electron-scattering depth of our fiducial model ($\tau=0.15$) was sufficient to make multiple scattering important which led to deviations from the analytical model. Models with lower mass-loss rates showed excellent agreement with the \cite{1991ApJ...379..663F} solution.

We also tested our method using dipolar magnetospheric accretion models, based on the geometry described by \cite{2001ApJ...550..944M}. We computed Paschen~$\beta$ and Bracket~$\gamma$ line profiles for one of the publicly available Muzerolle models and found satisfactory agreement \citep{2005MNRAS.356.1489S}.

\subsection{Example}

\begin{figure}
    \hspace{-0.3cm}
    \includegraphics[width=9cm]{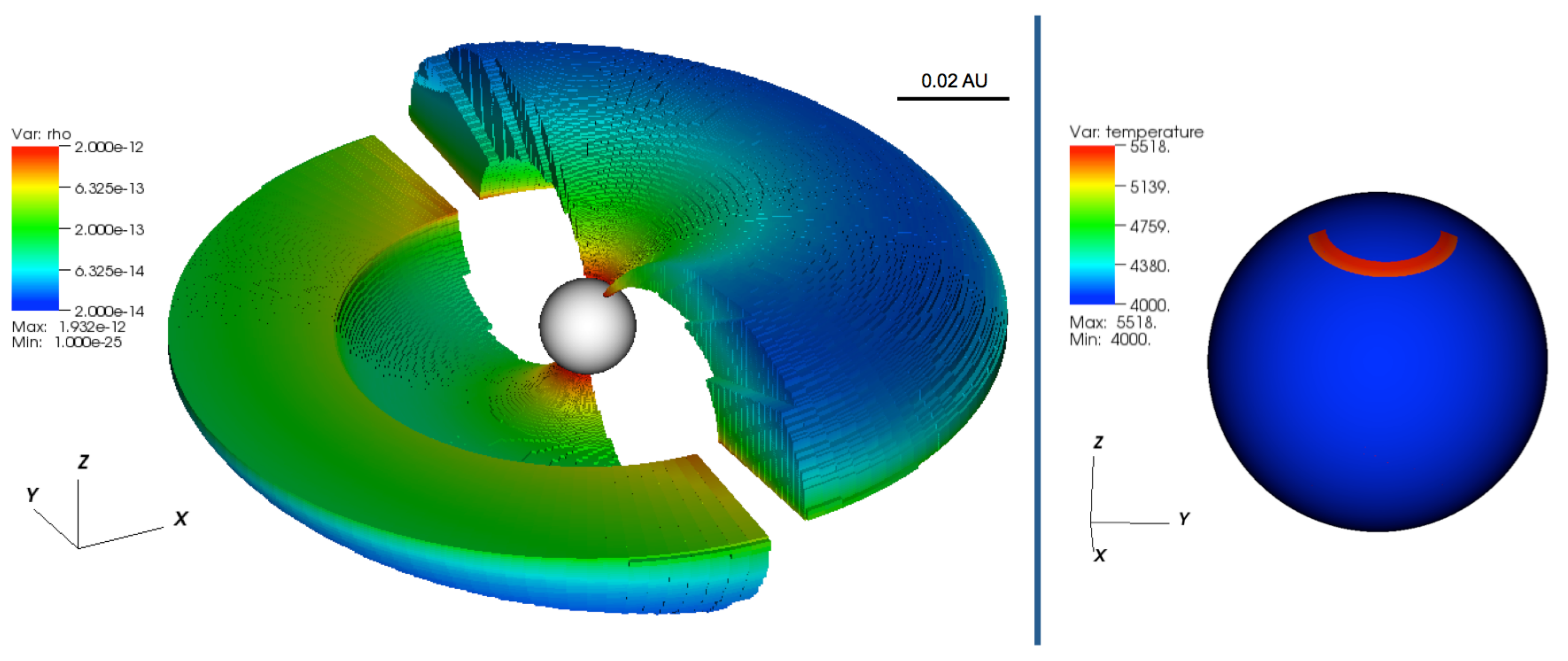}

    \hspace{-0.3cm}
    \includegraphics[width=9cm]{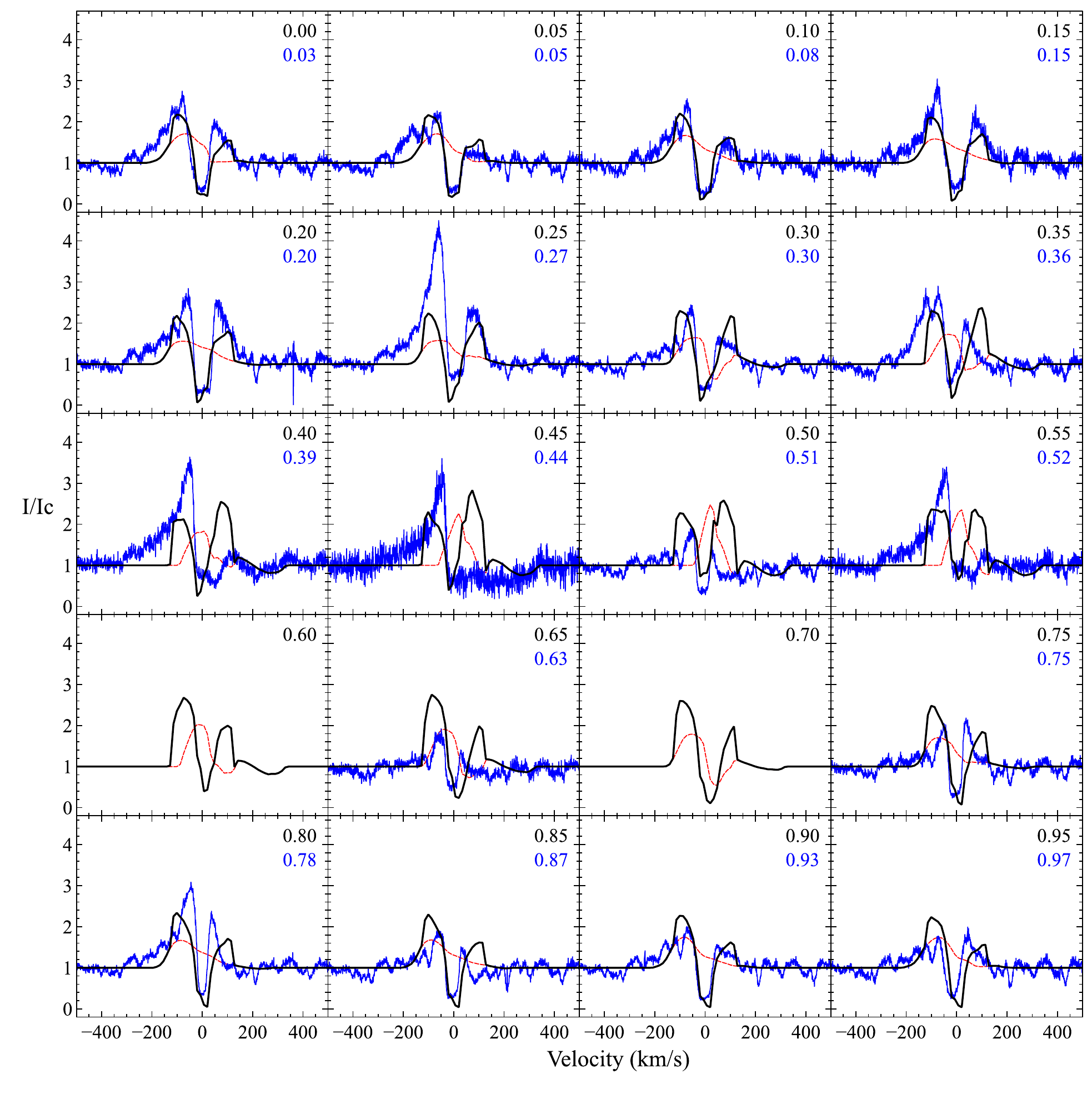}
    \caption{An example of an atomic line transfer calculation applied to the case of the classical T Tauri star AA Tau, taken from Esau et al. (2014). The upper panel shows the model geometry, with the star and gas density distribution of the inner disc on the left and an illustration of the stellar temperature, including the magnetospheric accretion hotspot, on the right. The lower panel shows observational line profiles of the H$\beta$ line, alongside the best \textsc{torus} model. The frames of the lower panel are ordered by rotational phase. }
    \label{fig:EsauAtomicLine}
\end{figure}

An example of the atomic line transfer in \textsc{torus} is given in Figure \ref{fig:EsauAtomicLine}.  This is a model of the classical T Tauri star AA Tau, taken from \cite{2014MNRAS.443.1022E}. The upper panels illustrate the density and temperature structure of the inner disc and stellar surface respectively. The lower panels compare observations of the  H\,$\beta$ lines profiles computed by \textsc{torus} with real observations taken by the  ESO
La Silla 3.6 m telescope with the HARPS high-resolution echelle spectrograph by \cite{2007A&A...463.1017B}. Different frames in this lower panel correspond to different orbital phases.  With these models \cite{2014MNRAS.443.1022E} constrained the geometry, accretion rate and outflow properties of the inner disc of AA Tau. 

\section{Hydrodynamics}
\label{sec:hydro}
\textsc{torus} is first and foremost a radiation transport code which is reflected in the breadth of microphysics available in the code. However the addition of a relatively basic hydrodynamics scheme allows \torus\ to offer a sophisticated radiation hydrodynamics framework in non-magnetised, non-relativistic regimes (see sections \ref{sec:radhydro}, \ref{sec:pdr}). The radiation hydrodynamics scheme has been applied on a range of physical scales from the formation of a single massive star up to Galactic star formation regions. 
Since the hydrodynamics scheme used contains well known components we provide only a brief overview of the central algorithm, expanding our discussion where appropriate.

\subsection{Algorithm overview}

\textsc{torus} solves hydrodynamic problems on its mesh using a finite volume hydrodynamics scheme that is 2nd order in space and 1st order in time.  It is total variation diminishing (TVD) which means that shocks are captured whilst minimising unphysical oscillations \citep{1997JCoPh.135..260H}. It also uses a Rhie-Chow interpolation scheme to avoid decoupled pressure fields that can result in a ``checkerboard effect'' \citep{1983AIAAJ..21.1525R}. Flux limiters include the Van Leer \citep{vanleer}, superbee, minmod and mc, with the Van Leer being the default as it offers low diffusivity without oversteepening as a superbee scheme tends to. Grid quantities are stored at cell centres and advection proceeds by constructing fluxes at cell interfaces.

For purely hydrodynamic scenarios an adiabatic equation of state is used. In radiation hydrodynamic models (see section \ref{sec:radhydro}) an isothermal equation of state is used with temperature/pressure set by the radiative transfer scheme. Gravity and radiation pressure can also appear as source terms (see sections \ref{sec:gravity} and \ref{sec:radpressure} respectively). 

\textsc{torus} does not currently include a prognostic turbulence closure scheme but viscosity can be included in the core hydrodynamics module using either a von Neuman-Richtemeyer viscosity \citep{1950JAP....21..232V} or a viscous stress tensor scheme. For 2D cylindrical simulations of discs the viscosity in terms of an $\alpha$ parameter at sound speed $c_s$ and disc scale height $H$ can also be specified, i.e. 
\begin{equation}
\alpha = \frac{\nu}{c_sH}
\end{equation}
which parameterises the viscous behaviour required to produce observed accretion rates through the disc that molecular viscosity alone cannot induce

A series of constraints on the time step are implemented depending on the physics included. The most basic is the Courant-Friedrichs-Lewy (CFL) criterion with a default CFL parameter of 0.3 \citep{1928MatAn.100...32C}. This prevents material being advected by more than one cell in a given time step (doing so would not guarantee conservation laws are upheld). Extra constraints are also available for calculations with additional physics. For example requiring that the time step be less than the ionisation/recombination timescales and small enough that radiation pressure doesn't drive material accross more than one cell. 

At this stage in the development of \textsc{torus}, dust is assumed to be dynamically coupled to the gas. In practice decoupled dust-gas dynamics is becoming more commonplace (particularly in disc applications), but there is currently a lack of a general, universally accepted scheme that is guaranteed to solve such dynamics in arbitrary scenarios \citep[see e.g.][for further information]{1992ARA&A..30...11D,2007A&A...462..355P, 2008A&A...479..883L, 2011MNRAS.415.3591J, 2011MNRAS.418.1491L,  2015P&SS..116...48G, 2016PASA...33...53H}

\subsection{Fluid dynamics and gravity}
\label{sec:gravity}
The dynamics can be influenced by both the gravity from point sources (such as stars) and, if there is enough mass, from non-stellar matter (i.e. gas and dust) on the grid. The $i^{th}$ point source of mass $M_i$ at $r_{star, i}$ adds to the potential at point $r$ through simple $N-$body physics
\begin{equation}
	\phi_{stars} = \sum_{i}  \frac{G M_i}{\sqrt{|r-r_{star, i}|^2 + \delta^2}}
\end{equation}
where $\delta$ is the smallest cell size on the grid, to avoid this term yielding floating point exceptions for cells within which a stellar source resides. The equation of motion of the sink particles is integrated over a timestep by using  the Bulirsch-Stoer method \citep{Press:1993:NRF:563041}, using the cubic spline softening of \cite{2007MNRAS.374.1347P}.

The gas self gravity requires solution of Poisson’s equation
\begin{equation}
    \nabla^2\phi = 4\pi G \rho
\end{equation}
which we do for a linearised version using a V-cycling multigrid Gauss-Seidel method.
Dirichlet boundary conditions are employed which are calculated using a multipole expansion with Legendre polynomials of the matter interior to the boundary.

The total potential from point potentials and gas is hence
\begin{equation}
	\phi_{tot} = \phi_{stars} + \phi_{gas}. 
\end{equation}
This feeds into the hydrodynamics as a source term.  

\subsection{Sink particles and N-body physics}

The collapse of a gas cloud under gravity can only proceed so far before the calculation is rendered intractable by the increasing density and diminishing time step. Sink particles \citep{1995MNRAS.277..362B} resolve this issue by, under certain conditions, transforming collapsing gas into mass reservoirs of finite radius that interact with the fluid field only through gravity (unless the sink is also emits radiation, magnetic field or feedback in the form of winds or jets). 

We adopt the same criteria for sink particle creation as \cite{2010ApJ...713..269F}.
For the cell under consideration we define a control volume that contains all cells within a predefined radius $r_{\rm acc}$.  Before a sink particle is created
 a number of checks on the hydrodynamical state of the gas in the control volume must be passed, briefly:
\begin{itemize}
\item The central cell of the control volume must have the highest level of AMR refinement.
\item The density of the central cell must exceed a predefined threshold density $\rho_{\rm thresh}$, thus ensuring that ${\bf \nabla} \cdot (\rho {\bf v}) < 0$ for that cell.
\item Flows in cells along the principle axes must be directed towards the central cell.
\item The gravitational potential of the central cell must be the minimum of all the cells in the control volume.
\item The control volume must be Jeans unstable i.e. $|E_{\rm grav}| > 2E_{\rm th}$.
\item The gas must be in a bound state i.e. $E_{\rm grav} + E_{\rm th} + E_{\rm kin} < 0$, where $E_{\rm kin}$ is the kinetic energy of the gas in the control volume where the speeds are measured relative to the velocity of the centre of mass of the control volume.
\item The control volume must not overlap with the accretion radius of any pre-existing sink particles.
\end{itemize}
If these tests are passed then a sink particle is created at the centre of the control volume, and accretes gas according to the method detailed below.

The gravitational influence of the gas on a given sink particle is found by summing up the gravitational forces from all the cells in the AMR grid. For all cells except that containing the sink particle we use
\begin{equation}
{\bf F}_j = \sum_i G M_j \rho_i V_i  s(r_{ij},h){\bf \hat{r}}
\label{equation_gasforces}
\end{equation}
where $M_j$ is the mass of the sink particle, and $r_{ij}$ and ${\bf \hat{r}}$ are the distance and direction vector between the cell centre and the sink particle respectively, $h$ is the gravitational softening length, and $s(r_{ij},h)$ is a cubic spine softening function (see equation~A1 of \citealt{2007MNRAS.374.1347P}).

%For the cell containing the sink particle we instead perform a sub-grid calculation by splitting the mass of the cell into $8^3$  subcells and sum the force over the subcells in an analogous manner equation~\ref{equation_gasforces}.

\subsection{Boundary conditions and ghost cells}
A number of dynamic boundary conditions are available. These are imposed by ghost cells, which are the 2 layers of cells at the grid edge. Ghost cells do not evolve in any way not permitted by the boundary conditions.  

The boundary conditions are 
\begin{itemize}
\item{reflecting: Material encountering this boundary sees a reflection of itself. That is, if the $i$ and $i+1^{th}$ cell contain $A$ and $B$ respectively, the next two cells which are the ghosts will contain $B$ and $A$ respectively. }
\item{periodic: Material leaving one side of the grid (e.g. at the $+x$ boundary) will re-emerge at the other ($-x$).  }
\item{free outflow, no inflow: material is permitted to flow freely of of the grid, but cannot return onto the grid.}
\item{inflow: A flow of material onto the grid is specified. For example, if a model considers a clump propagating with given speed through an infinite medium of constant density, an inflow of material with speed set by the clump speed is required to maintain the ambient medium. Inflow boundary conditions can also be specified in such a way that there are gradients or time variations in the properties of the inflowing material.   }
\item{fixed: The ghost cells maintain constant conditions regardless of what is going on in the grid.  }
\end{itemize}

\begin{figure}
    \centering
    \includegraphics[width=8.5cm]{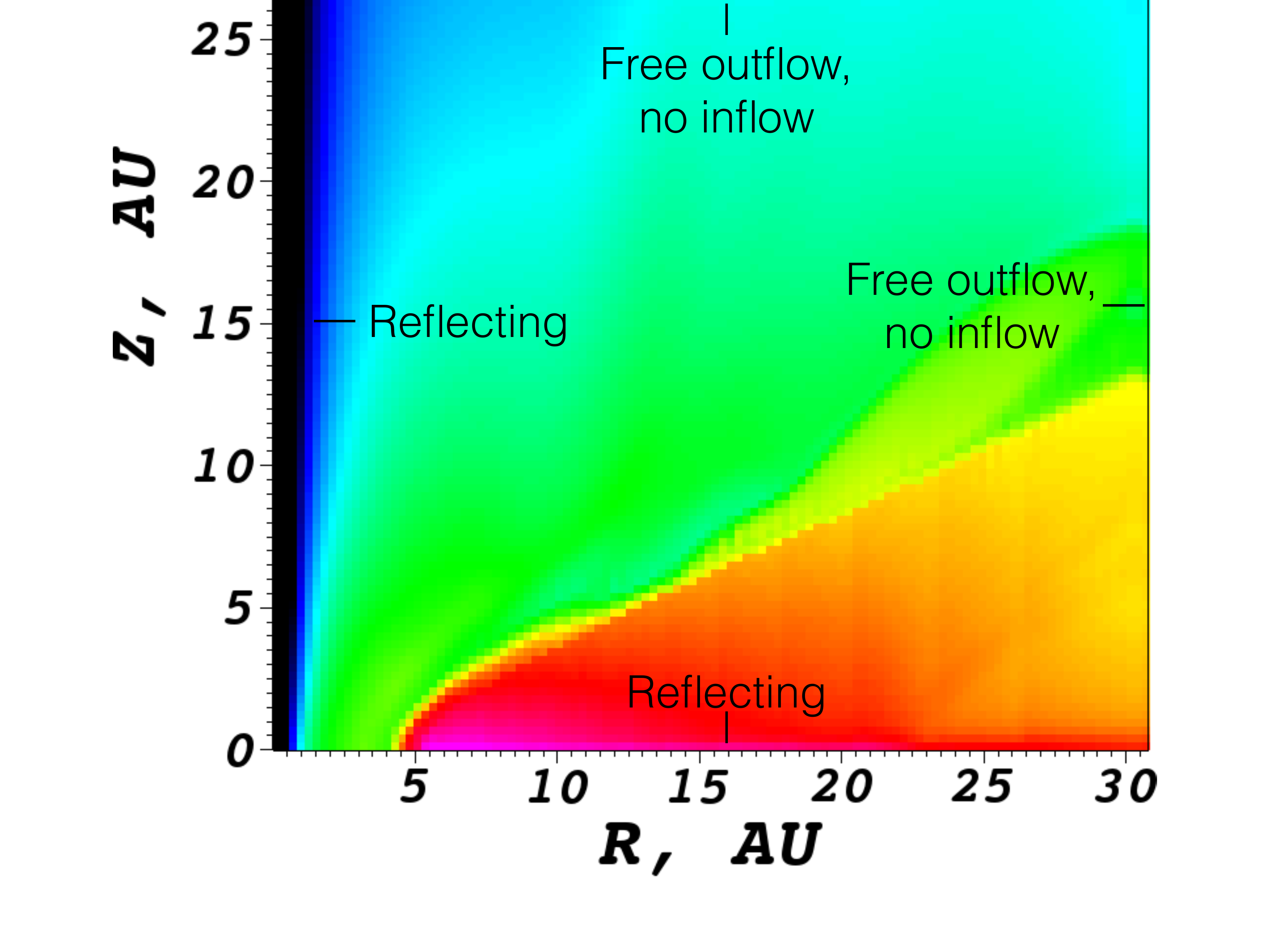}
    \caption{An example of using different boundary conditions at different boundaries. Here a disc quadrant is being modelled on a 2D cylindrical grid, so the boundary at the minimum $x$ and $z$ boundary is reflecting. The other boundaries permit material to stream freely away. }
    \label{fig:boundaries}
\end{figure}

The boundary conditions in \textsc{torus} can all be specified separately, for example if modelling a disc quadrant one could have a reflecting minimum $x$ and $z$ boundary and free outflow from the other boundaries (see Figure \ref{fig:boundaries})

%``reflecting'', ``periodic'', ``free outflow no inflow'' and ``inflow''. The outer 2 layers of cells on the grids are ghost cells, which serve no purpose other than to impose the boundary. 

\subsection{Validation}
\subsubsection{Hydrodynamics only tests}
\citet{2012MNRAS.420..562H} tested the hydrodynamics implementation in \textsc{torus}. It reproduces solutions consistent with the Sod shock tube solution \citep{Sod-1978}, which includes a shock and contact discontinuity \citep[Figure 2 of][]{2012MNRAS.420..562H}. It also reproduces self-similar solutions for a dynamically extreme Sedov-Taylor \citep{Sedov-1946, 1950RSPSA.201..159T} blast wave  \citep[Figure 3 of][]{2012MNRAS.420..562H}. Furthermore it has been shown to produce Rayleigh-Taylor \citep{Rayleigh1883, Taylor1950} and Kelvin-Helmholtz \citep{Kelvin-1871,Helmholtz-1871} instabilities \citep[Figures 4 and 5 of][]{2012MNRAS.420..562H}. The Sod shock tube test is part of the daily test suite, which will be discussed in section \ref{sec:testsuite}.

\subsubsection{Self-gravity tests}
\citet{2012MNRAS.420..562H} also presented the result of a collapse of a uniform density sphere, which results in the $n=1$ polytropic profile expected analytically.

\subsubsection{n-body gravity tests}
In \cite{2015MNRAS.448.3156H}, sink-sink interactions were tested using the three-body Pythagorean test problem of \cite{1913AN....195..113B}, which is also detailed by \cite{2011A&A...529A..27H}.  \textsc{torus} was shown to give excellent agreement with previous numerical solutions for this problem, the first of which was by \cite{1967AJ.....72..876S}. 

\cite{2015MNRAS.448.3156H} also tested the gas-sink physics in a \cite{1952MNRAS.112..195B} spherical accretion scenario, as well as a Bondi-Hoyle accretion test based on that done by \cite{2004ApJ...611..399K} in which comparable accretion rates were found.

 \subsection{Example}
\begin{figure*}
    \hspace{-0.7cm}
    \includegraphics[width=7.2cm]{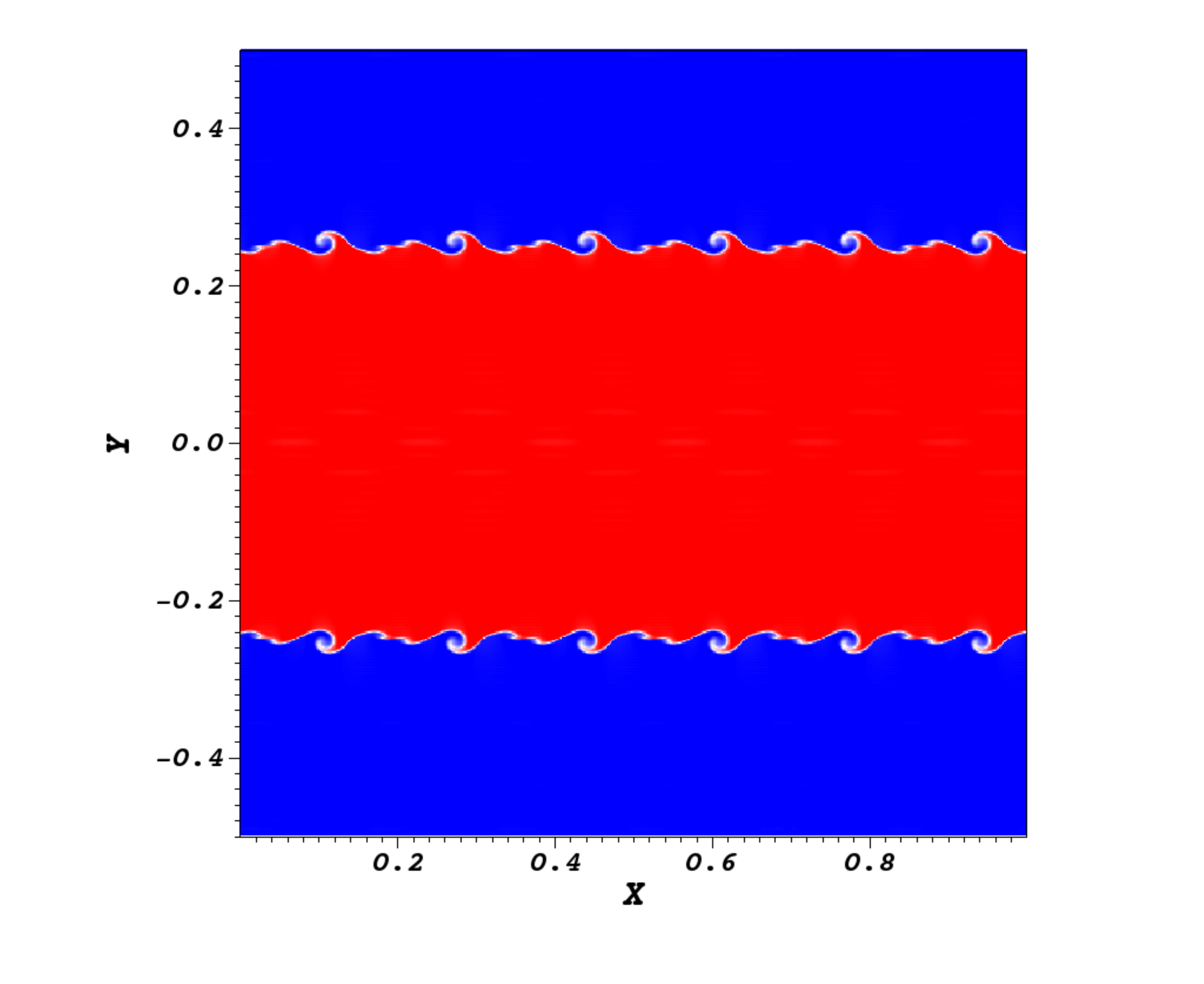} 
    \hspace{-1.3cm}
    \includegraphics[width=7.2cm]{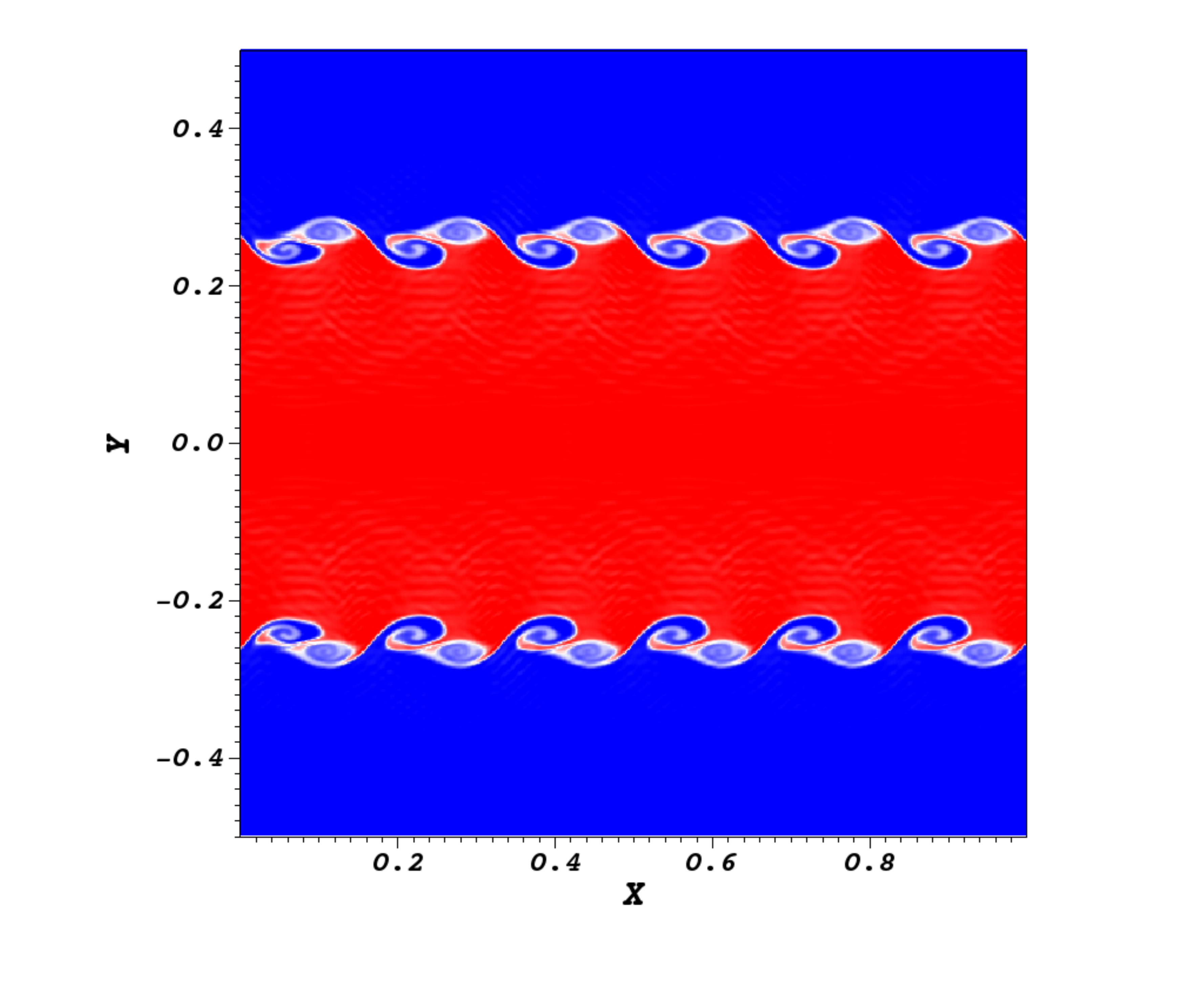}
    \hspace{-1.3cm}
    \includegraphics[width=7.2cm]{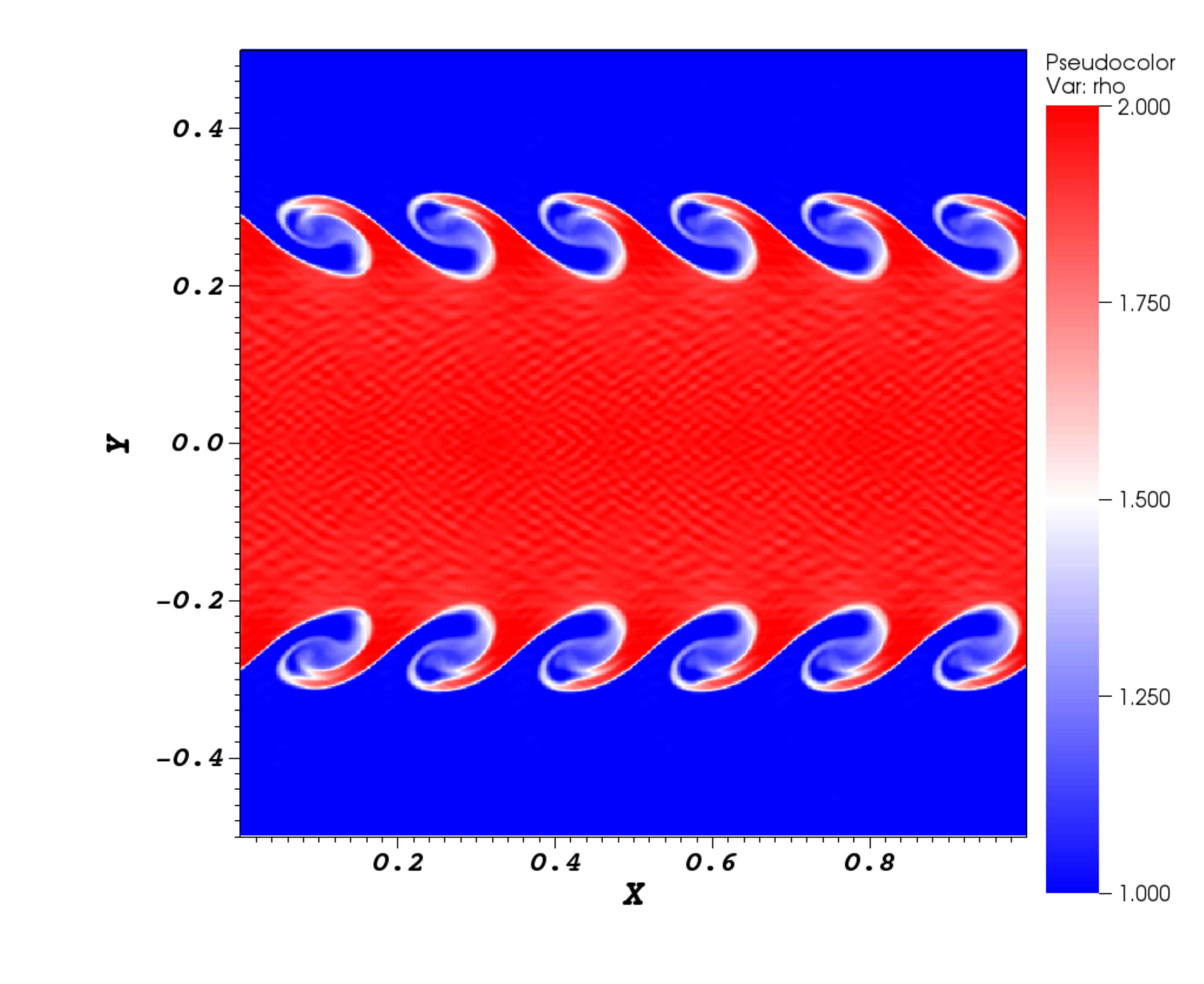}
    \caption{Snapshots at of the density distribution at three different times from a Kelvin-Helmholtz instability calculation based on the test model from \cite{2012MNRAS.420..562H}. }
    \label{fig:KH}
\end{figure*}

 In Figure \ref{fig:KH} we show three snapshots of the density distribution from a Kelvin-Helmholtz instability simulation based on the test model in \cite{2012MNRAS.420..562H}. This calculation is hydrodynamics only (no gravity), and employs a uniform cartesian mesh of $512^2$ cells. The higher density (red) material  is initially propagating to the right, and the lower density material to the left. The $\pm x$ boundaries are periodic, such that material leaving the right hand side re-enters on the left. The $\pm y$ boundaries are free outflow. A periodic perturbation to the initial velocity is applied of the form  
	\begin{equation}
		u = \left\{
    		\begin{array}{l l}
    			\frac{A}{6}\textrm{sin}(-2 \pi (x+1/2)), & |z-0.25| < 0.025 \\
			 & \\
    			\frac{A}{6}\textrm{sin}(2 \pi (x+1/2)), & |z+0.25| < 0.025. \\
    		\end{array} \right.
		\label{perturbation}
	\end{equation}
with $A=2.5\times10^{-2}$. The shearing vortices that form are symmetric about the mid-plane and the periodic boundaries are not inducing artefacts. Furthermore the vortices appear within the analytic characteristic Kelvin-Helmholtz timescale.

\section{Radiation hydrodynamics}
\label{sec:radhydro}

The native grid-based hydrodynamics scheme described above enables \textsc{torus} 
to perform radiation hydrodynamics calculations. \textsc{torus}
can also be coupled with the \textsc{sph-ng} smoothed particle hydrodynamics code to provide an alternative numerical method for the hydrodynamic component of a  radiation hydrodynamics calculation.

\subsection{Native radiation hydrodynamics}
The native radiation hydrodynamics scheme all takes place on the \textsc{torus} grid. Hydrodynamics (section \ref{sec:hydro}) and radiation transport/thermal (e.g. photoionisation, section \ref{sec:photo}) calculations take place iteratively via operator splitting, with the thermal structure of the radiative transfer calculation setting the pressure distribution in the hydrodynamics step. In terms of photoionisation, \textsc{torus} can include different levels of detail in the microphysics, from a simplified microphysics scheme analogous to other radiation hydrodynamics codes, incrementally through to the full photoionisation framework available in the code. It can also include radiation pressure and photodissociation region physics, which we discuss below.

\subsubsection{Simplified photoionisation for dynamics}
\label{sec:simpleRadHydro}
For dynamical applications the primary concern is computing a sufficiently accurate measure of the gas pressure, rather than the composition of the gas itself. To this end, some radiation hydrodynamics codes make a series of approximations that reduce the computation time significantly for a small decrease in thermal accuracy, for example in the series of papers by \citet{2007MNRAS.375.1291D, 2012MNRAS.422.1352D, 2013MNRAS.431.1062D} and the \textsc{ivine} code \citep{2009MNRAS.393...21G}. These approximations are as follows. 
\begin{enumerate}
    \item{The gas is hydrogen only}
    \item{Only a single photon frequency (energetic enough to ionise atomic hydrogen - 13.6eV) is considered}
    \item{Photons are not allowed to scatter, nor are recombination photons treated; the so-called ``on-the-spot'' approximation. Under this approximation the case B recombination coefficient is required - that which considers hydrogen recombinations into all states except the ground state \citep[the latter of which would result in the emission of an ionising photon for a hydrogen only gas, see][for further information]{2006agna.book.....O}. For the case B recombination coefficient we use the following function of gas temperature}
    \begin{equation}
        \alpha_{\textrm{B}} = 2.7\times10^{-13}\left(\frac{T_g}{10^4}\right)^{-0.8}\,\textrm{cm}^{3}\,\textrm{s}^{-1}
    \end{equation}
    \item{The thermal state of the gas is assumed to be a simple function of the hydrogen ionisation fraction $\chi$}
    \begin{equation}
        T_g = T_n + \chi\left(T_i - T_n\right)
    \end{equation}
    where typically $T_n=10$\,K and $T_{i}=10000K$ are the fully neutral and ionised gas temperatures respectively \citep[see Figure 4 of][]{2011MNRAS.413..401E}
\end{enumerate}
This hydrodynamic scheme represents the simplest available in \textsc{torus}.

\subsubsection{Full radiation hydrodynamics}
\label{sec:fullradhydro}
One of the main strengths of torus as a radiation hydrodynamics code is that it can incorporate detailed microphysics into dynamical applications, moving beyond the simplified prescription discussed in section \ref{sec:simpleRadHydro}. In ``detailed radiation hydrodynamics’’ the full photoionisation scheme discussed in \ref{sec:photo} is solved iteratively with the hydrodynamics. That is, we can include multiple gas species, thermally decoupled dust and gas, as well as a fully polychromatic radiation field that includes diffuse field (recombination) photons. The full thermal balance is also solved rather than approximating temperature as a function of hydrogen ionisation fraction. A comparison of the simplified and full approaches for the geometrically simple 1D calculations of the D-type expansion of an H\,\textsc{ii} region was made by \cite{2015MNRAS.453.2277H}, who found that typically the simplified scheme overestimates the extent of an H\,\textsc{ii} region over time. 

However using the detailed photoionisation scheme is much more computationally expensive, particularly in high resolution 3D calculations. Fortunately, Monte Carlo radiation transport scales very efficiently,  as we discuss in section \ref{sec:parallelisation}.

\subsection{Radiation pressure}
\label{sec:radpressure}
In addition to photoionisation setting the temperature, pressure and therefore gas dynamics, photons also impart a direct force upon matter through radiation pressure. \cite{2015MNRAS.448.3156H} showed that this radiation pressure force can be estimated using Monte Carlo radiative transfer, meaning that it is obtained essentially for free as part of a photoionisation calculation. The radiation pressure force per unit volume is
\begin{equation}
	f_{rad} = \frac{1}{c}\int\kappa_\nu \rho F_{\nu}d\nu,
\end{equation}
which in terms of Monte Carlo estimators is, in cell $j$,
\begin{equation}
	f_{\textrm{rad}, j} = \frac{1}{c}\frac{1}{\Delta t}\frac{1}{V_j}\sum \epsilon_i \kappa_\nu \rho \ell \hat{u}
	\label{radfequn}
\end{equation}

The $\kappa_\nu$ term in equation \ref{radfequn}  is the frequency dependent total opacity from all the mircophysics included in the simulation, not only the dust opacity. This allows radiation pressure to be applied not only in regions with dust in them, but also ionised regions which can feel radiation pressure through Thompson scattering and atomic line scattering.

For the opacity due to atomic line scattering we use the \citet{1975ApJ...195..157C} formulation of the radiation force being a multiple of the local Sobolev optical depth in the direction of photon propagation, and parametrisation of force approximately valid for stars with effective surface temperatures between 10,000K and 50,000K given by \citet{1982ApJ...259..282A}

\begin{align}
  F_{line}=&F_e M(t)\\
  t=&\sigma\!_e n v_{\rm{th}} \left|\frac{d \boldsymbol{u} \cdot \hat{\boldsymbol{l}}} {dl}\right|^{-1},\\
  M(t) =& 0.28 t^{-0.56}\left(\frac{N_e}{10^{-11}cm^{-3}}\right)^{0.09}
\end{align}
where $\sigma_e$ is the Thompson scattering cross section for an electron, $v_{th}$ is the thermal velocity of the gas and $l$ is a line in the direction of photon propagation at the gas element. \smallskip

In order to include this naturally in the radiation pressure calculations we multiply the electron scattering opacity in the Monte-Carlo step by a factor of $(1+ M(t))$, where $M(t)$ is calculated on a per photon basis.

In radiation hydrodynamics applications the radiation pressure force is included alongside photoionisation very efficiently, since the sum over path lengths through cells is already computed. 

In certain regimes, such as D-type expansion of HII regions around small numbers of massive stars, radiation pressure is of relatively minor in importance compared to photoionisation \citep{2014MNRAS.439.2990S, 2015MNRAS.453.2277H}. However in some scenarios, such as the early phases of massive star formation, radiation pressure can drive powerful outflows \citep[e.g.][]{2012A&A...537A.122K, 2016MNRAS.463.2553R, 2017MNRAS.471.4111H}. 

\subsection{Path length history method}

In order to increase the efficiency of the Monte-Carlo estimates for the radiation field we use a scheme where each of the previous estimates of the radiation field in a cell are weighted according to how many time steps ago they occurred and then averaged.
The weighting for each estimate of the radiation field is given by

\begin{equation}
    \label{eq:MCweights}
     W_i = \exp\frac{-i \Delta t}{t_{\rm{rad}}}
\end{equation} where $i$ is the number of time steps ago the estimate was made, $\Delta t$ is the time step of the simulation and $t_{\rm{rad}}$ is the radiation timescale used, for the simulations presented here the radiation time scale is assumed to be $0.6 v_{max}/\Delta x$ where $v_{max}$ is the largest velocity present in the simulation and $\Delta x$ is the size of a particular cell. This results in cells of different sizes having differing amounts of time averaging, small cells near the centre of the model will only have significantly less time averaging than the outer large cells. For a sufficiently large number of previous estimates the total of the weights can be approximated to the infinite sum:

\begin{align}
    \sum_{i=0}^{\infty} e^{-ai} &= \frac{e^a}{e^a -1},\\
        a &= \frac{\Delta t}{t_{\rm{rad}}},
\end{align}

\begin{equation}
  \label{eq:MCweightSum}
  \mathbf{\it{f}}_{\rm{sum}} = \sum_{i=0}^{\infty} \left[ \mathbf{\it{f}}_i e^{-ai} \right].
\end{equation}

Using this formulation of the weights allows us to retain all the information for the previous radiation history as a single value (equation \ref{eq:MCweightSum}). In order to calculate the weighted radiation value for the next timestep from the instant estimate of that time step and the weighted sum from the previous time steps we can use the fact all the weights from the previous timestep are simply multiplied by $e^{-a}$ to give the weights for the next time step, allowing us to calculate the new weighted radiation value using 
\begin{equation}
    \label{eq:MCcalc}
    \mathbf{\it{f}}_{n,\rm{weighted}} = \left( \mathbf{\it{f}}_n + \mathbf{\it{f}}_{\rm{sum}} e^{-a} \right) \frac{e^a-1}{e^a},
\end{equation}

Once this has been done $\bf{\it{f}}_{\rm{sum}}$ is set to the new value of $\bf{\it{f}}_n + \bf{\it{f}}_{\rm{sum}} e^{-a}$ for the next time step. For the value of $\bf{\it{f}}_{\rm{sum}}$ at $t=0$ we assume the radiation field has been static for a long time so that $\bf{\it{f}}_{\rm{sum}} =\bf{\it{f}}_0 \frac{e^a}{e^a - 1}$. \smallskip

This method gives an improved estimate of the radiation field by drawing on more information at the cost of introducing some time lag into the radiation field as it changes. We minimise this by using a final value for the radiation field of

\begin{align}
    \label{eq:MCshotNoise}
    \mathbf{\it{f}}_{n,\rm{final}} &= \mathbf{\it{f}}_{n,\rm{weighted}} X + \mathbf{\it{f}}_n (1-X)\\
        X &= \left(1+\frac{n_{\rm{cross}}}{400}\right)^{-0.5},
\end{align} where X is a factor  of the number of photon events in a cell to weight the final value towards the instantaneous estimate in well sampled regions and towards the weighted estimate in poorly sampled regions based on the number of photon events in that cell.

\subsection{Radiation hydrodynamics with SPH-NG}

The first use of \torus in a radiation hydrodynamic context was to couple \torus to the smooth particle hydrodynamics (SPH) code \textsc{sph-ng} \citep{1990ApJ...348..647B, 1990nmns.work..269B} as presented by \citet{2010MNRAS.403.1143A}. \torus can be compiled to produce a \textsc{fortran} module contained in a library which can be accessed from an SPH code via a subroutine call. Although the interface was designed to work with \textsc{sph-ng} it would also work with a different SPH code if passed suitable arguments. 

The particle  positions, densities, temperatures and smoothing lengths are passed through the subroutine interface and used to construct the \torus grid using the method described in section~\ref{sec:gridfromsph}. \textsc{sph-ng} performs calculations using internal energy, rather than temperature. The conversion is performed in \textsc{sph-ng} so that \torus receives and returns temperature values. The temperatures in the grid cells are initialised from the SPH particles and subsequently converged to equilibrium values using the dust radiative equilibrium algorithm described in section~\ref{sec:radeq}. The radiation source properties can either be specified in the \torus parameter file, or sources can be constructed from sink particles which are included in the list of SPH particles passed to \torus. When the temperature calculation has converged the temperature of each SPH particle is updated with the value from the grid cell which contains the particle, and the updated temperature is passed back to \textsc{sph-ng} via a subroutine argument. 

The frequency of coupling between the SPH calculation and \torus radiative equilibrium calculations is determined by how frequently the \torus subroutine is called. The required coupling frequency depends on the physics of the calculation, for example in  \citet{2010MNRAS.403.1143A} the \torus radiative equilibrium calculation was run every 4th time step of the SPH calculation as this matched the sound crossing time (and time scale for scale height variations) at the inner edge of the cirumstellar disc.

\subsection{Validation}
The tests of hydrodynamics and photoionisation are direct tests of the components of the radiation hydrodynamics scheme. The actual radiation hydrodynamic implementation was shown to agree well with 5 other codes in the starbench comparison D-type test \citep{2015MNRAS.453.1324B}. This follows the extent over time of a 1D expanding H\,\textsc{ii} region. Historically codes have compared with one of the analytic expressions for such expansion \citep[e.g. that of][]{1998ppim.book.....S}, however the starbench comparison project showed that all codes (including \textsc{torus}) obtained a consistent solution that was different to any of the analytical solutions. 

The dynamical radiation pressure scheme was tested by \cite{2015MNRAS.448.3156H} for a 1D radiation pressure driven expanding shell. This expansion has an analytical solution for the extent over time, which \textsc{torus} was found to give excellent agreement with. 

\citet{2010MNRAS.403.1143A} present tests where \torus calculates equilibrium temperatures in a circumstellar disc benchmark. They find that the inner edge of the disc is not clearly represented with SPH particles, which limits the ability to reproduce benchmark SEDs. However the temperature in the majority of the disc is within 20~percent of the benchmark value even with a relatively coarse SPH resolution ($10^6$ particles). When using \torus with grids generated from SPH particles it is important to consider the influence of sharp gradients which may not be well represented by SPH. This may be a significant source of error (e.g. in the case of generating an SED) or may be insignificant compared to other uncertainties (e.g. in a radiation-hydrodynamics calculation). This limitation is not inherent to \torus but is rather a consequence of the SPH representation and emphasises the importance of choosing  suitable numerical methods to solve the problem in hand. 

\subsection{Example}
\textsc{torus} has recently been used for cluster-scale models of massive star feedback in star-forming regions in 3D. \cite{2018MNRAS.477.5422A} modelled a $10^3$\,M$_\odot$ cloud initially being evolved under self-gravity and a seeded turbulent velocity field, such that the total kinetic energy equalled the gravitational potential energy. Stars were inserted into dense locations, with the most massive star being 34\,M$_\odot$ and the second being 11\,M$_\odot$. Photoionization and thermal equilibria, as well as radiation pressure, were calculated for every hydrodynamics time step. The full radiation hydrodynamical treatment detailed in section~\ref{sec:fullradhydro} was included in this model. This included atomic species up to sulphur, as listed in section~\ref{sec:photoionization_species} and Table~\ref{tab:photo_abundances}. Silicate dust grains \citep{1984ApJ...285...89D} were also included, having a median grain size of 0.12\,$\normalfont\mu$m, with dust temperatures being calculated separately from gas temperatures, except for a term accounting for collisional heat transfer between the two.  The diffuse radiation field was also included.

\begin{figure*}
    \centering
    \includegraphics[width=18cm]{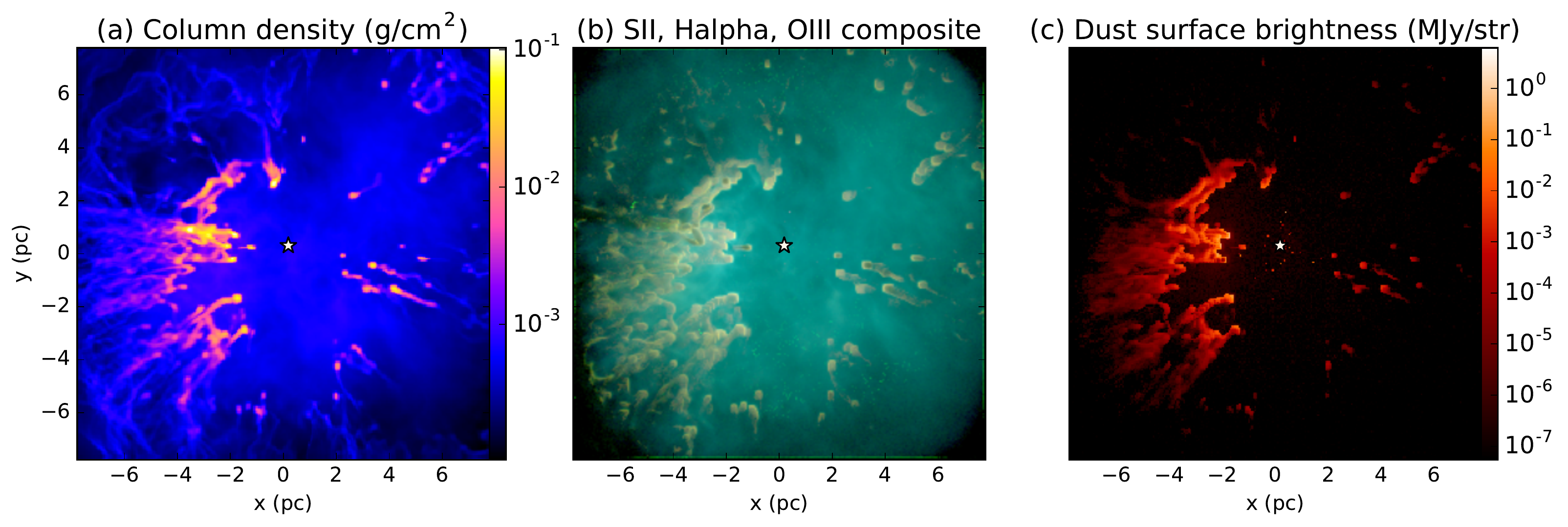}
    \caption{Snapshot of massive star feedback using the full RHD photoionization treatment by  \cite{2018MNRAS.477.5422A}. This is a 1000 solar-mass cloud which contains a 34 solar-mass star, indicated by the point near the centre of each frame. (a) Column density. (b) Three-colour composite of synthetic surface brightnesses of [S \textsc{ii}] 6731 $\angstrom$ (red), H$\alpha$ at 6563 $\angstrom$ (green), and [O \textsc{iii}] 5007 $\angstrom$ (blue). (c) Synthetic 24 $\normalfont\mu$m dust continuum emission.}
    \label{fig:clusterfeedback}
\end{figure*}

Expansion due to photoionization heating efficiently dispersed the cloud, with all mass being removed from the grid within 1.6\,Myr (0.74 cloud free-fall times). Outward mass fluxes at the grid boundary peaked at $2\times10^{-3}$\,M$_\odot$\,yr$^{-1}$. Self-consistent synthetic observations were also produced, using the temperatures and ionization states calculated during the radiation hydrodynamics evolution. Examples are in Figure~\ref{fig:clusterfeedback}, which shows column density alongside forbidden lines, recombination lines and dust continuum, at a snapshot 0.6 Myr after feedback was initiated. \cite{2018MNRAS.477.5422A} also used 20\,cm free-free continuum to estimate the production rate of Lyman continuum photons, finding that the diagnostic always underestimated the actual rate -- the discrepancy reached orders of magnitude as the nebula became more density-bounded and gas and photons escaped the domain.

\section{TORUS-3DPDR}
\label{sec:pdr}
Media are typically more optically thin to far ultraviolet (FUV, 912$ <\lambda < $2400\AA) photons than the extreme ultraviolet (EUV) photons considered in section \ref{sec:photo}. Unlike the EUV, FUV photons are not energetic enough to predominantly ionise the gas, however they do still influence the chemical composition and thermal properties of gas optically thick to the EUV.  Solving for the conditions in regions where FUV photons determine the gas properties, so called photon dominated or photodissociation regions (PDR's), is decidedly non trivial. The composition is set by large, complicated chemical networks that are temperature sensitive. Furthermore the temperature in turn depends upon the chemical composition.  For example heating contributions include polycyclic aromatic hydrocarbon (PAH) heating and  H$_2$ formation, and cooling contributions are dominated by lines of CII, CI, O and CO.  

A particular difficulty arises because of the importance of the cooling lines. These require an estimate of the escape probability, which in 3D means that an estimate of the column along each direction into $4\pi$ steradians is required. For this reason PDR models have historically been limited to 1D, where radiation is assumed to excite and cool along a single path, with infinite optical depth in other directions. This difficulty was resolved by \textsc{3d-pdr} \citep{2012MNRAS.427.2100B} which uses a healpix scheme \citep{2005ApJ...622..759G} to estimate the escape probability into $4\pi$ steradians. It does not, however, directly model the exciting UV field, assuming instead a spherical, planar or isotropic radiation field strength of arbitrary magnitude. In the case of embedded sources, such as in a turbulent star forming region with young stars, a more robust means of estimating the exciting radiation field from those stars is required. 

This problem has since been surmounted by coupling \textsc{3d-pdr} with \textsc{torus}. Since \textsc{3d-pdr} is a separate entity, we will not detail that algorithm further here. \textsc{torus-3dpdr}
\citep{2015MNRAS.454.2828B} is the resulting code. It is a direct incorporation of routines from the \textsc{3d-pdr} code \citep{2012MNRAS.427.2100B} into \textsc{torus}. That is, it is written in the same style and using the same octree grid (rather than the evaluation point scheme of \textsc{3d-pdr}). 

During a Monte Carlo radiative transfer calculation, the FUV field is obtained essentially for free. In Draine units, it is
\begin{equation}
    \chi = \int_{912\angstrom}^{2400\angstrom}  \frac{J_{\lambda} \textrm{d}\lambda}{1.71G_0}
\end{equation}
where $G_0$ is the \cite{1968BAN....19..421H} unit ($1.6 \times10^{-3}$ erg cm$^{-2}$ s$^{-1}$). In cell $j$ this is given by
\begin{equation}
    \chi = \frac{\epsilon_j}{\Delta t G_0 1.71 V_j}\sum_i \ell_i \left[912\angstrom<\lambda<2400\angstrom\right].
\end{equation}
in terms of Monte Carlo estimators.
Note that this can also be computed as a vector or tensor field. 
This Draine field is then fed into the modules that are based on the \textsc{3d-pdr} code, which solve for the PDR composition and thermal structure. This enables computation of detailed PDR models in 3D with accurate input radiation fields from the Monte Carlo radiative transfer.

\textsc{torus-3dpdr} currently uses a reduced version of the \textsc{umist 2012} chemical network of 33 species and 330 reactions \citep{2013A&A...550A..36M}. This network was chosen because it gives accurate temperatures (to within $\sim10$\,per cent of a much larger, comprehensive and costly network) for more reasonable computational cost. An earlier version of this network was also used in the \citet{2007A&A...467..187R} benchmarking project. A summary of the species included is given in Table \ref{tab:torus3dpdr_spec}. A summary of heating heating and cooling mechanisms considered by the code is given in Table \ref{tab:torus3dpdr_hc}. \textsc{torus-3dpdr} is capable of simultaneous modelling photoionisation and PDR physics by flagging cells which are dominated by photoionisation (i.e. hotter than some threshold value) and not including them in the PDR calculation.  

\begin{table}
    \centering
    \begin{tabular}{c}
    \hline
  H, H$_2$, 
  He , C+ 
  O, Mg+  \\
   e$^-$, H+, H$_2$+, H$_3$+, He+, O+, O$_2$, O$_2$+, \\
   			   OH+, C, CO, CO+, OH, HCO+, Mg, H$_2$O,  \\
			    H$_2$O+, H$_3$O, CH, CH+, CH$_2$, CH$_2$+, \\
			    CH$_3$, CH$_3$+, CH$_4$, CH$_4$+, CH$_5$+, \\
%			  Cosmic rays, PAHs, Dust\\    
    \hline
    \end{tabular}
    \caption{The 33 gas species included in the reduced network of \textsc{torus-3dpdr}. }
    \label{tab:torus3dpdr_spec}
\end{table}

\begin{table}
    \centering
    \begin{tabular}{|c | c|}
    \hline
    Heating processes    & Cooling processes \\
    \hline
      Photoelectric heating  & CO line emission \\
      C ionization  & CI line emission \\
      H$_2$ formation  & CII line emission \\
      H$_2$ photodissociation  & OI line emission     \\  
      FUV pumping  & gas-grain collisions   \\
      cosmic rays & \\
      turbulence & \\
      chemical heating & \\
      gas-grain collisions & \\
    \hline
    \end{tabular}
    \caption{Heating and cooling processes in \textsc{torus-3dpdr}}
    \label{tab:torus3dpdr_hc}
\end{table}

\subsection{Photochemical hydrodynamics}
PDR physics is not just important for determining the composition of gasses for chemical modelling and synthetic observations. Temperatures in these regions can be as high as a few thousand Kelvin, so they can be of dynamical importance. Indeed, the dynamics of some systems is dominated by the thermal properties of PDRs. For example the irradiation of surface layers of protoplanetary discs, away from the disc inner edge, is dominated by PDR heating. This has historically only been possible to study with 1D semi-analytic \citep[e.g.][]{2004ApJ...611..360A} or 2D hydrostatic models \citep[e.g.][]{2009ApJ...690.1539G}. \textsc{torus-3dpdr} means that scenarios such as this can now be directly modelled with photochemical-dynamical simulations. The approach used is very similar to the radiation hydrodynamics via operator splitting discussed in section \ref{sec:radhydro}, only with the addition of the PDR calculation. All other aspects of the model, such as boundary conditions and time stepping criteria, are the same. 

\subsection{Validation}
\citet{2015MNRAS.454.2828B} have shown that \textsc{torus-3dpdr} produces thermal and chemical profiles consistent with those in the code comparison project of \cite{2007A&A...467..187R}. We have also compared \textsc{torus-3dpdr} and \textsc{cloudy} in an extension of the HII40 Lexington benchmark and found good agreement in the thermal and compositional properties of the PDR.

Additional validation in a photochemical-dynamical context came from comparison of TORUS-3DPDR with semi-analytic solutions for flow profiles and mass loss rates from externally irradiated protoplanetary discs \citep{2016MNRAS.463.3616H}. These latter tests offer future benchmarks for developers of photochemical-dynamical codes.

\section{Synthetic observations overview}
\label{sec:synthetic_obs}

So far we have discussed the ways in which \textsc{torus} computes the composition and thermal structure of astrophysical media, and how this can couple with the dynamical evolution. However one of the main applications of radiative transfer codes is the production of synthetic observations - computing a theoretical model of the way in which an astrophysical system would appear to an observer \citep[for a review on synthetic observations, see][]{2018NewAR..82....1H}. Synthetic observations in \torus\ typically take one of three forms: a spectral energy distribution (SED), an image or a position-position-velocity (PPV) data cube. A PPV data cube comprises two spatial axes and one spectral axis (e.g. wavelength or frequency) and is the generalised case of a observation with both spatial and spectral resolution. In principal A PPV data cube can be converted into an image or spectrum by collapsing over the spectral or spatial axes. However \textsc{torus} can generate SEDs and images directly; in practice these are frequently the observational data products which the synthetic observation is being compared to. We now summarise the procedures by which \torus\ produces synthetic observations, covering SEDs (section~\ref{sec:seds}), images (section~\ref{sec:pp}) and PPV data cubes (section~\ref{sec:ppv}) before briefly discussing instrumentation effects (section~\ref{sec:instrumentation}).

\subsection{Spectral energy distributions (SEDs)}
\label{sec:seds}
Synthetic spectra in \textsc{torus} are computed using a Monte Carlo approach, similar to the dust radiative equilibrium and photoionisation calculations (see sections \ref{sec:radeq} and \ref{sec:photo} respectively). The number of photon packets used in computing the SED is a code parameter, and should scale with the number of wavelength bins used (i.e. the spectral resolution). The SED is computed one wavelength bin at a time. The probability of a photon being produced by the photon sources, rather than dust, is given by
\begin{equation}
    p_{\rm source} = \frac{\sum_{i=1}^{N_{\rm sources}} L_{{\rm source},i}}{\sum_{i=1}^{N_{\rm sources}} L_{{\rm source},i} + \sum_{i=1}^{N_{\rm cells}} j_{{\rm dust},i}  V_i }
\end{equation}
where $L$ is the source luminosity and $j_{\rm dust}$ is the cell emissivity. A uniform random deviate is chosen and if this is less than $p_{\rm source}$ then the photon position and direction are found according to the algorithms detailed in section~\ref{sec:sources}. The appropriate ($i^{th}$) cell for packets produced by the dust are found from the probability density function of dust emissivities
\begin{equation}
    p_i = \frac{\sum_{k=1}^i j_k V_k}{\sum_{k=1}^{N_{\rm cells}} j_k V_k}
\end{equation}
Once the cell is identified a random position and isotropic direction for the photon packet is selected.

In order to improve the signal-to-noise of the SED we adopt the peel-off method of  \cite{1984ApJ...278..186Y}. At each interaction (emission, scattering) the probability of the photon packet reaching the observer is calculated. The photon packet is then forced towards the observer (and ``detected''), but with flux weighted by the probability of propagating towards the observer. The same overall flux is therefore retrieved as for a completely random sampling, only with better signal-to-noise for a given number of photon packets (and hence computational expense). 

\textsc{torus} also employs a forced first scattering scheme, which requires photon packets to be scattered by dust before escaping the grid \citep[e.g.][]{1999ApJ...525..799W}. This ensures that optically thin components of the grid contribute to the resulting SED, rather than being missed due to the low (but finite) probability of scattering. 

\subsection{Synthetic images }
\label{sec:pp}

The basis for a synthetic image is defined in terms of the number of pixels per side (e.g. $512\times512$), the observer's distance from the centre of the simulation grid, the viewing vector (e.g. defined by inclination and position angles) as well as any offset from the grid centre. What results is an array of bins analogous to an array of CCD pixels. In the Monte Carlo radiative transfer framework one can immediately see that the random walk of photon packets would result in some ending up within this pseudo CCD array, but the probability of an individual packet doing so is low. The peel off technique described in \ref{sec:seds} is therefore employed to dramatically improve the signal to noise. By counting the photon packets in each spatial bin   \textsc{torus} computes the predicted spatial intensity distribution.

At present \textsc{torus} can compute 2D emission maps in the continuum (both from the dust and Bremsstrahlung processes), as well as recombination and forbidden lines. Molecular line data cubes (which we discuss next) can also be collapsed to produce 2D integrated intensity maps.

\begin{figure}
   \centering
    \hspace{-0.0cm}
    \includegraphics[width=8.8cm]{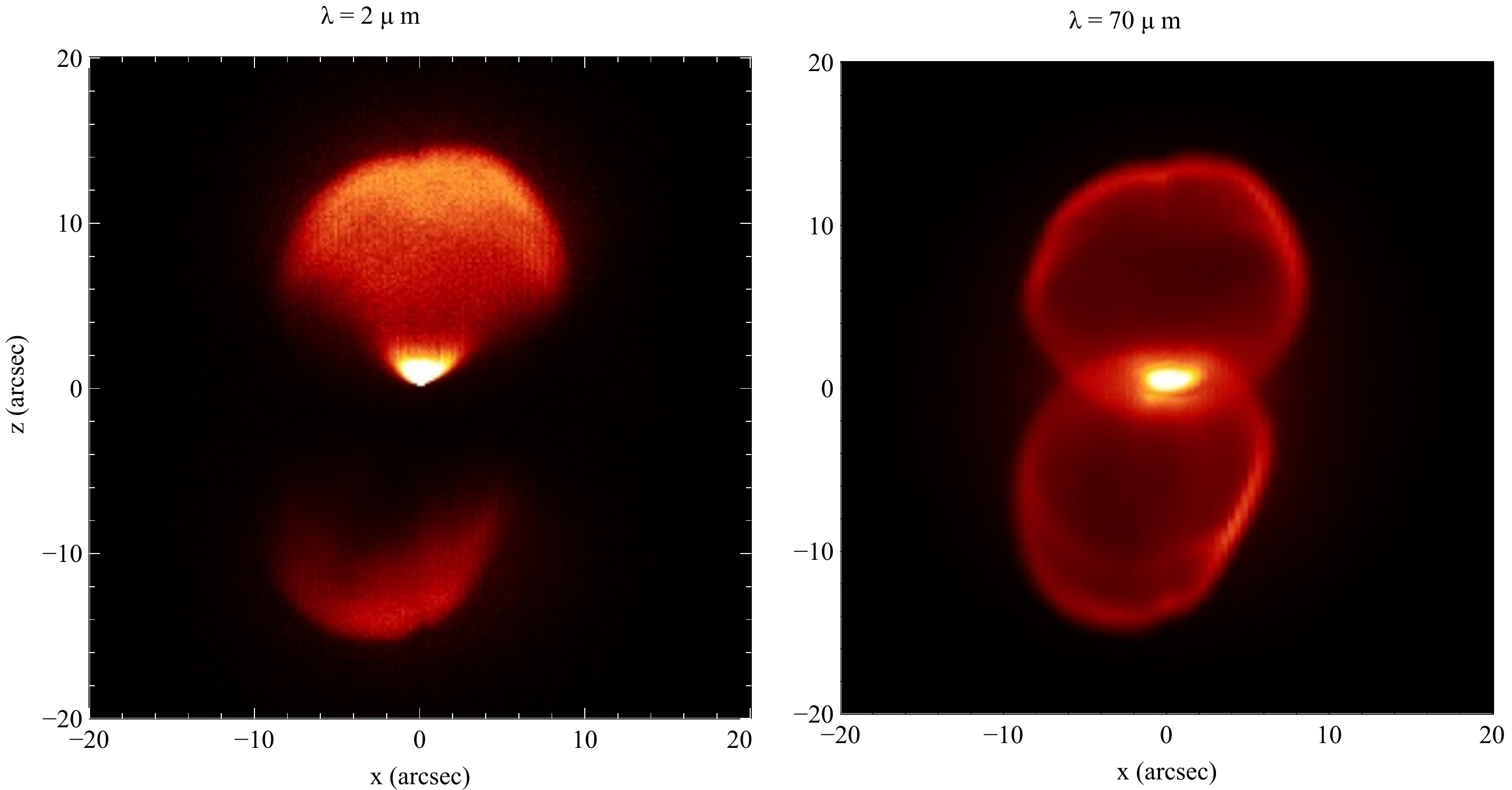}

    \caption{Simulated images (2\,$\mu$m and 24$\mu$m) of the radiation-hydrodynamical model of massive star formation at a time of 35\,kyr presented by \cite{2017MNRAS.471.4111H} viewed at an inclination of 60$^\circ$ assuming a distance of 1 kpc. The linear colour scales represent the surface brightness in MJy/str.} 
    \label{fig:bipolar_outflow}
\end{figure}

We present example dust continuum synthetic images of the bipolar cavities formed during the formation of a massive star in Figure~\ref{fig:bipolar_outflow} based on the calculations presented in \cite{2017MNRAS.471.4111H}. The 2\,$\mu$m image is formed predominantly by scattered light from the central object, and forced first-scattering and peel-off are both used to reduce the variance on the resulting image. The 70\,$\mu$m image is dominated by thermal emission from the warm inner surfaces of the bipolar cavities. 

\subsection{Data cubes}
\label{sec:ppv}
Often observations have both spatial and spectral coverage (e.g. MUSE in the optical and ALMA/JCMT in the sub-mm).  \textsc{torus} can compute PPV data cubes for atomic recombination lines and molecular lines. The cube basis is defined in a similar manner to the 2D images discussed in \ref{sec:pp}, with the addition of a third spectral axis that is set by a maximum and minimum velocity and a number of channels that determine the spectral resolution. 

In contrast to the Monte Carlo approach used for 2D images atomic and molecular line data cubes are produced using a ray tracing scheme. Ray tracing is more appropriate in these cases as scattering is not significant and optically thick regions can be better treated. There are two main modes of operation for molecular line data cubes. The first assumes that the observer is far from the object being imaged and generates the data cube by tracing a number of parallel rays through the AMR grid. The second mode of operation is used to generate synthetic Galactic plane surveys and traces a number of diverging rays outwards from the observer's position. The two modes of operation are explained in more detail in this section. 

\subsubsection{Far field data cubes}

When generating a molecular line data cube in the far field case the ray trace starts from the observer's position and finishes when the ray exits the grid on the far side. As the ray trace proceeds the total optical depth between the current point and the observer is accumulated ($\tau_{total}$) and the intensity is updated from $I_{\nu}^{\rm{old}}$ to $I_{\nu}^{\rm{new}}$ according to
\begin{equation}
    I_{\nu}^{\rm{new}} = I_{\nu}^{\rm{old}} + S_{\nu} \left( 1 - e^{-d\tau_\nu} \right) e^{-\tau_{\rm{total}}}
\end{equation}
where $S_{\nu}$ is the local source function at frequency $\nu$ given by
\begin{equation}
    S_{\nu} = \frac{j_\nu}{k_\nu}
\end{equation}
The local emission and absorption coefficients, $j_\nu$ and $k_\nu$ are calculated for gas only by default but dust opacity and emissivity can optionally be included. 
The transition specific optical depth $d\tau_\nu$ along the current line segment $ds$ is given by
\begin{equation}
    d\tau_\nu = \int k_\nu ds
\end{equation}

The local emission and absorption coefficients are affected by the local velocity, which Doppler shifts local emission and absorption relative to the frame of reference of the calculation. This effect is included via the line profile $\phi\left(v\right)$ given in equation~\ref{eqn:lineProfileFunction} which multiplies both the emission and absorption coefficients. The source function is unaffected, as the line profile is included in both $j_\nu$ and $k_\nu$ and cancels out, but the optical depth is dependent on the local velocity. There can be significant velocity gradients present which result in large variations in the emission and absorption coefficients within an individual grid cell. To ensure that the line profile is well sampled the ray trace through a cell is broken down into smaller segments $N_{seg}$ according to the local velocity gradient and turbulent line width. The number of line segments used to trace a single grid cell $N_{seg}$ is given by 
\begin{equation}
    \label{eqn:cubeVelSample}
   N_{seg} = \min\left(N_{seg,max},\max\left(N_{seg,min}, 5 \frac{\Delta \vec{V} \cdot \widehat{ds}}{v_{\rm{turb}}}\right)\right)
\end{equation}
where $v_{\rm{turb}}$ is the turblent velocity line width from equation~\ref{eqn:lineProfileFunction} which includes both thermal and turbulent broadening. By default $N_{seg,min}=5$ and $N_{seg,max}=100$. $\widehat{ds}$ is a unit vector in the direction of the ray trace and $\Delta \vec{V}$ is the change in velocity across the grid cell. 
Determining $N_{seg}$ in this manner ensures that the line width is typically sampled by 5 points during the ray trace. At each point the velocity is linearly interpolated to the required location using velocity values at the corners of the grid cells. Each octal holds 27 corner velocities so that the interpolation can be carried out without referencing another octal. The density within the cell is either assumed to be constant or is linearly interpolated to the current position (referred to as density sub-sampling). Density sub-sampling results in smoother features in the data cubes at the expense of greater run time for the calculation. If density sub-sampling is used then $N_{seg,max}=1000$ in equation~\ref{eqn:cubeVelSample}.  

By default one ray trace is performed through the centre of each pixel in the data cube. If the spatial resolution of the data cube significantly under samples the AMR grid then important structures may be missed. This problem can be mitigated by instructing \torus to perform more than one ray trace per data cube pixel. The number of rays per pixels can either be specified as an input parameter or can be determined at run time. In the latter case the number of rays is increased until the variance of the ray intensities ($\sigma_{\rm{ray}}^2$) is less than a given limit specified by a tolerance value $t_{\rm{ray}}$
\begin{equation}
    \sigma_{\rm{ray}}^2 < n_{\rm{ray}} \left( t_{\rm{ray}} I_{\nu} \right)^2
\end{equation}
where by default $t_{\rm{ray}}=0.01$. The location of each ray trace through the pixel is determined according to a \cite{SOBOL196786} quasi-random number sequence in which the ray origins tend to avoid each other giving better sampling properties. 

Generating atomic line data cubes follows a similar process to that for molecular line data cubes outlined above with some minor modifications. To determine the number of line segments through a cell  $N_{seg}$ is set to 2 but is doubled (and the trace through the cell repeated) if $d\tau_\nu>0.1$ and $\tau_{\rm{total}}<20$. If the ray trace is at a frequency close to a line resonance, or the velocity gradient in the cell would cause a line resonance to be traversed, then $N_{seg}$ is initially set to 20. 

\subsubsection{21cm line data cubes}
\label{sec:21cm_datacubes}

The 21~cm atomic hydrogen line is handled as a special case within the molecular physics module. This enables the molecular physics ray tracing procedures to be re-used but allows simplifications relevant to the 21cm line to be employed. For the 21~cm line there is no need to calculate level populations as the emissivity and opacity depend only on the local temperature and number density of hydrogen atoms. The opacity $k_{\nu}$ is given by
\begin{equation}
k_{\nu} = \frac{3 c^2 h A_0}{32 \pi k \nu_0} \frac{n(\rm{H})}{T} \phi
\left( v \right)
\label{eqn:opacity}
\end{equation}
\citep{rohlfs} where $c$ is the speed of light, $h$ is Planck's
constant, $A_0$ is the Einstein probability emission coefficient, $k$
is the Boltzmann constant, and $\nu_0$ is the frequency of the
H\,{\sc{i}} transition. The number density of hydrogen atoms $n(\rm{H})$ and the temperature $T$ are taken from values in the current grid cell. The emissivity is calculated from the opacity using Kirchoff's law
($j_{\nu}=k_{\nu} B_{\nu}$ where $B_{\nu}$ is the Planck
function) and the Rayleigh-Jeans approximation giving
\begin{equation}
j_{\nu} = \frac{3 \nu_0 h A_0}{16 \pi
  } {n(\rm{H})} \phi
\left( v \right).
\label{eqn:emissivity}
\end{equation}
Furthermore the line width does not include a turbulent component and comprises thermal broadening only.

In addition to 21~cm specific simplifications \torus also generates additional data products commonly produced from H\,{\sc{i}} observations. H\,{\sc{i}} observations of external galaxies are often presented as moment maps, for example from the THINGS survey \citep{2008AJ....136.2563W}. When generating H\,{\sc{i}} data cubes \torus will also calculate zero, first and second moment maps from the data cube. The zero moment map $M_0$ is an  intensity map integrated over all $n_v$ velocity channels 
\begin{equation}
    M_0 = \sum_{v=1}^{n_v} I_v \Delta v
\end{equation}
where $I_v$ is the intensity in velocity channel $v$ and $\Delta v$ is the channel width. The first moment is an emission weighted velocity given by 
\begin{equation}
    M_1 = \frac{\sum_{v=1}^{n_v} I_v v \Delta v}{\sum_{v=1}^{n_v} I_v}
\end{equation}
The second moment quantifies the velocity dispersion and is given by 
\begin{equation}
    M_2 = \sqrt{\frac{\sum_{v=1}^{n_v} I_v \left(v - \bar{v} \right)^2 \Delta v }{\sum_{v=1}^{n_v} I_v}}
\end{equation}
where $\bar{v}=M_1$ is the emission weighted velocity. 

To facilitate comparisons with observations H\,{\sc{i}} intensity values are converted into brightness temperatures ($T_B$) using
\begin{equation}
    T_B = \frac{I_v \lambda^2}{2k}
\end{equation}
where $\lambda$ is the wavelength of the emission and $k$ is Boltzmann's constant. The moment maps are then expressed in the same units as the observations e.g. first moments are in units of $\rm{K.km.s}^{-1}$

\begin{figure*}
   \centering
    \hspace{-0.0cm}
    \includegraphics[width=16cm]{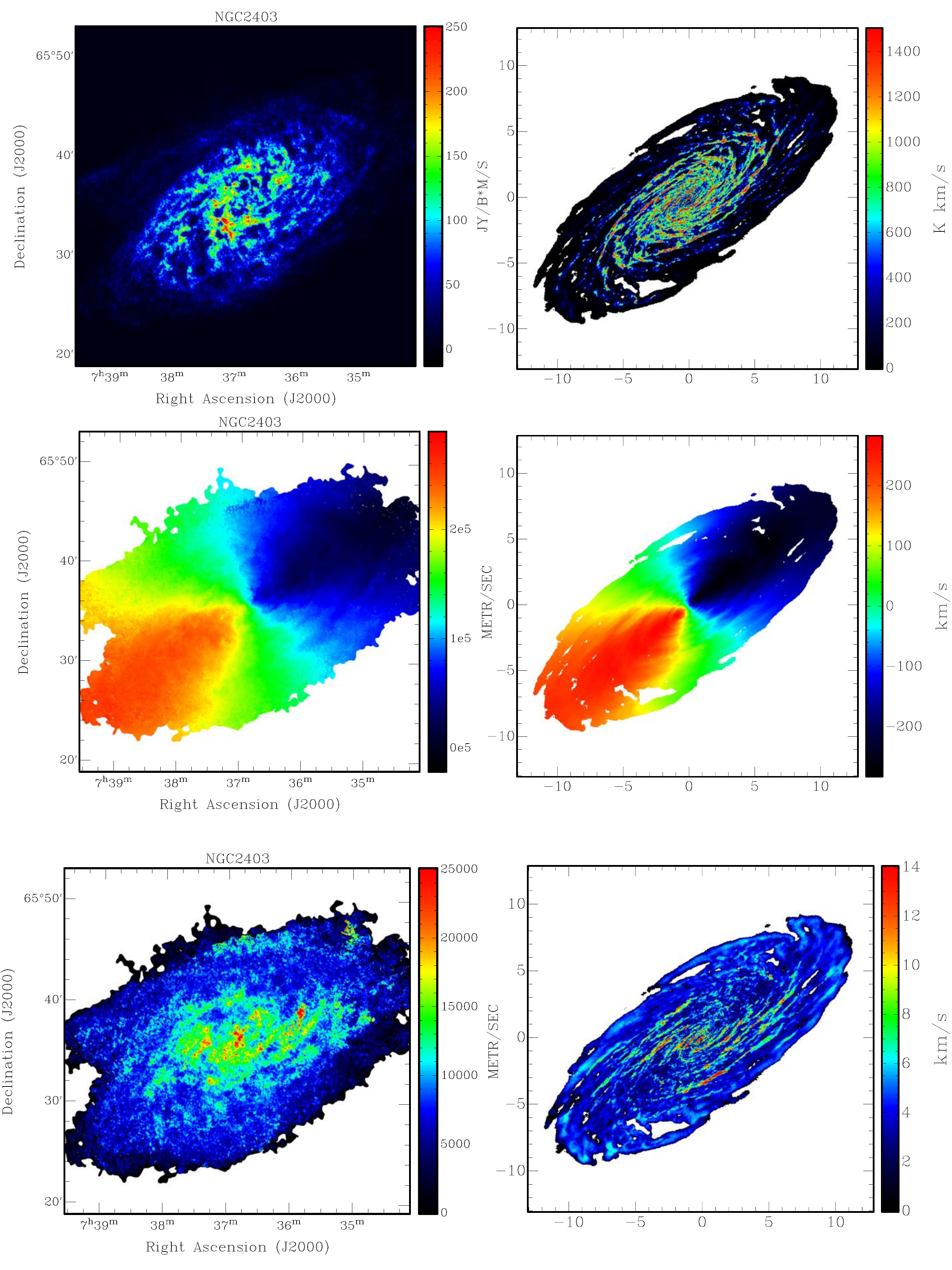}
%    \includegraphics[width=5cm, angle=270]{./NGC2403_mom0_torus-eps-converted-to.pdf}

  %  \hspace{-0.5cm}
 %   \includegraphics[width=5.cm]{./NGC2403_mom1.pdf}
  %  \includegraphics[width=5.cm, angle=270]{./NGC2403_mom1_torus-eps-converted-to.pdf}
    
  %  \hspace{-0.5cm}
  %  \includegraphics[width=5.cm]{./NGC2403_mom2.pdf}
 %   \includegraphics[width=5.cm, angle=270]{./NGC2403_mom2_torus-eps-converted-to.pdf}
    \caption{H{\sc{i}} moment maps from observations (left) and a \torus model (right). The zero moment (top row) is the integrated intensity, the first moment (middle row) is the emission weighted velocity, and the second moment (bottom row) measures the velocity dispersion. The observations are NGC2403 from the THINGS survey \citep{2008AJ....136.2563W}. The axes on the plots are in units of kpc.} 
    \label{fig:THINGS_moment_maps}
\end{figure*}

Figure~\ref{fig:THINGS_moment_maps} shows \torus moment maps and moment maps for NGC2403 from the THINGS survey. The \torus maps were generated from an SPH simulation by \cite{Dobbs2012}. By choosing a suitable SPH model, viewing angle and data cube resolution it is possible to generate synthetic moment maps which reproduce the main features of the observations, including enhanced velocity dispersion (second moment) in the spiral arms. 

\subsubsection{Galactic Plane Surveys}

A second mode of operation was developed for generating synthetic Galactic plane surveys from SPH simulations of spiral galaxies. In this mode \torus produces data cubes in Galactic latitude-longitude co-ordinates which can be compared with real Galactic plane surveys
\citep{2010MNRAS.407..405D,2012MNRAS.422..241A,2015MNRAS.447.2144D}.  
To achieve this the observer is placed inside the grid, where the observer's position is set as an input parameter and should be chosen such that the observer is in a location comparable to our location in the Galaxy. The observer's velocity is by default taken from the observer's location in the \torus grid but can be modified if required.  

Molecular line or H\,{\sc{i}} data cubes can be generated using a similar ray tracing process to the far field case but with some important modifications. When generating a Galactic plane survey a number of diverging rays are traced from the far side of the grid to the observer's position (i.e. the direction of the ray trace is reversed). By reversing the direction of the ray trace it is possible to
identify whether an individual grid cell contributes net emission or absorption to the ray. This is of scientific interest as it enables cells associated with H\,{\sc{i}} self absorption (HISA) to be identified; HISA has been proposed as a tracer of the early stages of molecular cloud formation \citep{2006AAS...208.4902G}. \torus can optionally generate positive and negative data cubes which contain contribution from net emitting and absorbing cells (the sum of the positive and negative cubes is the intensity cube). 

If the \torus grid was set up from an SPH simulation and an H\,{\sc{i}} data cube is generated then information from the data cube is mapped back onto the SPH particles. Each SPH particle is tagged with the Galactic latitude and longitude of the cell in the data cube which samples it; this enables particles contributing to a specific pixel in the data cube to be identified. The net contribution to the intensity from the grid cell in which the particle resides is also recorded; this enables particles associated with HISA to be identified so their properties can be examined \citep{2012MNRAS.422..241A}. 

\subsection{Instrumentation}
\label{sec:instrumentation}
In many instances the instrumentation itself leaves a non-negligible characteristic imprint on an observation, for example through noise, finite beam size, or spatial filtering inherent to interferometry. This can have important consequences for the observability of predicted characteristic signatures, and for testing observational diagnostics \citep{2018NewAR..82....1H}. \textsc{torus} does not directly account for these effects, in part because a substantial body of resources for doing so already exists which can be applied to a raw synthetic image. The main constraint on \textsc{torus} is therefore that it is required to produce outputs in a format that can be postprocessed, which it ensures by conforming to the FITS standard, as is discussed further in section \ref{sec:fits}. 

%\textsc{torus} can produce synthetic observations in a format compatible with the \textsc{casa} software, used to emulate ALMA data. This is very important for making a convincing proposal, since it demonstrates that interferometric effects and exposures times have been properly considered. \textsc{torus} has been used to support observing proposals for ALMA that have been successful despite fierce competition. For example data towards the tidally interacting star-disc system RW Aur has now been obtained (Rodriguez et al., in prep) with support from \textsc{torus} (Figure \ref{fig:RWAur_ALMA}). Each proposal assessment, whether successful or not, has given praise to the synthetic observational support. 

%Being able to process synthetic observations in this manner is also important for properly interpreting real data. 

%In cycle 3 Haworth used \textsc{torus} to support 7 proposals across 3 executives, of which two were awarded time (one with a EU PI, one NA). In cycle 4 Haworth also used the code to support 4 proposals, one of which was awarded priority A status. 

\begin{figure*}
    \centering
    \includegraphics[width=4.3cm]{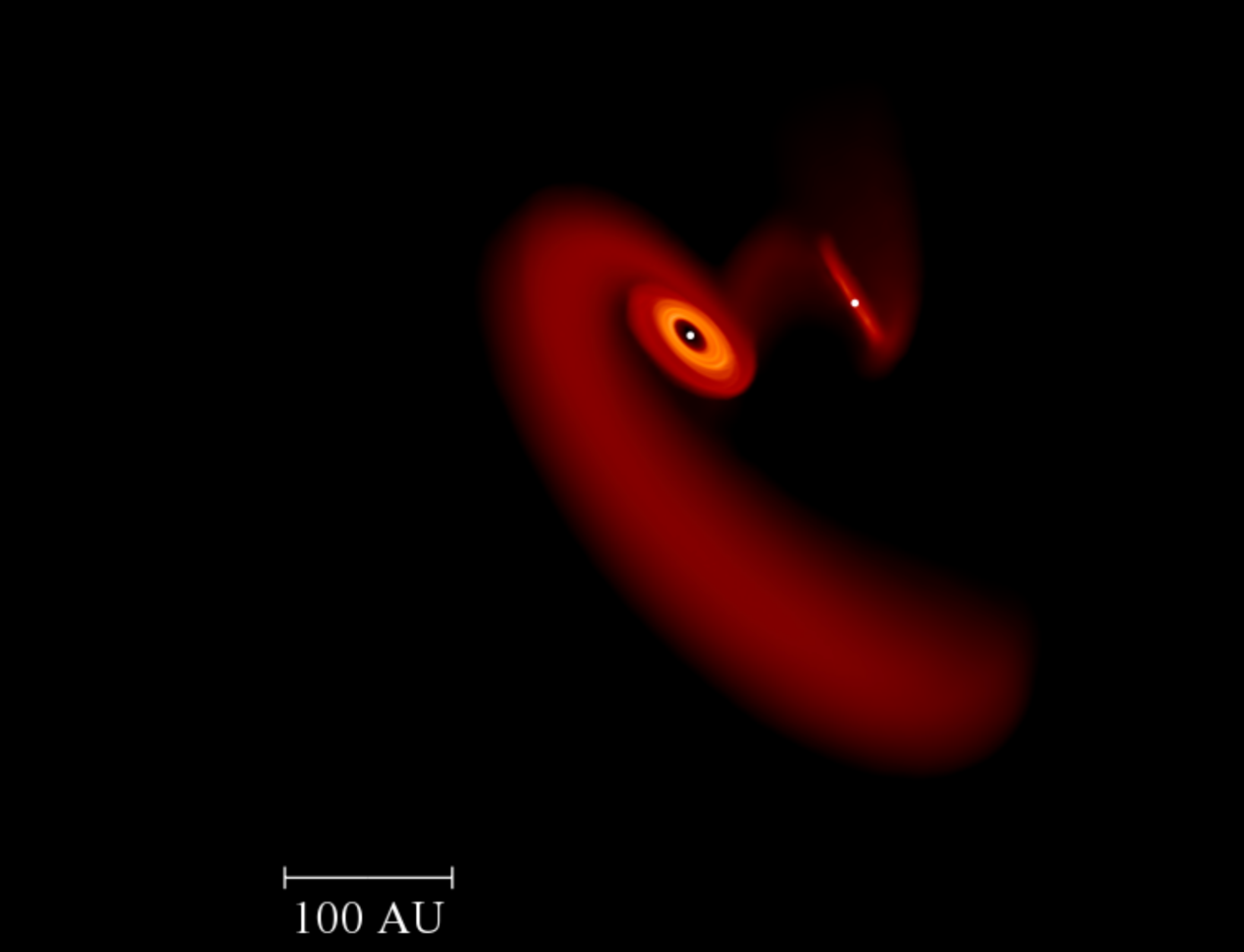}
    \includegraphics[width=12cm]{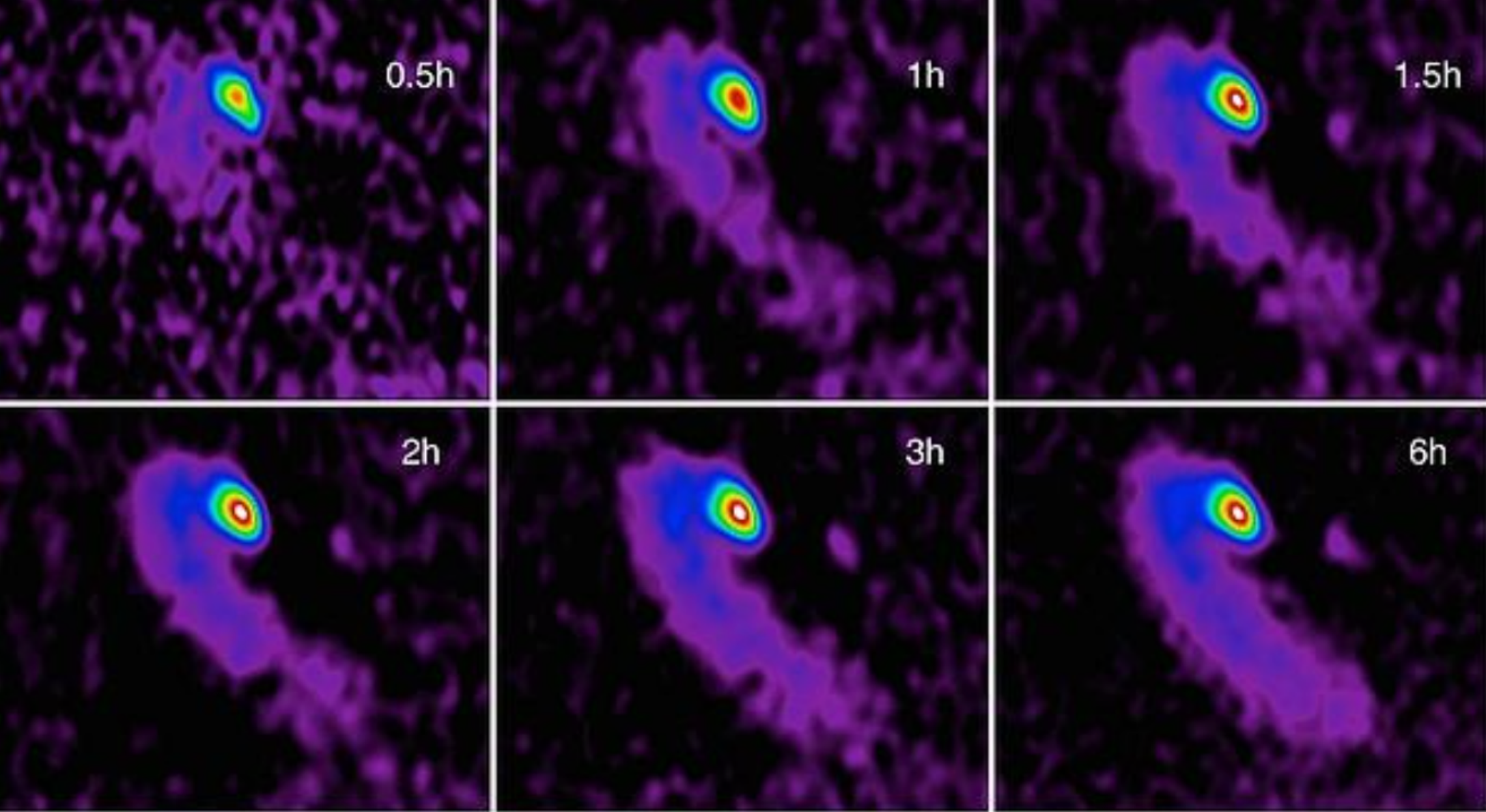}
    \caption{Synthetic CO observations of the tidally disrupted RW Aur system for different amounts of time on source with ALMA. The dynamical SPH models of Dai et al. (2015), left panel, were postprocessed with \textsc{torus} to produce data cubes in fits format. These data cubes  were then processed with CASA for different ALMA configurations/observing times. Synthetic observations such as these have helped make a successful case for ALMA observations of this system, despite the very high competition.  }
    \label{fig:RWAur_ALMA}
\end{figure*}

%I think torus has now proved itself quite good at getting alma time. I think it beats the rejection rate. Could perhaps provide a plot illustrating this, I think it would make the code seem very attractive. 

%\subsubsection{Future instrumentation}
%\textsc{torus} also has the capability to make predictions for upcoming instruments such as the James Webbb Space Telescope (JWST) and the Square Kilometer Array (SKA). 

\section{I/O and visualisation}

\subsection{The binary dump}
\label{sec:gridio}

The AMR mesh is written to disc as a compressed binary file. Some important quantities describing the model (in particular the revision identifier of the code, the date and time the model was run) are written to the binary dump, followed by the AMR mesh. The octal components of the tree are derived types, with dynamically allocated arrays of different dimensionality and variable types. Each component of the derived type is written in a self-describing manner, with a text flag describing what the component is, a flag describing the variable type (e.g. double precision floating point number, logical variable, character string, integer etc) and integer values describing the array shape. The values associated with the derived type are then written. The file writing routine works recursively though the tree.

This format not only allows backwards compatibility of binary dumps from older versions of \torus\, but also allows models run with different physics to be reprocessed with new physics (i.e. a dust radiative equilibrium model can then be read in to a model that computes molecular statistical equilibrium).

On reading the binary dump, a warning is written if the revision of the code reading the file differs from the revision of the code that wrote the binary dump. As the code works its way through the binary dump it dynamically allocates the appropriate components of the octal derived types. If a component is found that it does not recognise (for example a deprecated component) then a warning is given and the code continues.

Note if the code is running under MPI grid I/O occurs only through the zeroth thread. If the grid is domain decomposed (i.e. the code is running a hydrodynamics model) octal grid output information is sent from the domain decomposed threads to the zeroth thread, which recreates the tree from the root and writes the dump to disc. Hence the whole tree structure is retained in a single binary file, rather than having a set of parallel dump files. This allows the code to work on HPC systems that have a restricted number of I/O capable nodes (such as the BlueGeneQ), and also enables users to either change the domain decomposition 'on the fly', or read in a domain decomposed model on a single process.

The binary files can be substantial in size, but they do compress well. By default the code compresses the binary dumps as they are written out. The compression is done using the zlib library which uses the DEFLATE algorithm. It is possible to set a flag to write out uncompressed binary files.

\subsection{FITS files}
\label{sec:fits}
Some modules in the code produce FITS format \citep{2010A&A...524A..42P} images or datacubes. These files are written using the CFITSIO library \citep{2010ascl.soft10001P}, and contain standard WCS (world co-ordinate system) header keywords. The FITS files are compatible with standard FITS viewers, such as \textsc{ds9} \citep{2000ascl.soft03002S} and \textsc{gaia} \citep{2014ascl.soft03024D}, as well as observation preparation tools such as \textsc{casa}\footnote{\url{https://casa.nrao.edu/}} \citep{2007ASPC..376..127M}. Post-processing with software such as \textsc{casa} enables FITS files generated by \torus to emulate observations with interferometers such as ALMA \citep[e.g.][]{2012ASPC..461..849P}. Using \textsc{casa}, \textsc{torus} has produced synthetic observations to support a number of successful ALMA observing proposals (e.g. PIs: Ilee, Rodriguez, Pani\'{c}). An example of the kind of synthetic observations produced by \textsc{torus} to support ALMA observing proposals is given in Figure \ref{fig:RWAur_ALMA}, which shows an SPH model of the RW Auriga system, and synthetic ALMA CO integrated intensity maps for different total times on source. New ALMA data for this system was obtained with the support of \textsc{torus}, appearing in \cite{2018ApJ...859..150R}. 

For single dish instruments noise and Gaussian convolution of an image can also be added using a range of tools, for example \textsc{ciao} \citep{2006SPIE.6270E..60FB}. The FITS files produced by \textsc{torus} are also compatible with the \textsc{FluxCompensator} suite of single dish emulation tools by \cite{2017ApJ...849....3K}. 

\subsection{Visualisation}

We have not attempted to provide any graphical visualisation natively, and instead the code produces VTK format files \citep{Schroeder:1996:DIO:244979.245018, Schroeder:1998:VTO:272980}. By default the code uses the binary VTK format \citep{VTKuserguide} with the  {\tt .vtu} suffix. In summary this is an xml file that has a component that is a compressed, base64 encoded binary  array containing the octree itself. 
The VTK writing module enables the developer to select which octal derived type components should be written to the VTK file

There is a host of excellent open source 3-D visualisation tools available, as well as proprietary software. We typically use {\sc visit} \citep{HPV:VisIt}, or {\sc paraview} \citep{Paraview} to view the grid itself. 
% \footnote{\url{https://wci.llnl.gov/simulation/computer-codes/visit/}}
% \footnote{\url{https://www.paraview.org}}

If the model has stellar photon sources, these can also be written to VTK files, and potentially superimposed on the grid visualisation within {\sc visit} or {\sc paraview}. This is particular useful for those running hydrodynamics simulations that include sink particles.

%%%%%%%%%%%%%%%%%%%%%%%%%%%%%%%%%%%%%%%%
% Code performance and parallelization %
%%%%%%%%%%%%%%%%%%%%%%%%%%%%%%%%%%%%%%%%

\section{Code performance and parallelization}
\label{sec:parallelisation}

\subsection{Parallelisation techniques/optimisation}

The main drawback of Monte Carlo radiative transfer is that it can be computationally expensive, requiring each cell on the grid to be sufficiently sampled to give a converged result. This is particularly problematic for radiation hydrodynamics models where a large number of radiative transfer calculations are required (see section \ref{sec:radhydro}). Fortunately MCRT is the sampling of a large number of independent random events, so lends itself extremely well to large scale parallelization. 

In \torus the loops which calculate MCRT are parallelised with both OpenMP (shared memory parallelism) and MPI (distributed memory and shared memory parallelism). When MPI is used to parallelise a MC calculation each MPI process stores its own copy of the grid (hereafter referred to as ``multiple grid copies''). 

In a photoionisation calculation the computational grid can be domain decomposed by splitting the grid over MPI processes at a given level of refinement, resulting in a number of sub-domains which is a power of 2. Furthermore in a domain decomposed photoionisation calculation we use two additional features that can further reduce the computation time: load balancing processes and photon packet stacks. These features are discussed in more detail in the remainder of this section. 

\subsubsection{Multiple grid copies}
\label{sec:multiple_grid_copies}

An efficient way of scaling MCRT is to make $N_{grids}$ copies of the entire computational grid and for each copy to propagate a fraction $N_{monte}/N_{grids}$ of the total number of photon packets. Each copy of the grid counts path lengths through each cell, which are collated using \verb+MPI_ALLREDUCE+
once all packets are propagated. The sums of all the path lengths are then used to compute the ionisation and thermal balance.

When the grid is domain decomposed the sub-domains can be further parallelised using multiple grid copies. This can help to improve performance, by avoiding excessive overheads associated with small sub-domains, and relaxes the constraint that the number of sub-domains is a power of 2 (when $N_{grids}$ is an odd number).

\subsubsection{Hybrid MPI+OpenMP parallelisation}
\label{sec:hybrid_parallelsim}

Even with multiple grid copies and domain decomposition there are some limitations to the flexibility of MPI parallelism. With multiple grid copies each MPI process stores the whole grid which can lead to an excessively large memory footprint. This can be avoided by using OpenMP parallelism as each OpenMP thread accesses the same copy of the grid.

In a domain decomposed calculation the number of MPI processes is the number of sub-domains multiplied by the number of grid copies. This may not correspond neatly to the number of processors (cores) available on a given platform. Cores which would otherwise be idle can be utilized by running \textsc{torus} in a hybrid MPI+OpenMP configuration. A set number of cores will be used for the domain decomposition/grid copies and the remaining cores populated with threads generated in OpenMP parallelised regions.

\subsubsection{Load balancing}
\label{sec:load_balancing}
When a photoionization calculation is domain decomposed, copies of each sub-domain (not of the entire grid) can be created. Each sub-domain is put on a single processor. There can be an arbitrary number of such copies. These are allocated to regions of the grid which are predicted to do the most work, thus speeding up the Monte Carlo propagation. For example, load balancing processes can be allocated according to weighting factors including number of cells in each sub-domain, number of photon interaction events, number of emitting sources, or luminosity. At the very beginning of a radiation calculation, when no packets have been propagated, a good proxy for the work-load is the number of sources in each sub-domain. Afterwards, it is well characterised by the number of events.  Once the photon propagation is complete, the load balancing processes communicate their Monte Carlo estimators back to the associated domain using \texttt{MPI_ALLREDUCE}. 

The load balancing weighting changes for the thermal balance calculation, as each sub-domain will do approximately the same amount of work (in which case, the number of cells is the most appropriate weighting factor).

The benefit of this load balancing method over creating multiple grid copies is that processors are only allocated where they are required, instead of copying an entire domain (which would also include regions where very few events occur).

\subsubsection{Photon packet stacks}
\label{sec:photon_packet_stacks}

The efficiency of communicating photon packets between grid domains in a domain decomposed calculation can be improved by passing the photon packets in stacks, rather than individually, to reduce the latency associated with performing a large number of small communications. In this scheme the domain will propagate packets, keeping count of how many cross into each neighbouring domain and storing those ready to cross in a stack. Once a domain is ready to receive $N_{stack}$ packets, the stack is communicated via MPI using a single \verb+MPI_SEND+ / \verb+MPI_RECV+. There is a tradeoff in the improvement between the reduction in the number of MPI communications (which one wants to reduce) and the time threads spend idle while waiting to receive a stack, hence the value of $N_{stack}$  requires tuning for the specific configuration being run (see section~\ref{sec:performance_HIIregion}). 

\subsection{Performance and scaling}
\label{sec:performance_and_scaling}

At the time of writing the majority of High Performance Computing (HPC) systems used for computational astrophysics are clusters built from a number of individual compute nodes joined by an interconnect. The compute nodes typically comprise two physical chips each with multiple cores (each core acts as an individual processor). The CPUs are housed in ``sockets'' on the motherboard hence a compute node with two physical CPU packages is referred to as a two socket compute node. Cores in the same compute node share memory and can use OpenMP or MPI parallelism, whereas parallelism between compute nodes can use MPI but not OpenMP. The use of different types of parallelism and the mapping of the workload onto the hardware can have a significant impact on performance (e.g. distribution of OpenMP threads across sockets). In this section we present performance and scaling results with a view to informing the efficient use of computational resources on current hardware platforms. 

The following tests were run on the University of Exeter HPC system using two socket compute nodes with Intel\textregistered Xeon\textregistered E5-2640 v4 CPUs with a 2.40GHz clock speed. Each CPU has 10 cores so one compute node provides a total of 20 cores which can directly address 128GB of RAM. 

\subsubsection{3D disc benchmark}

The first test is a strong scaling study (i.e. fixed problem size) of the 3D disc benchmark (see section~\ref{sec:radeq_validation}). The grid is not domain decomposed but multiple grid copies are used with MPI (see section~\ref{sec:multiple_grid_copies}). We also investigate the performance of hybrid MPI+OpenMP parallelism (see section~\ref{sec:hybrid_parallelsim}). 

The 3D disc benchmark uses the dust radiative equilibrium algorithm, described in section~\ref{sec:radeq}, with 3\,200\,480 photon packets traversing a  cylindrical polar grid. This is a sufficiently large problem size that it can scale beyond a single compute node and is representative of many circumstellar disc applications using \torus. The configuration is identical to the \verb+disc_cylindrical+ benchmark except that file I/O has been switched off unless otherwise stated. The time taken to perform an iteration of the Lucy algorithm is recorded in the \verb+tune.dat+ file (shown as ``One Lucy Rad Eq Itr''). This represents the vast majority of the run time more than 99.9\% in a serial run without I/O) so we measure the average time for one iteration of the Lucy algorithm as our performance metric for this case.

We begin by testing performance with different compilers and optimisation flags for a pure-OpenMP configuration on a single compute node. We need to understand the node-level performance before scaling up to larger configurations and it also informs how to run \torus on a standalone server or desktop machine. The two compilers tested were the GNU Fortran compiler (gfortran 5.4.0) and the Intel Fortran compiler (ifort 16.0.3) with O2 and O3 optimisation levels. The results are shown in table~\ref{tab:compilers_disc_test} as run1--run4. Best performance is achieved with the Intel compiler and O3 optimisation which is 6-9\% faster than the other runs. We also tested the architecture specific optimisation flag (-xCORE-AVX2) which enables the AXV2 vector instructions in the target processor to be used. However this run did not perform as well as the run with generic O3 optimisation. The performance of \torus does not depend on working efficiently with large blocks of data held contiguously in memory so it is unsurprising that enabling vector instructions does not help performance.
\begin{table*}
    \centering
    \begin{tabular}{|c|c|c|c|c|c|c|}
    \hline
     Run ID & Compiler & Optimisation flags & No. of threads & Affinity & Time (s)   & Normlised time\\
     \hline
    run1 & gfortran  & -O2   & 20 & none    & 193.0 & 1.11 \\
    run2 & gfortran  & -O3   & 20 & none    & 191.7 & 1.11 \\
    run3 & ifort     & -O2   & 20 & none    & 195.7 & 1.13 \\
    run4 & ifort     & -O3   & 20 & none    & 179.9 & 1.04 \\
    run5 & ifort     & -O3 -xCORE-AVX2 & 20 & none & 195.5 & 1.13 \\
    run6 & ifort     & -O3   & 20 & compact & 173.2 & 1    \\
    run7 & ifort     & -O3   & 20 & scatter & 175.5 & 1.01 \\
    run8 & ifort     & -O3   & 10 & compact & 216.4 & 1.25 \\
    run9 & ifort     & -O3   & 10 & scatter & 231.3 & 1.36 \\
    \hline
    \end{tabular}
    \caption{Performance of the dust radiative equilibrium algorithm for different compilers, compiler flags and OpenMP thread affinities. The GNU Fortran compiler (gfortran) is version 5.4.0 and the Intel Fortran compiler (ifort) is version 16.0.3. The time is the average time taken for one iteration of the Lucy algorithm and the normalised time is the time divided by the shortest time. OpenMP thread affininty was set using the KMP_AFFINITY environment variable.}
    \label{tab:compilers_disc_test}
\end{table*}

Runs 6 and 7 investigate the effect of setting OpenMP thread affinity. If thread affinity is not set then OpenMP threads can migrate between CPU cores whereas if affinity is set this binds a thread to a specific core. Run6 (compact affinity) is 4\% faster than the corresponding run with no affinity set. Run7 is (scatter affinity) is indistinguishable from run6 but this is unsurprising in a fully populated node. To further illustrate the effects of affinity two runs were performed with 10~OpenMP threads (i.e. a half populated node) and either ``compact'' or ``scatter'' affinity. With compact affinity all the threads run on cores which on the same socket and the threads share a memory controller. With scatter affinity half the threads are on one socket and half are on the other socket in the compute node. The performance with compact affinity is significantly better than the performance with scatter affinity which is expected based on the way \torus allocates memory. The grid is initialised by thread~0 and with a first-touch memory allocation policy the grid will reside in memory which is owned by the memory controller of the first CPU on the node. Threads on the CPU in the second socket access this memory more slowly than threads on the CPU in the socket (even though they can directly address this memory) which is referred to as Non-Uniform Memory Access (NUMA). With compact affinity and 10 threads per node all the threads access the grid via the local memory controller; with scatter affinity half the threads have to access the grid via the memory controllor on the other CPU. For subsequent tests we will use the Intel compiler with O3 optimisation and where OpenMP is used the affinity will be set to compact to maximise performance. 

\begin{figure}
    \centering
    \includegraphics[width=9cm]{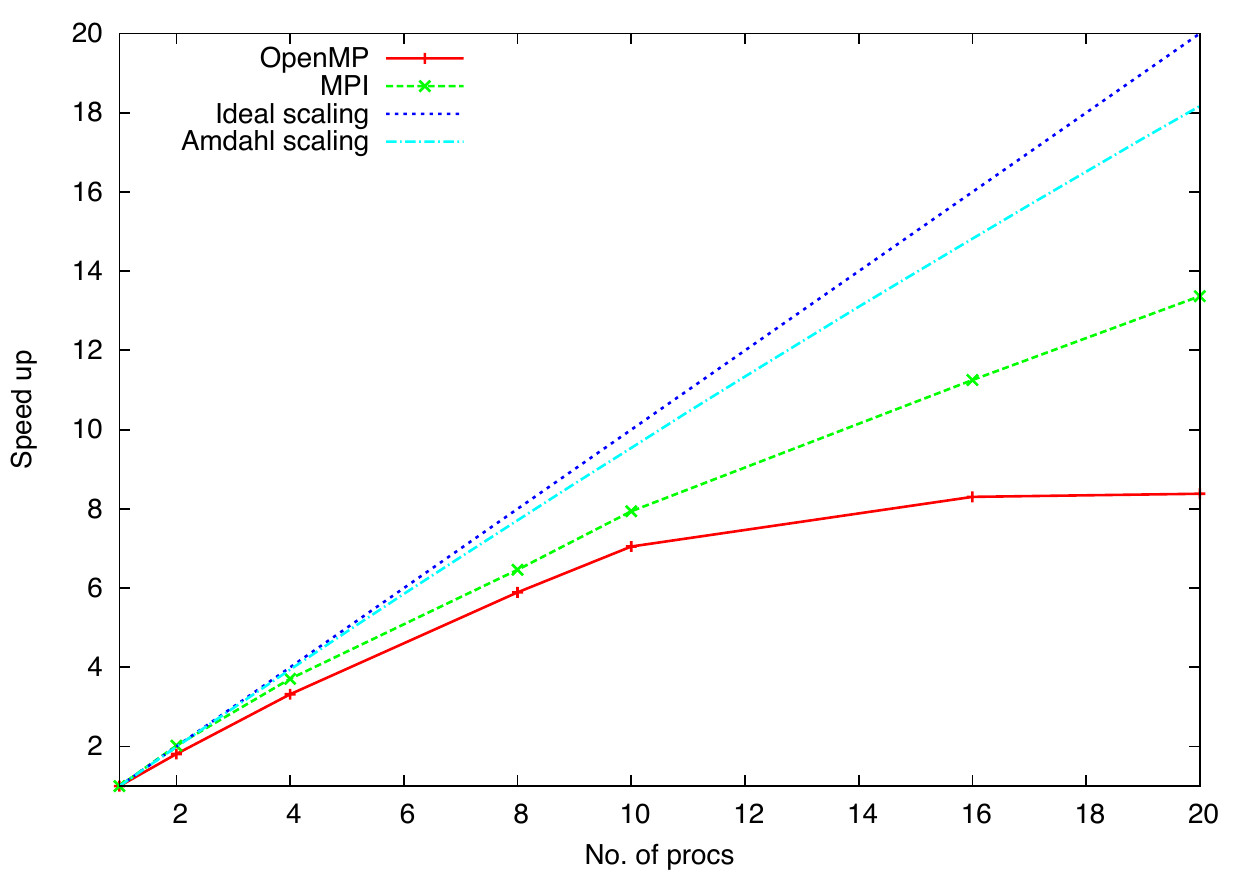}
    \caption{Parallel scaling on a single compute node for OpenMP and MPI parallelisation. The red-solid line shows OpenMP scaling, the green-dashed line shows MPI scaling, the blue-dotted line is ideal scaling and the light-blue-dot-dashed line is the scaling predicted by Amdahl's law.}
    \label{fig:single_node_scaling}
\end{figure}

Next we investigate single node OpenMP and MPI performance. Both OpenMP and MPI configurations were run using 2, 4, 8, 10, 16 and 20 cores (i.e. power of 2 core counts and fully/half populated node). A serial run, with no parallelism was performed to enable a parallel speed up to be calculated and the results are shown in fig.~\ref{fig:single_node_scaling}. The red-solid line shows OpenMP scaling, the green-dashed line shows MPI scaling and the blue-dotted line is ideal scaling. 
Up to 10 cores the OpenMP and MPI performance is similar but OpenMP scaling beyond 10 cores is much less efficient than MPI scaling. In both cases the scaling is less than the idealised case and a number of factors contribute to real-world scaling being less than ideal. One contributing factor is the fraction of work which can be parallelised and the effect this has on strong scaling is referred to as Amdahl's law. The speed up $S_N$ using $N$ processors is limited to 
\begin{equation}
    S_N=\frac{1}{\frac{p}{N} + s}
\end{equation}
where $p$ is the fraction of parallel work and $s=1-p$ is the fraction of serial work. From the serial run we determine that the parallel fraction is 0.99467 which limits the maximum possible speed up to 188 (i.e. in the absence of any other factors scaling is limited to around 9 nodes with 20 cores per node). In practice the single node scaling is below that predicted by Amdahl's law (light-blue-dot-dash line) but factors such as parallel overheads, load balancing and resource contention are not accounted for by Amdahl's law. 

Resource contention can be illustrated by running with 10~MPI processes which are either placed all on the first socket or distributed evenly over both sockets. In the former case the time for one iteration is 1.12 times longer than the latter case. When the MPI processes are all placed on the same socket they are sharing the resources of the memory system and this contention is less severe when the processes are distributed across two sockets. Consequently the default process placement for MPI is typically to distribute processes across sockets to improve performance. OpenMP achieves better performance when threads share a memory controller, whereas MPI processes do not share a memory address space, so not benefit from sharing the memory system.

To conclude the single node scaling study we ran two hybrid OpenMP/MPI configurations. The  first uses 2~MPI processes and 10~OpenMP threads per process, and the second uses 4~MPI processes and 5~OpenMP threads per process. The run times for the two hybrid configurations, and the fastest OpenMP and MPI runs, are shown in table~\ref{tab:single_node_best_buy}. 
\begin{table}
    \centering
    \begin{tabular}{|c|c|c|c|}
    \hline
    Configuration     & Memory use & Time    & Normalised \\ 
                      &      (GB)  &  (s)    &   time  \\
    \hline
     OpenMP           & 1.73       &  179.9  & 1.59    \\
     MPI              & 34.9       &  112.8  & 1       \\
     Hybrid (2x10)    & 3.47       &  120.8  & 1.07    \\
     Hybrid (4x5)     & 6.92       &  122.5  & 1.09    \\ 
     \hline
    \end{tabular}
    \caption{Comparison of single node performance for OpenMP, MPI and hybrid configurations. The OpenMP time is taken from the fastest 20-core run. Hybrid (2x10) uses 2 MPI processes and 10 OpenMP threads per MPI process. Hybrid (4x5) uses 4 MPI processes and 5 OpenMP threads per MPI process. The memory use is the maximum resident size reported by the GNU time command.}
    \label{tab:single_node_best_buy}
\end{table}
Each of the three possible parallelisation methods (OpenMP, MPI and hybrid) has its merits. OpenMP is the easiest to use both in terms of building \torus (support for OpenMP is in the compiler and does not require a separate MPI library) and in terms of running a job. However OpenMP performance is significantly below the corresponding MPI performance on a two socket HPC node (this will also be the case on a two socket standalone server). MPI has a much larger memory footprint than OpenMP, as each MPI process stores its own copy of the grid. However if sufficient memory is available MPI offers the best performance and configuring a single node MPI job is typically only slightly more difficult than configuring an OpenMP job. A hybrid configuration can deliver performance close to the pure-MPI case with a much smaller memory footprint. However hybrid configurations can be more difficult to configure correctly to ensure the correct placement of MPI processes and OpenMP threads onto CPU cores.

\begin{figure}
    \centering
    \includegraphics[width=9cm]{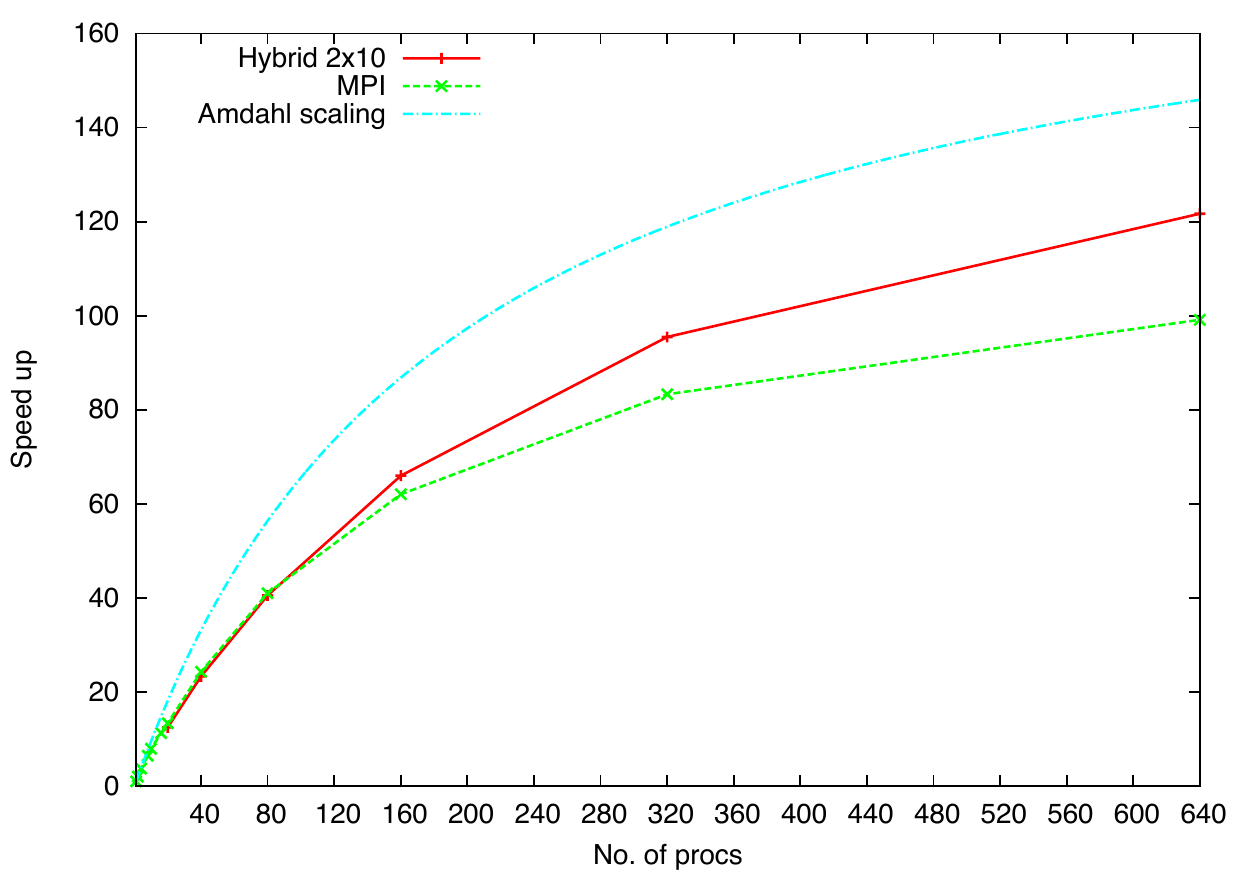}    
    \caption{Parallel scaling on up to 32 compute nodes (640 processors) for MPI and hybrid parallelisation. The red-solid line shows hybrid scaling, the green-dashed line shows MPI scaling, and the light-blue-dot-dashed line is the scaling predicted by Amdahl's law.}
    \label{fig:multi_node_scaling}
\end{figure}

We conclude the study of the dust radiative equilibrium performance with results from multiple node scaling tests. Tests were run on 2, 4, 8, 16 and 32 compute nodes (40, 80, 160, 320 and 640 processors) using MPI and hybrid (2x10) parallelism. The scaling result are shown in fig.~\ref{fig:multi_node_scaling}.
Speed up continues to increase up to 640 processors (32 nodes) although below the idealised rate. Initially the best performance is achieved with pure-MPI however the hybrid OpenMP/MPI configuration scales better and is faster beyond 160 processors. Achieving substantially better parallel scaling to high processor counts would require parallelisation of the remaining $\sim0.5$\% of the workload which is serial. 

In practice the dust radiative equilibrium calculation is often used in parameter space studies in which parallelism also comes from running multiple instances of \torus with different input parameters. In this case it is preferable to use a smaller number of processors and run more instances concurrently. Improvements to performance can still be found by considering compilers and the associated flags and an appropriate form of parallelism (OpenMP, MPI, hybrid) for the calculation size and available resources. It is also important for performance to switch off writing of restart dumps if possible. When \torus is scaled to higher processor counts this can become a significant fraction of the iteration time and is only required if the radiative equilibrium calculation needs to be restarted. 

\subsubsection{3D HII region expansion}
\label{sec:performance_HIIregion}

The second performance study uses a domain decomposed radiation hydrodynamics calculation of an expanding D-type ionisation front. In this section we investigate the performance impact of using load balancing processes (see section~\ref{sec:load_balancing}) and varying the photon packet stack size (see section~\ref{sec:photon_packet_stacks}).

\begin{figure}
    \centering
    \includegraphics[width=9cm]{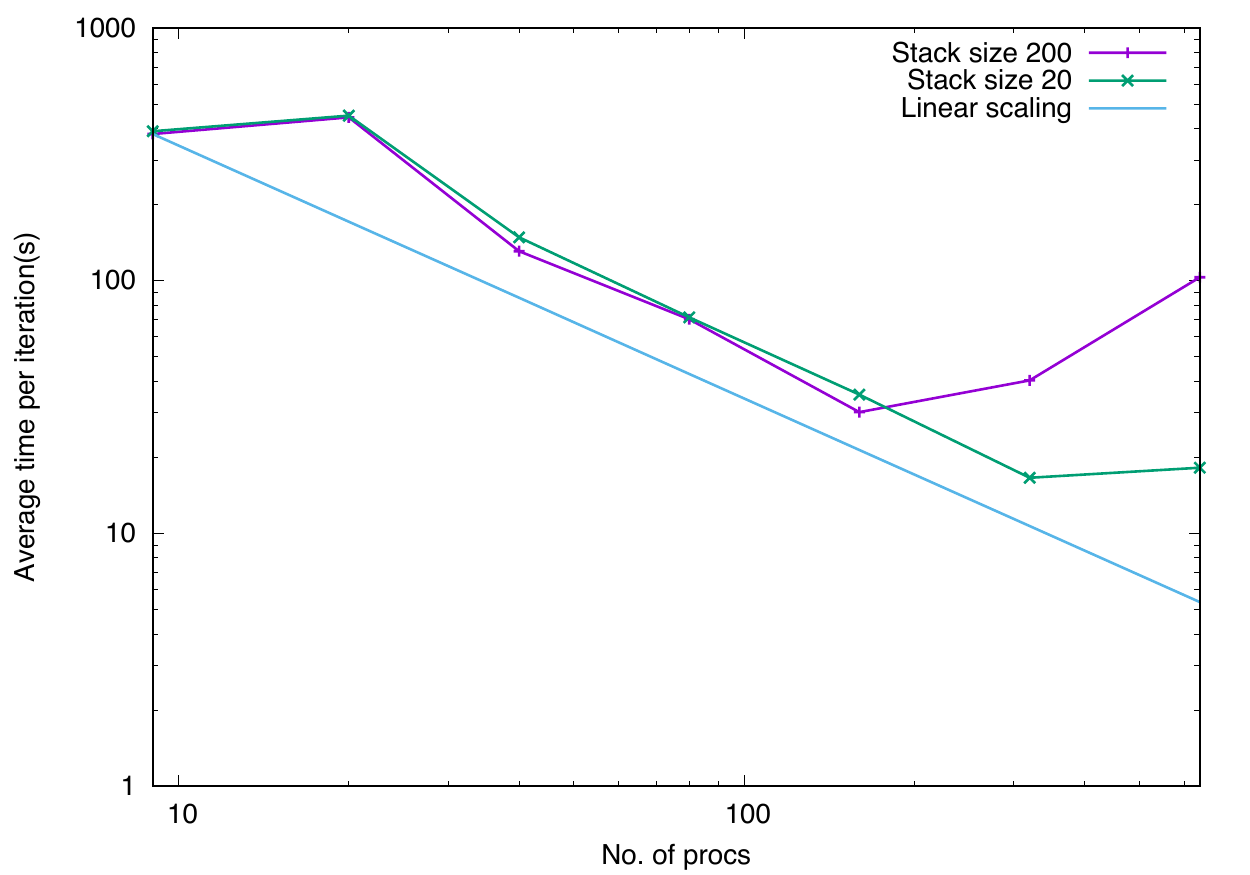}
    \caption{Average iteration time for the photoionisation step in a radiation hydrodynamics calculation. The purple line shows timings with a stack limit of 200 and the green line with a stack limit of 20. The blue line shows idealised linear scaling from the 9~processor run.}
    \label{fig:rh_scaling}
\end{figure}

A single star, with properties identical to the Lexington benchmark star (see section~\ref{sec:photoionisation_validation}), is located at the origin of the grid. The star is surrounded by a uniform density medium into which an ionisation front propagates. The calculation includes full photoionisation microphysics with a negligible dust abundance. The grid has a Cartesian geometry with $64^3$ cells which is sampled with 26~214~400 photon packets. The performance results are shown in table~\ref{tab:radhydro_timings}. In each case there are 8~MPI processes calculating the hydrodynamics step with a variable number of load balancing processes contributing to the photoionisation calculation. The time recorded is the average time for all iterations of the photoionisation loop (shown as ``One photoionization itr'' in the \verb+tune.dat+ file) except for the first iteration which is not representative of subsequent iterations. The hydrodynamics step takes only 4-8 seconds to run and is not included in the timings in table~\ref{tab:radhydro_timings}. 
\begin{table*}
    \centering
    \begin{tabular}{|c|c|c|c|c|r|}
    \hline
         Hydro procs & Balancing procs & Total procs & No. of nodes & $N_{stack}$ & Av. time (s) \\
    \hline
         8           &    0            &   9         & 1            &  200 & 380.2 \\
         8           &   11            &  20         & 1            &  200 & 442.9  \\
         8           &   31            &  40         & 2            &  200 &  130.8 \\
         8           &   71            &  80         & 4            &  200 &  70.3  \\
         \textbf{8}  &  \textbf{151}   & \textbf{160}& \textbf{8}   &  \textbf{200} & \textbf{30.2}  \\
         8           &  311            &  320        & 16           &  200 &  40.3  \\
         8           &  631            &  640        & 32           &  200 &  103.1 \\
                     &                 &             &              &      &  \\
         8           &    0            &   9         & 1            &  20  & 390.2 \\
         8           &   11            &  20         & 1            &  20  & 449.8 \\
         8           &   31            &  40         & 2            &  20  & 148.5 \\
         8           &   71            &  80         & 4            &  20  &  71.6 \\
         8           &  151            &  160        & 8            &  20  &  35.4 \\
         \textbf{8}  &  \textbf{311}   & \textbf{320}& \textbf{16}  &  \textbf{20}  & \textbf{16.6} \\
         8           &  631            &  640        & 32           &  20  &  18.2 \\
    \hline
    \end{tabular}
    \caption{Timings for a radiation hydrodynamics calculation of a D-type ionisation front expansion. The average time is the time taken for all iterations of the photoionisation loop excluding the first iteration. The fastest run for a given $N_{stack}$ is shown in bold.}
    \label{tab:radhydro_timings}
\end{table*}
The average iteration time is plotted in fig.~\ref{fig:rh_scaling} as a function of number of processors. Timings with a stack size of 200 are plotted as a purple line, timings with a stack size of 20 are plotted as a green line, and an idealised linear scaling from the 9~ processor run is shown as a blue line. 

Adding a small number of load balancing processes actually slows the calculation down, but as the number of load balancing processes is increased the load balancing process becomes much more efficient, and between 2 and 8 nodes the calculation scales well for both stack sizes. With the larger stack size of 200 the calculation slows down beyond 8 nodes. However decreasing the stack size to 20 enables scaling to 16 nodes with approximately half the run time of the fastest 8~node run. In general lower node counts perform slightly better with a larger stack size, whereas larger jobs work significantly better with a smaller $N_{stack}$. A smaller $N_{stack}$ enables load balancing to work more effective which outweighs the higher parallel overheads for a large number of processors. 

\section{Operating framework}

\torus has been developed for a number of years at the University of Exeter where many of the developers are, or were, based. The \torus code is not open source but is available on a collaborative basis by contacting the developers. Over time the user and developer base has expanded and the operating framework has evolved accordingly as described in this section. 

\subsection{Version control}
\label{sec:version_control}

Using a version control system is a vital when developing a large code such as \torus particularly when there are multiple developers working on a common code base. The version control system not only safely stores the latest version of the code in a remote repository but also records the history of changes made to the code and a log message which describes why the change was made. 

The \torus code and data files use the version control system git\footnote{\url{https://www.git-scm.com}}, with repositories hosted on the bitbucket\footnote{\url{https://bitbucket.org}} service. 

Developers are strongly encouraged to carry out major development on a branch which is then merged back into the main line of development when it has been tested and shown to be working correctly. Minor changes can be developed directly on a working copy of the trunk (main line of development) which is tested before being committed to the repository. 

\subsection{Build system}

\torus is compiled using a Makefile which correctly handles re-building dependent modules when code changes are made. The Makefile contains a number of different configurations which are selected by specifying the \verb+SYSTEM+ environment variable. Generic configurations are provided for popular compilers and in addition to configurations which are tailored to specific platforms (typically HPC systems). At the time of writing the most popular compilers used to build \torus are the GNU Fortran compiler (gfortran) and the Intel Fortran compiler (ifort). Generic configurations are provided for these two compilers and \torus is routinely built and run with these compilers. The configurations also specify compiler flags to be used for debugging or optimisation, and these are typically tailored to the specific platform and compiler being targeted. 

Many of the source files contain pre-processor directives which are used to include or exclude code from the compilation. The most important application of pre-processor keys is to exclude lines of code specific to builds using MPI parallelism. As MPI involves making subroutine calls which are provided by an external library these calls need to be excluded from non-MPI builds. When \torus runs it reports the pre-processor options and parallelisation method which were selected when the executable was built to make sure that the built is suitable for its indended use. \torus also reports the number of MPI processes and/or OpenMP threads (according to the parallelisation in use) and the host name where each MPI process is running. Knowing the location of each MPI process can be used to check that a multi-node job has been started correctly, which is of particular significant in hybrid configurations where the number of MPI processes per node is less than the number of available cores.

In addition to using the Makefile directly there is a build script which can be used to compile \torus for the full range of parallelisation options (serial, OpenMP, MPI and hybrid OpenMP/MPI), using either the Intel or GNU compilers. The script will check for the availability of both compilers and build with the Intel compiler if available or the GNU compiler otherwise. The Intel compiler has been found to give better performance than the GNU compiler (see section~\ref{sec:performance_and_scaling}) hence it is preferred when available.

\subsection{Testing framework}
\label{sec:testsuite}

A number of benchmark and test cases are available in the code repository and these are summarised in table~\ref{tab:benchmarks}. A number of tests are run daily to ensure that key functionality continues to work correctly in the main line of development. These are marked with a ``D'' in table~\ref{tab:benchmarks}. The suite of tests exercise different modules and the tests are repeated using different parallelisations methods (OpenMP, MPI, hybrid) to detect possible bugs in parallelisation. Each configuration is built with debug flags which enable run time checks (e.g. checking for out of bounds array access) which are not normally included as they slow down execution. 
\begin{table*}
    \centering
    \begin{tabular}{|l|l|l|c|c|c|c|}
    \hline
     Name                  & Description                    & Functionality   & Domain          & OpenMP & MPI   & Hybrid  \\
                           &                                & tested          & decomposed?     &        &       &         \\
     \hline
     HII_region            & HII region                     & Photoionisation   & No              &  D,S   & D,S   &   D,S   \\
     HII_regionMPI         & HII region                     & Photoionisation   & Yes             &  -     & D,S   &     -    \\
     angularImageTest      & l-b-v data cube generation     & Data cubes        & No              &  D,S   & D,S   &   D,S    \\
     cylinder_image_test   & Image generation               & Images            & Yes             &  -     & D,S   &     -    \\
     disc                  & 2D circumstellar disc          & Dust rad. eqn.    & No              &  D,S   & D,S   &   D,S     \\
     disc\_cylindrical     & 3D circumstellar disc          & Dust rad. eqn.    & No              &  S     & S     &     S   \\
     gravtest              & 3D multigrid gravity solver    & Self gravity      & Yes             &  -     & D,S   &     -     \\
     gravtest_2d           & 2D multigrid gravity solver    & Self gravity      & Yes             &  -     & D,S   &     -     \\
     hydro                 & Shock tube                     & Hydrodynamics     & Yes             &  -     & D,S   &     -     \\
     molebench             & Molecular physics              & Molecular physics & No              &  D,S   & D,S   &     D,S  \\ 
     nbody                 & N body solver                  & N-body dynamics   & No              &  D,S   & D,S   &     D,S \\ 
     restart               & Restart disc (2D)              & Restart           & No              &  D,S   & D,S   &     D,S  \\
     sphToGridBinary       & SPH to grid (binary dump)      & Grid generation   & No              &  D,S   & D,S   &     D,S  \\
     sphbench              & SPH to grid (ASCII dump)       & Grid generation   & No              &  D,S   & D,S   &     D,S  \\
    \hline
    \end{tabular}
    \caption{Benchmarks and test cases. Domain decomposed configurations must use MPI. D indicates tests which run in the daily test suite and S indicates tests which run in the stable version tests. The disc and disc\_cylindrical benchmarks are from \citet{2004A&A...417..793P}; the HII\_region and HII\_regionMPI benchmarks are from \citet{1995aelm.conf...83F};molebench is described in \citet{2010MNRAS.407..986R}; sphbench is based on tests in \citet{2010MNRAS.403.1143A}.}
    \label{tab:benchmarks}
\end{table*}
The compiler used in the daily tests is gfortran 4.8.2 and the following flags are used: 
 \begin{verbatim}
-ggdb -fbacktrace -fcheck=all -Wunderflow -Wall 
-Werror -Wno-surprising 
-Wno-error=maybe-uninitialized 
-pedantic-errors -std=f2008 -fall-intrinsics 
-ffpe-trap=invalid,zero,overflow
 \end{verbatim}
Extensive run time error checking and floating point exception trapping are enabled and, where practical, warnings are converted to errors to help maintain coding standards. In order to improve portability the Fortran 2008 standard is enforced which disallows compiler specific extensions. For MPI and hybrid runs the OpenMPI 1.8.1 library is used to provide MPI functionality. The MPI versions of the tests are run with the \verb+gcov+ coverage analysis tool\footnote{\url{https://gcc.gnu.org/onlinedocs/gcc/Gcov.html}} to show which parts of \torus are being exercised and how many times each line of code is executed.

When a stable version of the code is being prepared an exanded series of tests is run, marked with an ``S'' in table~\ref{tab:benchmarks}. The stable version tests include cases which are too computationally expensive to run every day and also includes a repeat run of all tests with optimisation flags enabled. The test suite script is included in the code repository to enable full testing of changes before they are committed to the repository.

\subsection{Automatic BibTex generation}

One of the difficulties of running any large scale numerical modelling tool is that it is always underpinned by a sizeable set of papers spanning RT/hydrodynamics methods, numerical algorithms (e.g. array sorting, multi-dimensional optimisation, integrators), to science inputs such as atmosphere models, atomic or molecular rates, or grain optical properties. It is very important that these key papers get cited, both morally and scientifically. 

Often however it is difficult to work out where the data or algorithms used in a particular model come from. We have therefore implemented a scheme in which the BibTex entries for all the papers that contributed to a particular model is constructed automatically by the code itself, along with a brief description of what each paper provided. The log from the model run urges the user to cite these papers in any resulting output.

This functionality is provided by a module that allows developers to insert an ADS  bibliography code \citep{2000A&AS..143...41K} at the appropriate point in the source code, e.g.: \\
{\tt call addBibcode("1999A\&A...344..282L", \\
"Leon Lucy's radiative equilibrium algorithm")}\\
The above adds the ADS code to the list that needs to be cited. A database of BibTex entries is hosted in the \torus\ data directory, and the appropriate BibTex entry is added to the output list. When a new ADS code is added to the source (i.e. one that does not appear in the database of BibTex entries) an ADS search is automatically invoked (if an internet connection is available, and the unix {\tt curl} command is present) that finds the appropriate BibTex text and inserts it into the database.
This means that the developer only has to add the appropriate ADS bibliography code to the source code, and not the full BibTex entry.

\section{Summary}
%perhaps should mention to users to cite this paper and, for feature X, paper Y. 
We have reviewed the features and applications of the \textsc{torus} Monte Carlo radiation transport and hydrodynamics code, including the implementation and validation of the numerical methods.

\textsc{torus} has the power to compute dust radiative equilibrium, photoionisation and atomic/molecular statistical equilibrium and photodissociation. It can also do this in a time dependent manner. Furthermore it has a hydrodynamics solver that can be coupled with the radiative transfer/composition microphysics to perform state of the art radiation hydrodynamics calculations. 

This paper should be regarded as the primary reference for people using the code. However we would encourage users to also cite the original work that presented the specific features of the code used, in order that they are properly credited. The underlying algorithms, and microphysical quantities (cross sections, rates etc) that underpin the code should also be cited where appropriate, and the references to those papers are generated automatically at run time.

The code is available on a collaborative basis, and we  encourage potential users of {\sc torus} to contact the authors for access to the git repository and for help in applying the code to their particular astrophysical problem.

\section*{Acknowledgments}

The authors wish to acknowledge the contributions of {\sc torus} developers who are no longer working in the field of astrophysics, in particular Neil Symington, Ryuichi Kurosawa, David Rundle, and Chris Reeves.

TJH acknowledges funding from Exeter's STFC Consolidated Grant (ST/M00127X/1). T. J. Haworth is funded by an Imperial College Junior Research Fellowship. 

The authors would like to acknowledge the use of the University of Exeter High-Performance Computing (HPC) facility in carrying out this work. Some of the calculations for this paper were performed on the University of Exeter Supercomputer, a DiRAC Facility jointly funded by the Science and Technology Facilities Council (STFC), the Large Facilities Capital fund of BIS and the University of Exeter. This work also used the DiRAC Complexity system, operated by the University of Leicester IT Services, which forms part of the STFC DiRAC HPC Facility (www.dirac.ac.uk). This equipment is funded by BIS National E-Infrastructure capital grant ST/K000373/1 and STFC DiRAC Operations grant ST/K0003259/1. DiRAC is part of the National E-Infrastructure. 

In addition, some of the calculations in this paper were run on COSMOS Shared Memory system at DAMTP, University
of Cambridge operated on behalf of the STFC DiRAC
HPC Facility. This equipment is funded by BIS National Einfrastructure capital grant ST/J005673/1 and STFC grants ST/H008586/1, ST/K00333X/1. \\

\noindent Part of this work used the DiRAC Data Analytics system at the
University of Cambridge, operated by the University of Cambridge
High Performance Computing Serve on behalf of the STFC DiRAC
HPC Facility (www.dirac.ac.uk). This equipment was funded
by BIS National E-infrastructure capital grant (ST/K001590/1),
STFC capital grants \\ ST/H008861/1 and ST/H00887X/1, and STFC
DiRAC Operations grant ST/K00333X/1. \\

The authors thank the Royal Astronomical Society for hosting a writing retreat at Burlington House, during which much of this paper was written.

\bibliographystyle{mn2e}\biboptions{authoryear}
%\bibliographystyle{elsarticle-num-names}\biboptions{authoryear}
%\bibliographystyle{elsarticle-harv}\biboptions{authoryear}
%\bibliographystlye{model2-names}\biboptions{authoryear}
\bibliography{bibliography2}

\bigskip

\newpage
\appendix

\clearpage
\onecolumn
\section{TORUS in the refereed literature}

%\onecolumn    
%\begin{table}
%    \begin{tabular}{|l | l|}
%    \hline

\begin{longtable}{|l|l|}
    \caption{A list of refereed papers that have made use of \torus. The first column provides the paper reference, the second a sentence outlining how the code was used in that paper. The reference are grouped by the type of application, and within each group the papers are ordered chronologically.}
    \label{torusPapers}\\
%    \toprule
    \hline
    Reference & Summary of \torus\ use  \\
    \hline
 %   \toprule
    %\midrule
    \hline
    \multicolumn{2}{|c|}{Atomic line transfer} \\
    \hline
    \cite{2000MNRAS.315..722H} &  Original code description and spectropolarimetry of H$\alpha$ in O-supergiants\\
    \cite{2001MNRAS.326.1265D} & Electron-scattering in discs applied to $\beta$~Cep \\
    \cite{2002MNRAS.333...55D} & Line profile models of $\theta^1$~Ori C \\
    \cite{2002MNRAS.337..341H} & Spectropolarimetric H$\alpha$ profiles of O-supergiants \\
    {\cite{2005A&A...430..213V}} & Spectropolarimetric line profiles from scattering off discs \\
    \cite{2005MNRAS.356.1489S} & Line profiles from structured CTTS magnetospheres \\
    \cite{2005MNRAS.358..671K} & Modelling of magnetosphere of SU~Aur \\
    \cite{2006MNRAS.370..580K} & Magnetosphere plus wind H$\alpha$ models of CTTS \\
    \cite{2008MNRAS.385.1931K} & Line profile simulation post-processing of MHD magnetosphere models \\
    \cite{2011MNRAS.416.2623K} & H and He line profiles from CTTS \\
    \cite{2012MNRAS.426.2901K} & Line profile simulations of the winds of CTTS \\
     {\cite{2012A&A...541A.116A}} & Magnetospheric accretion models \\
    \cite{2014MNRAS.442.3643P} & Line profile modelling of pre-FUor star V1331 Cyg \\
     \cite{2014MNRAS.443.1022E} & Modelling of CTTS AA~Tau \\
     \cite{2016MNRAS.456..156G} & Near-IR line profile modelling of Herbig star VV~Ser \\
     \cite{2016MNRAS.457.2236K} & Line profile modelling of Herbig Be star HD~58647 \\

    \hline
    \multicolumn{2}{|c|}{Molecular line transfer} \\
    \hline
    \cite{2010MNRAS.407..986R} & Description of molecular line RT module and cluster simulation post-processing \\
    \cite{2016MNRAS.461..385B} & Molecular line models of the transition disc HD~163296 \\
    \cite{2019MNRAS.482.4673J} & Synthetic observations of self-gravitating discs around massive YSOs \\
   
    \hline
    \multicolumn{2}{|c|}{Dust continuum models} \\
    \hline
    
    {\cite{2000A&A...361..273H}} & Dust scattering in Wolf-Rayet/O-star binary WR137.\\
    \cite{2004MNRAS.350..565H} & Model images of Wolf-Rayet/O-star binary WR104 \\
    
     \cite{2007ApJ...661..374T} & Disc inner rim models including dust settling \\
    {\cite{2007A&A...468.1009H}} & Models of class~0 sources \\
       \cite{2008ApJ...677L..51T} & Inner disc models of MWC~275 and AB~Aur \\
    \cite{2008ApJ...689..513T} & Imaging and SED modelling of MWC~275 and AB~Aur \\
    {\cite{2009A&A...498..967P}} & Optically thick dusty disc benchmark models \\ 
  \cite{2010MNRAS.409.1307M} & Brown dwarf circumstellar disc models \\
    \cite{2011MNRAS.416.1500H} & Time-dependent radiative transfer method \\
     \cite{2012MNRAS.423.1775M} & Fitting brown dwarf photometry with \torus\ models \\
    {\cite{2013A&A...557A..35V}} & Disc and envelope radiative transfer \\
      \cite{2014Sci...345.1590C} & Disc radiative-equilibrium modelling \\
      \cite{2014ApJ...794..123C} & Radiation-equilibrium models of protostellar discs \\
       \cite{2015PNAS..112.8965B} & Radiative-equilibrium models of protostellar discs \\
        \cite{2015ApJ...799..204C} & Radiative-equilibrium models of TW~Hya protostellar disc \\
    \cite{2015ApJ...807....2C} & Disc RT models including embedded protoplanet \\
   \cite{2016ApJ...816L..21C} & Radiative equilibrium modelling of protostellar discs \\
   
    \cite{2016MNRAS.458..306H} & Synthetic observations of spirals in protostellar discs  \\
    \cite{2016ApJ...819...13C} & Disc radiative equilibrium modelling \\
     \cite{2017ApJ...838...20M} & Modelling of transitional discs \\
    \cite{2018MNRAS.473..317H} & Modelling sub-Keplerian rotation/observables of the disc about the AGB star L$_2$ Pup.\\
    \cite{2018arXiv180810762D} & Modelling of Herbig Ae star HD~142666 \\

    \hline
     \multicolumn{2}{|c|}{Radiation hydrodynamics} \\
     \hline
     \cite{2010MNRAS.403.1143A} & Dusty disc radiation hydrodynamics using SPH and \torus\ \\
    
     \cite{2012MNRAS.420..562H} & Radiation hydrodynamics method and triggered star formation \\
      \cite{2015MNRAS.448.3156H} & Radiation hydrodynamics methods including radiation pressure and sink particles \\
       \cite{2015MNRAS.453.1324B} & D-type H\,{\sc ii} expansion benchmarks \\
  \cite{2015MNRAS.453.2277H} & Models of H\,{\sc ii} region expansion \\
  \cite{2015MNRAS.454.2828B} & \textsc{torus-3dpdr} code paper \\
  \cite{2016ApJ...832..110C} & Radiative-equilibrium models of IM Lup's protostellar disc \\
  \cite{2016MNRAS.457.1905H} & RHD models of internal X-ray driven clearing of discs \\
  \cite{2016MNRAS.463.3616H} & Photochemical-dynamical models of externally FUV irradiated protoplanetary discs \\
  \cite{2017MNRAS.468L.108H}  & Models of external evaporation of IM Lup \\
 \cite{2017MNRAS.471.4111H} & RHD models of high-mass star formation \\
 \cite{2018MNRAS.477.5422A} & Photoionisation and radiation pressure feedback in clusters \\
 \cite{2018MNRAS.475.5460H} & RHD models of Trappist-1 precursor discs \\
 \cite{2018MNRAS.481..452H} & A large grid of RHD models of  externally irradiated protoplanetary discs \\

    \hline
  \multicolumn{2}{|c|}{Post-processing of hydrodynamical simulations} \\
   \hline
   \cite{2004MNRAS.351.1134K} & Synthetic images and photometry of an SPH cluster simulation \\
      \cite{2010MNRAS.407..986R} & Description of molecular line RT module and cluster simulation post-processing \\
      \cite{2012MNRAS.426..203H} & Synthetic observations of triggered star formation \\
       \cite{2013MNRAS.431.2673K} & Line profile post-processing of MHD simulations of CTTS \\
     \cite{2013MNRAS.431.3470H} & Molecular line synthetic observations of triggered star formation \\ 
       \cite{2014MNRAS.444..919P} & Synthetic CO maps from galaxy scale SPH simulations \\
       \cite{2015MNRAS.446.3608D} & Synthetic CO observations of cloud-cloud collisions from SPH simulations \\
        \cite{2015MNRAS.449.3911P} & Synthetic CO line observations of SPH galaxy simulations \\
    \cite{2015MNRAS.450...10H} & Synthetic molecular line observations of cloud-cloud collisions \\
    \cite{2015MNRAS.454.1634H} & Further synthetic molecular line observations of cloud-cloud collisions \\
      \cite{2015MNRAS.447.2144D} & Synthetic molecule line surveys from galaxy-scale SPH simualtions \\
       \cite{2015MNRAS.449.1996D} & Synthetic observations of SPH models of a star-disc encounter \\
     \cite{2016MNRAS.456..136A} & Photoionisation and bow-shocks from runaway stars \\
    {\cite{2016A&A...586A.114M}} & Photoionisation and stellar-wind bubbles \\
     \cite{2016MNRAS.458.3667D} & Molecular line post-processing of SPH simulations of giant molecular clouds \\
     \cite{2017MNRAS.466.1857G} & Photoionisation models of a wind blown bubble \\
     \cite{2018MNRAS.477.1004H} & Spiral morphology of the Elias 2-27 protostellar disc \\
     \cite{2018MNRAS.474..800Y} & SED models of SPH first-core simulations \\
    \hline

\end{longtable}
\clearpage
\twocolumn
    
%    \end{tabular}
%    \caption{Caption}
%    \label{tab:literature}
%\end{table}
%\twocolumn    

\end{document}